\documentclass[openany]{report}
\usepackage{graphicx}
\usepackage{natbib}
\usepackage{xcolor}
\usepackage{hyperref}

\title{{\tt BIGSTICK}: A flexible configuration-interaction shell-model code \footnote{UCRL number: LLNL-SM-739926}} 

\author{Calvin W. Johnson\footnote{Department of Physics, San Diego State University, 5500 Campanile Drive, San Diego CA 92182-1233},
W. Erich Ormand\footnote{Lawrence Livermore National  Laboratory, P.O. Box 808, L-414,
Livermore, CA 94551},  
Kenneth S. McElvain\footnote{Department of Physics, University of California, Berkeley 366 Leconte Hall MC 7300, Berkeley, CA 94720},  \\
Ryan Zbikowski \footnote{Computational Science Research Center, San Diego State University, 5500 Campanile Drive, San Diego CA 92182},
and Hongzhang Shan\footnote{Computational Research Division, Lawrence Berkeley Laboratory, Berkeley, CA, 94720}}

\begin{document}
\maketitle


\begin{abstract}
We present {\tt BIGSTICK}, a flexible configuration-interaction open-source shell-model code for the many-fermion problem. 
Written mostly in Fortran 90 with some later extensions, {\tt BIGSTICK} utilizes a factorized on-the-fly algorithm for computing many-body matrix elements, 
and has both MPI (distributed memory) and OpenMP (shared memory) parallelization, and can run on platforms ranging from laptops to
the largest parallel supercomputers. It uses a flexible yet efficient many-body truncation scheme, and reads input files in multiple formats, 
allowing one to tackle both phenomenological (major valence shell space) and \textit{ab initio} (the so-called \textit{no-core shell model}) calculations.
{\tt BIGSTICK} can generate energy spectra, static and  transition   one-body densities, and expectation values of scalar operators.  
Using the built-in Lanczos algorithm one can compute transition probability distributions and decompose wave functions into components defined 
by group theory. 

This manual provides a general guide to compiling and running {\tt BIGSTICK}, which comes with numerous sample input files, as well as 
some of the basic theory underlying the code. This manual also provides some, though not all, details into the inner workings. 

This code is distributed under the MIT Open Source License. 
The source code and sample inputs are found at 
 {\tt github.com/cwjsdsu/BigstickPublick}.

\end{abstract}

\tableofcontents


\chapter{Introduction}


There are many approaches to the quantum many-body problem.
{\tt BIGSTICK} is a configuration-interaction many-fermion code,
written in Fortran 90. It solves 
for low-lying eigenvalues of the Hamiltonian of a many fermion system;
it does this by creating a basis of many-body states of Slater determinants (actually, the occupation representation
of Slater determinants). The Slater determinants are antisymmetrized products of single-particle states with 
good angular momentum, typically derived from some shell-model-like potential; hence we call this a shell-model basis.
The Hamiltonian is assumed to be rotationally invariant and to conserve parity, 
and is limited to two- and, optionally, three-body forces. Otherwise
no assumptions are made about the form of the single-particle states or 
of the Hamiltonian.

The capabilities of {\tt BIGSTICK} will be detailed below, but in addition to calculating the
energy spectra and occupation-space wavefunctions, it can compute particle occupations, 
expectation values of operators, and static and transition densities and strengths.
Most of the applications to date have been in low-energy nuclear physics, but in principle any
many-fermion system with two fixed `species' and rotational symmetry can be addressed 
by {\tt BIGSTICK}, such as the electronic structure of atoms and cold fermionic gases in a spherically symmetric
trap; although we have yet to publish papers, we have carried out demonstration calculations 
for such systems, with `spin-up' and `spin-down' replacing `proton' and `neutron.'  We apologize to 
any atomic physicist who will have to translate our terminology.

In this next chapter we review the basic many-body problem. Chapter \ref{CI} outlines  the configuration-interaction method and
discusses in broad strokes the principles of the algorithms in {\tt BIGSTICK}.  Chapter \ref{quickstart} gives
an introduction to how to compile and run {\tt BIGSTICK}, while Chapter \ref{detailruns} 
goes into running the code more detail. \textbf{If you are interested in running 
{\tt BIGSTICK} immediately, go directly to Chapter \ref{quickstart}}.

In this manual we do not give substantial information on the inner workings of the code, although some details are outlined in Sec.~\ref{mpi} on MPI paralleization.  Some of the terminology is explained in the glossary, Appendix \ref{glossary}
The code itself is heavily commented.   While internal information 
in {\tt BIGSTICK} is highly compressed through factorization, a technique outlined in Chapter \ref{CI}, it is possible to get out explicit representations of the many-body 
basis states and the many-body Hamiltonian matrix; see Chapter \ref{peek}. Chapter \ref{lanczos} discusses our use of the Lanczos algorithm.

Finally, parallel capabilities of the code is discussed in Chapter \ref{parallel}.

\section{Expectations of users}

Who do we expect to use {\tt BIGSTICK}, and how do we expect them to use it? 
We designed {\tt BIGSTICK} to be run on a variety of platforms, from laptops to leadership-class supercomputers. 
We also imagined, and tried to design, {\tt BIGSTICK}  for a spectrum of users, with various expectations of them.

A crucial point for any and all users: 
{\tt BIGSTICK} \textbf{requires at least two kinds of input files to run}, a description of 
the single-particle space and a file of interaction matrix elements. While we supply with the distribution a number of example 
input files. it is important for both novice and routine users to understand that such examples are just the \textbf{beginning} and 
not the sum of nuclear physics. \textbf{In general it is up to the user to provide interaction files.}  We can use the {\tt .int} interaction files usable by 
{\tt NuShell/NuShellX} as well as the interaction files used by {\tt MFDn} (Sec.~\ref{MFDinput}) and {\tt NuHamil} (Sec.~\ref{nuhamil}).

It is also equally important to not ask {\tt BIGSTICK} to be smarter than you are.  While {\tt BIGSTICK} 
employs many error traps to avoid or at least flag the most common  mistakes, the principle of ``garbage in, garbage out'' still applies.

While this manual provides a fairly comprehensive introduction to running  {\tt BIGSTICK},  it is 
\textit{not} a detailed tutorial in configuration-interaction methods, the atomic or nuclear shell models, 
or to basic nuclear physics. We  expect the reader to, above all, be comfortable with non-relativistic 
quantum mechanics (i.e., to fully understand the Schr\"odinger equation and with Dirac's bra-ket notation), and to be fluent
of the ideas and terminology of the shell model, especially the nuclear shell model, and to understand 
the basic principles of configuration-interaction methods. We review the latter in the opening of 
Chapter \ref{CI}, so that is a good place to start to check your level of comfort. We  suggest additional 
references in Appendix \ref{shellmodelrefs}.

\section{How to cite and copyright notices/licenses}
\label{CClicense}

 If you successfully use {\tt BIGSTICK} in your research, please use the following 
 citations:
 
\noindent $\bullet$ C. W. Johnson, W. E. Ormand, and P. G. Krastev,  Comp. Phys. Comm. \textbf{184}, 2761-2774 (2013). (You can also find 
this article at arXiv:1303.0905.)

\noindent $\bullet$  C. W. Johnson,W. E. Ormand, K. S. McElvain, R. Zbikowski, and H. Z. Shan, arXiv:1801.08432v2 (this updated report)

The first paper, \citet{BIGSTICK}, in particular discusses the underlying factorized on-the-fly algorithm.  This documents focuses instead
on how to run {\tt BIGSTICK}.

\bigskip

This code is distributed under the MIT Open Source License:

\bigskip

Copyright (c) 2017 Lawrence Livermore National Security and  the San Diego State University Research Foundation.

\medskip

Permission is hereby granted, free of charge, to any person obtaining a copy of this software and associated documentation files (the "Software"), to deal in the Software without restriction, including without limitation the rights to use, copy, modify, merge, publish, distribute, sublicense, and/or sell copies of the Software, and to permit persons to whom the Software is furnished to do so, subject to the following conditions:

\medskip

The above copyright notice and this permission notice shall be included in all copies or substantial portions of the Software.

\medskip

THE SOFTWARE IS PROVIDED "AS IS", WITHOUT WARRANTY OF ANY KIND, EXPRESS OR IMPLIED, INCLUDING BUT NOT LIMITED TO THE WARRANTIES OF MERCHANTABILITY, FITNESS FOR A PARTICULAR PURPOSE AND NONINFRINGEMENT. IN NO EVENT SHALL THE AUTHORS OR COPYRIGHT HOLDERS BE LIABLE FOR ANY CLAIM, DAMAGES OR OTHER LIABILITY, WHETHER IN AN ACTION OF CONTRACT, TORT OR OTHERWISE, ARISING FROM, OUT OF OR IN CONNECTION WITH THE SOFTWARE OR THE USE OR OTHER DEALINGS IN THE SOFTWARE.

\subsection{LAPACK copyright notice}

We use LAPACK subroutines in our code.  The following are the LAPACK copyright notices.

\smallskip

Copyright (c) 1992-2013 The University of Tennessee and The University
                        of Tennessee Research Foundation.  All rights
                        reserved.
\smallskip
                        
Copyright (c) 2000-2013 The University of California Berkeley. All
                        rights reserved.

\smallskip

Copyright (c) 2006-2013 The University of Colorado Denver.  All rights
                        reserved.

\smallskip

Additional copyrights may follow

\bigskip

Redistribution and use in source and binary forms, with or without
modification, are permitted provided that the following conditions are
met:

\smallskip

\noindent - Redistributions of source code must retain the above copyright
  notice, this list of conditions and the following disclaimer.

\smallskip

\noindent - Redistributions in binary form must reproduce the above copyright
  notice, this list of conditions and the following disclaimer listed
  in this license in the documentation and/or other materials
  provided with the distribution.
\smallskip

\noindent 

- Neither the name of the copyright holders nor the names of its
  contributors may be used to endorse or promote products derived from
  this software without specific prior written permission.

\bigskip

The copyright holders provide no reassurances that the source code
provided does not infringe any patent, copyright, or any other
intellectual property rights of third parties.  The copyright holders
disclaim any liability to any recipient for claims brought against
recipient by any third party for infringement of that parties
intellectual property rights.

\smallskip

THIS SOFTWARE IS PROVIDED BY THE COPYRIGHT HOLDERS AND CONTRIBUTORS
"AS IS" AND ANY EXPRESS OR IMPLIED WARRANTIES, INCLUDING, BUT NOT
LIMITED TO, THE IMPLIED WARRANTIES OF MERCHANTABILITY AND FITNESS FOR
A PARTICULAR PURPOSE ARE DISCLAIMED. IN NO EVENT SHALL THE COPYRIGHT
OWNER OR CONTRIBUTORS BE LIABLE FOR ANY DIRECT, INDIRECT, INCIDENTAL,
SPECIAL, EXEMPLARY, OR CONSEQUENTIAL DAMAGES (INCLUDING, BUT NOT
LIMITED TO, PROCUREMENT OF SUBSTITUTE GOODS OR SERVICES; LOSS OF USE,
DATA, OR PROFITS; OR BUSINESS INTERRUPTION) HOWEVER CAUSED AND ON ANY
THEORY OF LIABILITY, WHETHER IN CONTRACT, STRICT LIABILITY, OR TORT
(INCLUDING NEGLIGENCE OR OTHERWISE) ARISING IN ANY WAY OUT OF THE USE
OF THIS SOFTWARE, EVEN IF ADVISED OF THE POSSIBILITY OF SUCH DAMAGE.
 
 \section{Reporting bugs and other issues}
 
 If you run into trouble, \textit{first read this manual}. Most issues are caused by mistakes in setting up input files, in particular inconsistencies 
 between the single-particle space defined and the interaction file(s).  Second,  please  \textit{read the output carefully}: we have striven to write detailed error traps and often {\tt BIGSTICK} will 
 notify the user of problems.   Try running the sample cases and make sure they run to correct completion 
 and that you understand the inputs. 
 
 If, having exhausted all the resources found here, you still have a problem, you may send your issue to Calvin Johnson, 
 {\tt cjohnson@sdsu.edu}.   In particular send a 
 copy of the \textit{entire} output written to screen, which often contains important clues, the input files, and all output files with the 
 extensions {\tt .res}, {\tt .log} and {\tt .bigstick}.  Although we hope to be able to help, we cannot guarantee it. 
 
 As discussed elsewhere, {\tt BIGSTICK} is developed for Linux and Linux-like environments such as Mac OS X.  We have made no 
 attempt to adapt to a Windows environment. Although it has a user-friendly 
 menu-driven interface, it still assumes a reasonable facility with many-body physics and in particular low-energy nuclear physics.

 Development of {\tt BIGSTICK} is ongoing. We hope to release future versions of the code as additional major capabilities come on line.

 \section{What's new in each version}
 
 {\tt BIGSTICK} is highly versioned, with a three-number-code for each version, i.e., 7.11.4 (as of this writing). This allows us to 
 track bugs that may arise, especially those introduced accidentally.  Many of the output files, such as the {\tt .res} and {\tt .log} files, include the version number, and 
 if you want to report a problem, you should mention that too.

   In addition, the {\tt BIGSTICK} distribution includes  a text file, {\tt WHATSNEW.txt}, which lists the most recent developments in the code, as well as 
   reflecting the historical development. You should consult it especially if you received an update. 
  
 \section{A brief history of {\tt BIGSTICK}, and acknowledgements}
 
 In 1997, when two of us (Ormand and Johnson) were both at Louisiana State University, we decided to write 
our own many-fermion configuration-interactionI code 
 christened {\tt REDSTICK},  English for \textit{Baton Rouge}.
Over the next decade {\tt REDSTICK} evolved and improved.  Most important 
were the addition of three-body forces and 
parallelization.  As it approached the ten-year mark, we noticed certain limitations, 
particularly in the set-up, and starting in 2007 we began developing new algorithms. 

By this time, Ormand had moved to Lawrence Livermore National Laboratory and Johnson had left for
San Diego State University.
Working  first  with a student (Hai Ah Nam) and later a postdoc (Plamen Krastev) at San Diego State 
University,
we carefully studied bottlenecks in parallelization in the application of the Hamiltonian. 
These studies led us to break up the application of the Hamiltonian by basis sectors, defined by 
quantum numbers such as $J_z$ and $\pi$ (parity),  which 
had two useful outcomes. First, we rewrote our central application 
routines using simple arrays rather than the derived types used in {\tt REDSTICK}; this gave 
a speed-up of nearly a factor of 2.  Second, applying the Hamiltonian by quantum numbers  
allowed a more transparent factorization of the Hamiltonian and  better 
parallelization. 

With these improvements and dramatic speed-ups, we had an entirely new code,  {\tt BIGSTICK}. 

Starting around 2014, through the good graces of Wick Haxton we teamed up with UC Berkeley and 
Lawrence Berkeley Laboratory, and especially Haxton's graduate student Ken McElvain. 
Ken's background in the computer industry proved invaluable, and he was able to tweak the 
existing code into fantastic performance, especially with regards to parallelism.  Hongzhang Shan of Lawerence Berkeley 
wrote an improved algorithm for using OpenMP in matvec operations, and Ryan Zbikowski, a Ph.D student at 
SDSU's Computational Science Research Center,  prototyped and implemented 
the block Lanczos algorithm.

In addition to the help of Hai Ah Nam and  Plamen Krastev, 
we would also like to thank  Esmond Ng, Chao Yang, and Sam Williams, of Lawrence Berkeley 
National Laboratory,  James Vary and Pieter Maris of Iowa State University, 
and many other colleagues who have provided helpful discussions,
suggestions, feedback and insight over the years.  Jordan Fox helped find some bugs, and Stephanie 
Lauber helped find typos and confusing statements. Mark Caprio also contributed thoughtful and clarifying feedback on the manual. Dmitriy Rodkin also identified a number of bugs, for which I am grateful.
 
 Over the years our primary research funding has come through the U.S. Department of Energy, which has directly and often indirectly 
 supported the development of {\tt BIGSTICK}. We are deeply grateful for this support.  Support for this project came primarily from the U.S. Department of 
Energy, in the form of grants Grant  DE-FG02-96ER40985,  DE-FG52-03NA00082,  DE-FG02-03ER41272, as well as Louisiana State University, Lawrence Livermore National
Laboratory,  San Diego State University, University of California, Berkeley, and 
Lawrence Berkeley National Laboratory. 

\section{Available post-processing codes}

There are a number of codes available that mesh with {\tt BIGSTICK}, also available on github:

\begin{itemize}
    \item In the {\tt github.com/cwjsdsu/BigstickPublick/util/} folder, there are a number of utilities for processing transition densities. 

    \item For the same of efficiency, {\tt BIGSTICK} can only handle one basis at a time, with fixed proton and neutron number and $J_z$.
    A post-processing code, {\tt RHODIUM}, can compute one-body spectroscopic amplitudes, charge-changing one-body densities, and one-body densities between different initial and final $J_z$ and/or parity. It can be downloaded from the {\tt /cwjsdsu/Rhodium} repository, including a manual; the latter can also be found at arXiv:2510.23787.

    \item The {\tt tracer} code, found at {\tt /cwjsdsu/tracer}, efficiently computes configuration centroids; it can use this information to optimize the orbital weights for sophisticated many-body truncations.  A paper will be posted soon.
    
\end{itemize}

\section{This version}

{\tt BIGSTICK} versions are given a three-number code. This manual is (mostly) up-to-date for Version 8.0.0.

Here we summarize some of the changes.

\begin{itemize}
    \item The directory structure has been updated and modernized, as well the compiler directives for MPI. 

    \item A number of bugs have been fixed. Some additional bugs were introduced, then fixed. 

    \item Parity of converged states is now labeled (but be aware--if a state is not fully converged, the declared parity could be wrong).

    \item We have introduced block Lanczos as a diagonalization option. 

    \item We have introduced the capability to compute Green's functions (resolvants) with a choice of several options; see manual.

    \item Two-body densities have been introduced, and in particular, diagonal (same initial and final state) densities.

      \item {\tt BIGSTICK} can use interactions generated by the {\tt NuHamil} code, (T. Miyagi, EPJA \textbf{59}:150, (2023), or arXiv:2302.07962), using the MFDn format. See the {\tt BIGSTICK} manual for instructions.

    \item Options to create human-readable {\tt .trwfn} files from binary {\tt .wfn} files, and vice versa, added, as options `{\tt (tx)}', `{\tt (tw)};' useful for users who wish to add capabilities.

    \item Options to compute configuration probabities added: `{\tt (co)}' (in current run) and `{\tt (cx)}' (from previous {\tt .wfn} file).

    \item Some efficiencies improved, in particular when generating a very large number (hundreds or thousands) of converged states.

    \item A number of new options, mostly for specialized tasks, have been added.

\end{itemize}

In addition, two other codes have been released.  

{\tt RHODIUM} is a postprocessing code for {\tt BIGSTICK} wave functions. It can, for example, compute spectroscopic amplitudes, charge-changing densities, and so on. These are not possible in {\tt BIGSTICK}, which has a fixed basis that conserves proton and neutron number, for example.  The code is found at {\tt github.com/cwjsdsu/Rhodium}, and the full manual, available also with the {\tt RHDOIUM} distribution, is at arXiv:2510.23787.

The other code is {\tt tracer}, which computes configuration probabilities. It can also be used to optimize the single-particle orbital \textit{weights}, for an approximate centroid energy (ACE) truncation scheme.
The code is found at {\tt github.com/cwjsdsu/tracer}.
A paper is at arXiv:2511.03161.


\chapter{How we solve the many-body problem}
\label{CI}

In this chapter we  discuss the principles of configuration-interaction (CI) many-body 
calculations \citep{Sh98,BG77,br88,ca05,cook1998handbook,jensen2017introduction,PhysRev.122.1826,PhysRev.97.1474,sherrill1999configuration}, including some different classes of CI codes, and give an overview of its application in {\tt BIGSTICK}. 
 Configuration-interaction is sometimes called the interacting shell model, as (a) one typically 
builds the many-body basis from spherical shell-model single particle states and (b) to distinguish from the non-interacting shell model, sometimes also called 
the independent particle model. 

The key points here are:

\begin{itemize}

\item We represent the many-body Schr\"odinger equation as a  matrix eigenvalue problem, typically with very large basis dimensions. {\tt BIGSTICK} 
can compute problems with dimensions up to $\sim 10^7$ on a laptop, up to $\sim 10^8$ on a desktop machine, and up to $\sim 10^{10}$ on 
parallel supercomputers. 

\item The large-basis-dimension eigenvalue problem has two computational barriers. The first is how to solve the eigenvalue problem itself, especially given that we almost never need 
\textit{all} of the eigenvalues. The second 
is, despite the fact the matrix is typically very sparse,
the amount of data required is still huge. 

\item We address the first problem by using the Lanczos algorithm, which efficiently yields the low-lying eigenpairs. 

\item We address the second by not explicitly storing all the non-zero matrix elements, but instead invoking a on-the-fly algorithm. This on-the-fly 
algorithm, first implemented in the Strasbourg group's code {\tt ANTOINE} (\citep{ANTOINE}), exploits the fact that the interaction only acts on two- or three- particles 
at a time. The on-the-fly algorithm can be thought of as partially looping over spectator particles.

\item The on-the-fly algorithm explicitly depends upon the existence of two species of particles, for example protons and neutrons, or in the case of 
atoms, spin-up and spin-down electrons, so that both the many-body basis and the action of the Hamiltonian can be factorized into two components. 
This factorization is guided by additive/multiplicative quantum numbers, such as $M$, the $z$-component of angular momentum, and parity.
This factorization efficiently and losslessly ``compresses'' information; we outline the basic concepts below.

\item In order to implement many-body truncations, we have an additional additive pseudo-quantum number, which we call $W$ . This allows a general, though 
not infinitely flexible, ability to truncate the basis.  We discuss these truncations below, but include for example $n$-particle, $n$-hole truncations and 
the $N_\mathrm{max}$ truncation typical of the no-core shell model. 

\end{itemize}

With these efficiencies we can run both ``phenomenological'' and \textit{ab initio} or no-core shell model calculations, on machines ranging from 
laptops to supercomputers.   Although we do not discuss it in depth in this document, we rely heavily upon both factorization and use of quantum 
numbers in parallelization.

\section{Matrix formulation of the Schr\"odinger equation}

The basic goal is to solve the non-relativistic many-body Schr\"odinger equation for $A$ identical fermions of mass $M$,
\begin{equation}
\left ( \sum_{i=1}^A - \frac{\nabla_i^2}{2M} + \sum_{i < j} V(\vec{r}_i - \vec{r}_j) \right) 
\Psi(\vec{r}_1, \vec{r}_2,\ldots,\vec{r}_A) = E \Psi(\vec{r}_1, \vec{r}_2,\ldots,\vec{r}_A),
\label{Schrodinger}
\end{equation}
which we often will write using the more compact Dirac bra-ket notation
\begin{equation}
\hat{H} | \Psi \rangle = E | \Psi \rangle.
\end{equation}
Already even Eq.~(\ref{Schrodinger}) is simplified, as it leaves out explicit spin degrees of freedom, and the potential 
here is purely local and two-body. {\tt BIGSTICK} can handle nonlocal interactions without blinking. {\tt BIGSTICK} can also 
use
three-body forces, although the latter ups computational demands by nearly two orders of magnitude, and in the current release 
the three-body forces are not optimized. 


The basic idea of configuration interaction is to expand the wavefunction in some convenient many-body basis 
$\{ | \alpha \rangle \}$:
\begin{equation}
| \Psi \rangle = \sum_\alpha c_\alpha | \alpha \rangle
\end{equation}
Then, if the basis states are orthonormal, $ \langle \alpha | \beta \rangle = \delta_{\alpha, \beta}$, the Schr\"odinger equation becomes a matrix eigenvalue equation
\begin{equation}
\sum_{\beta} H_{\alpha,\beta} c_\beta = E c_\alpha.
\label{matrixeigenvalue}
\end{equation}

Because we typically deal with many-fermion systems, the wavefunction $|\Psi\rangle$ is completely antisymmetric under interchange 
of any two particles,
\begin{equation}
\Psi(\vec{r}_1, \vec{r}_2, \ldots, \vec{r}_i, \ldots, \vec{r}_j, \ldots) = - \Psi(\vec{r}_1, \vec{r}_2, \ldots, \vec{r}_j \ldots, \vec{r}_i, \ldots). 
\end{equation}
(One can use configuration-interaction methods for many-boson systems, but then the basis states would be totally symmetric, and a 
completely separate code would be required.) 
A useful many-body basis are therefore Slater determinants, which are antisymmetrized products of single-particle wavefunctions. 
(As we will note several times in this manual, it is often important to distinguish between \textit{single-particle} states and \textit{many-body} 
states, as well as between, for example, \textit{two-body} matrix elements and \textit{many-body} matrix elements.)  

We do not explicitly use Slater \textit{determinants} but rather  the \textit{occupation representation} of Slater determinants using fermion 
creation and annihilation operators, also known as second quantization.  We assume the reader is comfortable with Slater determinants and the algebra of fermion operators, 
and therefore give only a terse exposition in order to be clear about our terminology.

Suppose we have a set of $N_s$ single-particle states, $\phi_i(\vec{r})$ where $i$ describes each unique state by its quantum numbers.
{\tt BIGSTICK} assumes single-particle states with rotational symmetry, 
and the available quantum numbers are $n$, $l$, $j$, and $m$. Here $l$ is the orbital angular momentum, $j$ is the total 
angular momentum, and $m$ is the $z$ component of total angular momentum.  $n$ is the `radial' quantum number; it distinguishes different 
states with the same angular momentum quantum numbers but a different radial wavefunction. It plays no other internal role in {\tt BIGSTICK}, 
though it is relevant to calculating the value of matrix element input into the code.  {\tt BIGSTICK} 
can use single-particle states with arbitrary radial components, as long as they orthonormal; it is up to the user to keep track of what radial 
wavefunction is being assumed. In many cases, for example in so-called no-core shell model (NCSM) calculations, one uses a harmonic oscillator basis, 
but that is by no means mandatory.  
In the same way, $l$ really only gives the parity of each single-particle state.

Once a single-particle basis is defined, second quantization allows us to define many-body states.
Starting with a fermion vacuum state $| 0 \rangle$, 
the operator $\hat{a}^\dagger_i$ creates the single fermion state $\phi_i$.  Then the many-body state 
\begin{equation}
\hat{a}^\dagger_{i_1} \hat{a}^\dagger_{i_2} \hat{a}^\dagger_{i_3} \ldots | 0 \rangle
\end{equation}
is the occupation representation of the Slater determinant of the single particle states $\phi_{i_1}, \phi_{i_2}, \ldots$.   For succinctness we will 
refer to such many-body states as `Slater determinants' even when we mean the occupation representation.

Using  a one-body operator such as the kinetic energy $\hat{T}$ can be written using second quantization:
\begin{equation}
\hat{T}  = \sum_{ij}T_{ij} \hat{a}^\dagger_i \hat{a}_j, 
\end{equation}
where $T_{ij} = \langle i | \hat{T} | j \rangle = \int \phi_i^* \hat{T} \phi_j $ is the one-body matrix element of the operator; the actual value is determined through the integral sketched above. 
Two-body operators, e.g. interactions between two particles, can be similarly represented, though with two annihilation operators followed by two creation operators. 
It is useful to note that all {\tt BIGSTICK} and similar CI codes read in are  numerical values of the matrix elements. This means the actual form of the single-particle 
wavefunctions is hidden (although {\tt BIGSTICK}, like nearly all other nuclear CI codes, requires  single-particle states to have good angular momentum). 

The many-body matrix elements are thus exercises in fermion second quantization algebra: $H_{\alpha \beta} = \langle \alpha | \hat{H} | \beta \rangle$ 
where the basis states $|\alpha \rangle, |\beta\rangle$ and the Hamiltonian operator $\hat{H}$ are all expressed using creation and annihilation operators, 
given exactly in Appendix \ref{ops2ndquant}.

\section{Representation of the basis}

 The occupation representation is a 
natural one for the computer as a single particle state can either be occupied or unoccupied, represented by a 1 or a 0. Thus the state
$$
\hat{a}^\dagger_2 \hat{a}^\dagger_4 \hat{a}^\dagger_5 \hat{a}^\dagger_8 | 0 \rangle
$$
can be represented by the bit string
$$
01011001
$$
as the single particle states 2,4, 5 and 8 are occupied and the rest unoccupied. 
Of course, consistency in ordering is important as one has to pay careful attention to phases.

 In the 1970s Whitehead and collaborators used bit manipulation for fast calculation of matrix elements in the occupation scheme (\citep{lanczos}). The basic idea is 
simple: consider a creation operator, say $\hat{a}^\dagger_4$, applied to some Slater determinant represented by a bit string. If the 4th bit is 0, then 
the action of applying $\hat{a}^\dagger_4$ is to create a 1 in its place:
$$
\hat{a}^\dagger_3 |110001 \rangle = | 110101 \rangle,
$$
while if it is already occupied, then the state vanishes in a puff of digital smoke:
$$
\hat{a}^\dagger_3 |100101 \rangle =0.
$$
Similarly an annihilation operator such as $\hat{a}_2$ will destroy a state if the second bit is empty
$$
\hat{a}_2 |100111 \rangle =0,
$$
but will replace a 1 bit with a 0,
$$
\hat{a}_2 |110101 \rangle =- |100101 \rangle.
$$
The minus sign arises, of course, from fermion anticommutation relations. In this way one can almost trivially find the action of, say, a two-body 
operator on a state:
$$
\hat{a}^\dagger_3 \hat{a}^\dagger_5 \hat{a}_4 \hat{a}_1 | 1 1 0 1 0 1 1 \rangle = - \hat{a}^\dagger_3 \hat{a}^\dagger_5 | 0 1 0 00 11 \rangle = - | 0 1 1 0 1 0 11 \rangle
.
$$
Then one can search through the basis to find out what state $| 0 1101011\rangle$ is.


In general we work with Hamiltonians which are rotationally invariant. This
means one can find eigenstates of the Hamiltonian which are also simultaneous eigenstates of total angular momentum 
$\hat{J}^2 = \hat{J}_x^2+ \hat{J}_y,^2+\hat{J}_z^2$ and of (by convention) $\hat{J}_z$, that is, 
\begin{eqnarray*}
\hat{H} | \Psi \rangle = E |\Psi \rangle; \\
\hat{J}^2 | \Psi \rangle = \hbar^2 J(J+1) |\Psi \rangle; \\
\hat{J}_z | \Psi \rangle = \hbar M |\Psi \rangle.
\end{eqnarray*}
We say such states `have good angular momentum.' It is important to note that $E$ generally depends upon $J$, that is, except for special cases (usually involving 
additional symmetries) states with different $J$ are not degenerate, for a given value of $J$ the value of $E$ does not depend upon $M$. 
In practical terms, what this means is that the Hamiltonian is block-diagonal in $J$; it is also block-diagonal in $M$, but the blocks for the same $J$ but different $M$ 
have the exact same eigenvalues. 

Of course, whether or not the Hamiltonian is explicitly block diagonal depends upon the choice of basis. We call these different choices basis `schemes.'

{\tt BIGSTICK}, like most nuclear CI codes, constructs its many-body basis states using single-particle states which also have good angular momentum, i.e., 
have eigenvalues $j(j+1)$ and $m$ for $\hat{J}^2$ and $\hat{J}_z$, respectively. (Here and hereafter we set $\hbar =1$.)  The addition of total angular momentum 
is nontrivial, requiring Clebsch-Gordan coefficient, but as $\hat{J}_z$ is the generator of an Abelian subgroup, any product of single-particle states each with good 
$m_i$ has good total $M= m_1 +m_2 + \ldots$.  

What this means is it is both possible and easy to construct individual Slater determinants which have good $M$ (i.e., are eigenstates of $\hat{J}_z$).  These will almost never 
be states also of good $J$.  But because $\hat{H}$ commutes with both $\hat{J}_z$ and $\hat{J}^2$, if we take all states of a given $M$ and diagonalize $\hat{H}$, 
the eigenstates will be guaranteed to also have good $J$ (barring `accidental' degeneracies that rarely occur).  Taking states of fixed $M$ is called an \textit{M-scheme basis}.
It is the simplest shell-model basis. 

But the $M$-scheme isn't the only choice. As mentioned above, one can also make the many-body Hamiltonian matrix explicitly diagonal in $J$ as well as $M$. This is
a \textit{J-scheme basis}.  Such bases are significantly smaller in dimension, typically an order of magnitude smaller than the $M$-scheme.   Of course, there are obvious 
costs.  Almost always a state with good $J$ must ultimately be a superposition of $M$-scheme Slater determinants.  This means both the $J$-scheme basis states, and the many-body matrix elements in this basis, are more costly to calculate. 

(Historically, in chemical and atomic physics one used \textit{configuration state functions} with good angular momentum, which we would call 
the $J$-scheme. The use of simple Slater determinants in chemical and atomic physics seems to have been introduced by \citet{knowles1984new}
apparently unaware of Whitehead's innovation.)

One can go even further.  Many nuclei exhibit strong rotational bands, which can be reproduced using the group SU(3).   If the nuclear Hamiltonian commuted with 
the Casimir operators of SU(3), or nearly so, then the Hamiltonian would be block diagonal in the irreps of SU(3), or nearly so, and 
SU(3) would be \textit{dynamical symmetry}.   One can imagine other group structures as well.

Because of this, some groups use group-theoretical bases, also called \textit{symmetry-adapted bases}, 
such as a \textit{SU(3)-scheme basis} (\citep{draayer2012symmetry}), based upon calculations which suggest that nuclear wavefunction are dominated by a few group irreps. The $SU(3)$-scheme is just like the $J$ only more so: the basis is more compact, but the basis states and the many-body matrix elements
even more complicated to derive.  On the other hand, the $SU(3)$-scheme makes the origin of rotational motion more transparent and potentially offers a more 
compact representation and understanding of the wavefunctions. 
Each of these schemes offer advantages and disadvantages.  

\subsection[Factorization of the basis]{Use of quantum numbers: factorization of the basis}

One of the advantages of the $M$-scheme is that despite the fact it is the least compact of basis schemes, it can be represented very efficiently with 
factorization.  Factorization is an idea used throughout {\tt BIGSTICK}, and is  most easily illustrated in the basis.

We work in the $M$-scheme, which means every many-body basis state has the same definite value of 
$M$. If we have an even number of particles, $M$ is an integer, while for odd numbers it will be a half-integer ($1/2, 3/2, -5/2$, etc.). Internally {\tt BIGSTICK} 
doubles these so they can be represented by even or odd integers, respectively.  

Each basis state, however, is a simple tensor product of a proton Slater determinant and a neutron Slater determinant. Because the $m$ quantum numbers 
are additive, we have the total $M = M_p + M_n$, the sum of proton and neutron $M$-values. 

Absent other constraints, \textit{every} proton Slater determinant with $M_p$ not only can but \textit{must} be combined with \textit{every} neutron Slater determinant with 
$M_n = M - M_p$; this, in part, guarantees that rotational invariance is respected and that the final eigenstates will have good total $J$. This in turn leads to a 
shortcut.

Consider the case of the $^{27}$Al nucleus, using the $sd$ valence
space. This assumes five valence protons and six valence neutrons
above a frozen $^{16}$O core. The total dimension of the many-body
space is 80,115, but this is constructed using only 792 five-proton
states and 923 six-neutron states.

The  reader will note that $792 \times 923 = 731016 \gg 80115$.
Indeed, not every five-proton state can be combined with every
six-neutron state. The restriction is due to conserving certain additive quantum
numbers, and this restriction turns out to limit usefully the nonzero
matrix elements of the many-body Hamiltonian, which we will discuss more in the
next section.

For our example, we chose  total $M = + 1/2$ (though we could have
chosen a different half-integer value). This basis requires that
$M_p + M_n = M$; and for some given $M_p$, \textit{every} proton
Slater determinant with that $M_p$ combines with \textit{every}
neutron Slater determinant with $M_n = M - M_p$. This is illustrated
in Table \ref{al27}, which shows how the many-body basis is
constructed from 792 proton Slater determinants and 923 neutron
Slater determinants. Note we are ``missing'' a neutron Slater
determinant; the lone $M_n = -7$ state has no matching (or
`conjugate') proton Slater determinants.

\begin{table}
\caption{Decomposition of the $M$-scheme basis for 5 protons and 6
neutrons in the $sd$ valence space ($^{27}$Al), with total $M = M_p
+ M_n +1/2$. Here ``pSD'' = proton Slater determinant and ``nSD'' =
neutron Slater determinant, while ``combined'' refers to the
combined proton+neutron many-body basis states.  The subset of the
basis labeled by fixed $M_p$ (and thus fixed $M_n$) we label a
'sector' of the basis. \label{al27}}
\begin{tabular}{|r|r|r|r|r|}
\hline
$M_p$ & \# pSDs  & $M_n$ & \# nSDs & \# combined \\
\hline
+13/2 & 3 & -6 & 9 & 27 \\
+11/2 & 11 & -5 & 21 & 231 \\
+9/2 & 28 & -4 & 47 & 1316\\
+7/2 & 51 & -3 & 76 & 3876 \\
+5/2 & 80 & -2 & 109 & 8720 \\
+3/2 & 104 & -1 & 128 & 13,312\\
+1/2 & 119 &  0 & 142 & 16,898\\
-1/2 & 119 &  +1 & 128 & 15,232 \\
-3/2 & 104 &  +2 & 109 & 11,336\\
-5/2 & 80 &  +3 & 76 & 6080\\
-7/2 & 51 &  +4 & 47 & 2444\\
-9/2 & 28 &  +5 & 21 & 588\\
-11/2 & 11 &  +6 & 9 & 99 \\
-13/2 & 3 &  +7 & 1 & 3 \\
\hline
Total & 792 & & 923 & 80,115 \\
\hline
\end{tabular}
\end{table}

As a point of terminology, we divide up the basis (and thus any wavefunction vectors)
into \textit{sectors}, each of which is labeled by $M_p$, and any additional quantum numbers
such as parity $\Pi_p$; that is, all the basis states constructed with the same
$M_p$ ($\Pi_p$, etc.) belong to the same basis `sector' and have contiguous indices.
Basis sectors are also useful for grouping operations of the Hamiltonian, as described
below, and can be the basis for distributing vectors across many processors, although
because sectors are of different sizes this creates nontrivial issues for load balancing.

While we can represent the 80,115 basis states of $^{27}$Al in the $sd$ with 792 proton Slater determinants and 923 neutron 
Slater determinants, the storage is even more impressive for large systems. For example, in the $pf$ shell, $^{60}$Zn, with 10 valence protons and 
10 valence neutrons, has for $M=0$ a basis dimension of 2.3 \textit{billion}.  But these are represented by $\sim 185,000$ proton Slater determinants 
and the same number of neutron Slater determinants. (In principle with self-conjugate systems $N=Z$ systems one could gain further savings 
by keeping only one set of Slater determinants. Because that is a small number of nuclides, we chose not to do so.)
The savings are not as dramatic for no-core shell model calculations with $N_\mathrm{max}$ truncation. For example, $^{12}$C in a basis or 
$N_\mathrm{max}=10$ has a basis dimension of 7.8 billion, constructed from 1.8 million each proton and neutron Slater determinants. The 
reason for the lessened efficiency is the many-body truncation.

We note that factorization not only  provides dramatic lossless compression of data, it also accelerates the set up of data.  In the set up phase of any CI code, 
one of the major tasks is searching through long series of bitstrings and, when one uses quantum numbers to organize the data, sorting.  Factorization improves 
this by reducing the length of lists to be searched and sorted. Our second level of factorization further reduces those lists, making searches and sorts even faster.

While factorization of the Hamiltonian was, to the best of our knowledge, pioneered by \citet{ANTOINE} in the code {\tt ANTOINE}  and adopted as well by {\tt EICODE} (\citep{Toi06}), {\tt NuShell/NuShellX} (\citep{NuShellX}), and 
{\tt KSHELL} (\citep{shimizu2013nuclear}) (and possibly others we are unaware of), {\tt BIGSTICK} has uniquely implemented a second level of factorization. Because most users 
never see this level, we direct those interested to our  paper 
for more details. 

{\tt BIGSTICK} does provide some information about this. In normal runs, as well as in modeling runs, you will see
\begin{verbatim}
  .... Building basis ... 
  
  Information about basis: 
 there are           27  sectors            1
 there are           27  sectors            2
                38760  SDs for species            1
               184722  SDs for species            2
  Total basis =             501113392
  
  .... Basis built ... 
\end{verbatim}
The above example is for $^{56}$Fe in the $pf$ shell with $M=0$. The \textit{sectors} are the subsets of the proton and neutron Slater determinants 
(`{\tt SDs}') with fixed quantum number $M$, parity, and optionally $W$.   Here `species 1' refers to protons and `species 2' refers to neutrons.

\section{The Lanczos algorithm and computational cost}

\label{computational_cost}

With bit manipulation allowing one to quickly calculate matrix elements, one could address much larger spaces, spaces so large they were not amenable to complete 
diagonalization, e.g., through the Householder algorithm (\citep{parlett1980symmetric,numericalrecipesfortran}).  But in nuclear structure configuration-interaction one almost never wants all the  eigensolutions; instead 
one typically just wants the low-lying states.  Thus \citet{lanczos} introduced another innovation: use of the Lanczos algorithm to find the extremal eigenstates. 

The Lanczos algorithm is a subspecies of Arnoldi algorithms. We describe the Lanczos algorithm in Chapter \ref{lanczos}, 
but the key idea is that, starting from an initial vector often called 
the \textit{pivot}, one iteratively constructs a sequence of orthonormal basis vectors that form 
a \textit{Krylov subspace}, as well as the elements of the Hamiltonian in that subspace.  The genius of Arnoldi/Lanczos algorithms is that they use the matrix to 
be diagonalized to construct the basis vectors; by applying the Hamiltonian matrix to a given basis vector one constructs, after orthogonalization, the next basis vector. 
One can show via the classical theory of moments that the extremal eigenvalues of the Hamiltonian in the Krylov subspace quickly converge to those of the full space. 
Although it depends upon the model space, the Hamiltonian, and the choice of pivot (starting vector to kick off the Lanczos iterations), one can often reach a converged 
ground state energy in as few as twenty Lanczos iterations, and the lowest five states in as few as 100 iterations.  (Advanced versions of Lanczos, namely thick-restart Lanczos (\ref{thickrestart}) and block Lanczos (\ref{block}) are also implemented.)

Now let us think about the computational cost of carrying out CI, both in terms of operations (time) and memory (storage). Before doing so let us highlight a 
key point. Most often in discussing CI one cites the basis dimension. But, as we will argue below, the real measure of the computational cost is the number of 
nonzero matrix elements. Now, for any given scheme, the number of nonzero matrix elements scales with the basis dimension. However, for different schemes 
the proportionality is different: $J$-scheme is denser than $M$-scheme; furthermore, even within the same basis scheme, different truncations have different 
densities, e.g., the NCSM is much denser than `phenomenological' calculations. Therefore, for absolute comparison of the computational cost of a problem, 
the number of nonzero matrix elements is a much better measure than basis dimension.

That said, let us  look at the computational cost of matvec:
$$
w_\alpha = \sum_\beta H_{\alpha, \beta} v_\beta.
$$
Let the dimension of the vector space be $N$. 
If the many-body matrix \textbf{H} is a fully dense (but real, symmetric) matrix, the above matvec requires $~N^2$ operations as well as storage of $~N^2$ many-body 
matrix elements.  However, \textbf{H} is almost never fully dense.  This can be most easily understood in the $M$-scheme, where the fundamental occupation-space 
basis states can be represented as raw bit strings.  A two-body interaction can at most shift two bits. Therefore if two basis states $|\alpha \rangle$ and 
$|\beta \rangle$ differ by more than two bits, the matrix element between them \textit{must} be zero.  A typical `sparsity' of $M$-scheme Hamiltonians is 
$~ 2\times 10^{-6}$, that is, only two out of every million many-body matrix elements is nonzero. (Three-body forces, naturally, lead to denser matrices, 
roughly two orders of magnitude denser.) 

If one has a basis dimension of a million, then there are roughly a million nonzero matrix elements; because one needs not only the value of the matrix element 
but some index to local it in the matrix, in single precision this requires roughly 8 megabytes of memory.  When one goes up to a basis dimension of one billion, 
however, this number goes up by $(10^3)^2$ to 8 terabytes!  Reading 8 Tb of data even from fast solid state disks is a very slow proposition. If one stores 
the matrix elements in core memory across many processors, as the code {\tt MFDn} does, this requires a minimum of many hundreds if not thousands of processors. 

\section{Representation of the Hamiltonian}




As discussed above, $M$-scheme configuration-interaction calculations require a many-body vector space of very large dimensions, 
and the many-body Hamiltonian matrix, while very sparse, still in large cases nonetheless the nonzero matrix elements end requiring a huge amount of data. 

If you were to examine closely, say, the bit representation of the basis states, or the nonzero matrix elements, you'd find something  confounding: 
quite a lot of data is repeated, over and over.  The same proton bit strings (which we generally call proton Slater determinants, although technically they are 
\textit{representations} of said determinants) are repeated many times, sometimes many millions of times or more, and the 
same for the neutron bit strings (neutron Slater determinants).  In the same way, the same values appear, thousands and millions of times, in the non-zero 
many-body matrix elements, though with both positive and negative values.

This redundancy can not only be understood, it can be turned to our advantage through \textit{factorization}, both of the basis and of matvec operations. 

\begin{table}
\caption{Number of one- and two-body `jumps' and storage requirements for representative
atomic nuclei in different model spaces (described in Appendix B). For storage
of nonzero matrix elements (penultimate column) we assume each many-body matrix element is stored
by a 4-byte real number and its location encoded by a single 4-byte integer.  Storage of a single
jump (initial and final Slater determinants for a species, and matrix element and phase) requires
13 bytes.  All storage (final two columns) are in gigabytes (GB). 
\label{jumps}}
\begin{tabular}{|c|c|c|c|c|c|c|}
\hline
Nuclide  &  space  & basis  &  $\#$ 1-body & $\#$ 2-body  & Store   & Store \\
         &         & dim    &  jumps & jumps  & m.e.s   & jumps \\
\hline
$^{28}$Si  & $sd$ & $9.4 \times 10^4$ & $4.8 \times 10^4$
& $7.6 \times 10^3$  & 0.2  & 0.002 \\
$^{52}$Fe  & $pf$ & $1.1 \times 10^8$ &  $4.0\times 10^6$ & $8.5 \times 10^6$ & 700 & 0.16 \\
$^{56}$Ni  & $pf$ & $1.1 \times 10^9$ &  $1.5\times 10^7$ &
$4.0 \times 10^7$ & 9800 & 0.6 \\
$^{4}$He   & $N_\mathrm{max} =22$ & $9 \times 10^7$ &   $5.3 \times 10^8$  & $4.7 \times 10^9$ & 9300 &  69  \\
$^{12}$C   & $N_\mathrm{max} =8$ & $6 \times 10^8$ &  $6 \times 10^8$ & $3 \times 10^9$
& 5200 & 45 \\
$^{13}$C   & $N_\mathrm{max} =6$ & $3.8 \times 10^{7}$ & $7 \times 10^{7}$ & $3 \times 10^8$ & 210 & 4.3 \\
\hline
\end{tabular}
\end{table}

 The  idea is similar to the factorization of the basis. Any two-body Hamiltonian can be split into forces that 
act only on protons, forces that act only on neutrons, and interactions between protons and neutrons. Consider forces acting only on protons; in a 
factorized basis, the neutrons are spectators. If we write our basis states as a simple tensor product between a proton `Slater determinant' 
$ | i_p \rangle  $ and 
and a neutron Slater determinant, $|  j_n \rangle $, so that the basis state $ | \alpha \rangle =| i_p \rangle | j_n \rangle $  the pure proton Hamiltonian matrix element is 
\begin{equation}
\langle \alpha | \hat{H}_{pp} | \alpha^\prime \rangle = \langle i_p | \hat{H}_{pp} | i^\prime_p \rangle \delta_{j_n, j^\prime_n}.
\end{equation}
We therefore only have to store the proton matrix element $\langle i_p | \hat{H}_{pp} | i^\prime_p \rangle$, and can trivially loop over the neutron Slater determinants. You can see how you could get dozens, hundreds, or thousands of matrix elements with the same value, just with different $j_n$. 
Furthermore, because the neutron Slater determinants are frozen, the quantum numbers cannot change, which severely restricts the action 
of the proton-only Hamiltonian. 
The matrix elements $\langle i_p | \hat{H}_{pp} | i^\prime_p \rangle$ are called \textit{jumps} and we only need to store them (and know of the 
neutron indices $j_n$ over which to loop).  

For proton-neutron interactions the action is more complicated but the same basic ideas hold: one stores separate proton jumps and neutron 
jumps and reconstructs the value of the matrix element. Table \ref{jumps} shows the storage for nonzero matrix elements and of jumps needed for a number of representative nuclei, in 
both phenomenological and NCSM calculations. You can see there are at least two orders of magnitude difference. Thus, for example, $^{52}$Fe, which
would require 700 Gb of storage for just the nonzero matrix elements, only needs less than a Gb for storage in factorization (in this particular case, storage 
of the Lanczos vectors is much higher burden) and thus can be run on an ordinary desktop computer. 

The price one pays, of course, is the factorized reconstruct-on-the-fly algorithm is much more complicated.

\section{An incomplete  survey of other codes}

While this manual is about {\tt BIGSTICK}, it is appropriate to put it in the context of other (nuclear) configuration-interaction codes.  One can broadly classify them 
by (a)  basis scheme, (b) representation and storage of many-body matrix elements, (c) rank of interactions (i.e., two-body only or two- and three-body forces), (d) 
parallelism, if any, and finally (e) general area of applicability, e.g., primarily to phenomenological spaces, which usually means a frozen core, and interactions, or 
to \textit{ab initio} no-core shell model calculations.  Please keep in mind that most of these codes are unpublished or have only partial information published, and 
that many of the details have been gleaned from private conversations;  information on some codes, 
such as the powerful Japanese code {\tt MSHELL}, do not seem to be available.   
We apologize for any accidental misrepresentations.  All of these codes have powerful 
capabilities and have made and are making significant contributions to many-body physics.

Among the very earliest codes was the Oak Ridge-Rochester code from the 1950s and 1960s, which fully diagonalized the Hamiltonian after computing the 
many-body $J$-scheme matrix elements via coefficients of fractional parentage. 
 It was succeeded by the Whitehead (Glasgow) code and its descendents, which used bit manipulation 
to compute the many-body matrix elements in the $M$-scheme, and solved for low-lying eigenstates using the Lanczos algorithm, 
{\tt ANTOINE}
(\citep{ANTOINE}),  {\tt MFDn} (\citep{MFDN}), and {\tt KSHELL}  (\citep{shimizu2013nuclear}) are also $M$-scheme codes.  Examples of
$J$-scheme codes in nuclear physics include {\tt OXBASH} (\citep{OXBASH}) and
its successors {\tt NuShell} and {\tt NuShellX} (\citep{NuShellX}),  {\tt  NATHAN}
(\citep{PhysRevC.59.2033}), and {\tt EICODE} (\citep{Toi06}).
There have also been attempts to use group theory to construct so-called symmetry-guided bases. 
The main effort is in SU(3) \citep{draayer2012symmetry}.
Although this approach is very promising, only time will tell for sure if the advantages gained by group theory will outweigh the technical difficulties needed to implement, for although the bases are small, they are significantly denser, and furthermore the group theory is very challenging.

Regarding access to  the many-body Hamiltonian matrix element, the Oak Ridge-Rochester and Whitehead codes stored matrix elements on 
disk, as do {\tt OXBASH} and {\tt NuShell}.  The very successful code {\tt MFDn} \citep{MFDN}, used primarily but not exclusively  for no-core shell model calculations, stores the many-body Hamiltonian matrix elements in RAM, much faster to access than storing on disk, but
for all but the most modest of problems requires distribution across
hundreds or thousands of nodes on a parallel computer spread across many MPI processes.

Factorization methods, pioneered in {\tt ANTOINE} \citep{ANTOINE}, have been used in
several other major CI codes: {\tt NATHAN}\citep{PhysRevC.59.2033}, {\tt EICODE} \citep{Toi06},
{\tt  NuShellX} \citep{NuShellX}, {\tt KSHELL}  \citep{shimizu2013nuclear}, 
and our own unpublished codes {\tt REDSTICK} (so named because it was originated at Louisiana State University, located in Baton Rouge), 
and  of course {\tt BIGSTICK}.
Factorization has also been used in nuclear structure physics as a 
gateway to approximation schemes \citep{AP01,PD03,PJD04,PD05}.  The codes most widely used by people beyond their authors have been {\tt OXBASH} and 
its successor {\tt NuShell/NuShellX}, and {\tt ANTOINE}.  Because of their wide use, and because one of us (Ormand) heavily used {\tt OXBASH}, the default formats 
for our input {\tt .sps} and {\tt .int} files are heavily modeled upon the {\tt OXBASH/NuShell/NuShellX} formats.

Like {\tt BIGSTICK}, {\tt MFDn} has been parallelized with both MPI and OpenMP and has carried out some of the largest supercomputer runs in the field. 
{\tt NuShellX} has only OpenMP parallelization.  The parallelization of other codes is unknown.

In closing, we note that besides configuration-interaction there are many other approaches to the many-body problem, such as the coupled cluster method, the Green's function 
Monte Carlo method, the in-medium similarity renormalization group, density functional methods, and so on, each with their own advantages and 
disadvantages. 
There are also methods closely related to configuration interaction,  such as the `Monte Carlo shell model,' the `shell-model Monte
Carlo,' generator-coordinate codes, and the importance truncation shell model. 
The main weakness of configuration interaction is that it is not size extensive, which 
means unlinked diagrams must be cancelled and thus the dimensionality of the problem grows exponentially with particle number and/or single-particle 
basis. The advantages of CI is: it is fully microscopic; its connection to the many-body Schr\"odinger equation (\ref{Schrodinger}) is pedagogically transparent; 
it generates excited states as easily as it does the ground state; it can handle even and odd numbers of particle equally well and works well far from closed shells;
and finally places no restriction on the form of either the single-particle basis or on the interaction (i.e., local and nonlocal forces are handled equally well, 
because the occupation space basis is intrinsically nonlocal to begin with).  


\chapter{Getting started with {\tt BIGSTICK}}

\label{quickstart} 

{\tt BIGSTICK} is a configuration-interaction many-fermion code, 
written in Fortran 90. It solves 
for low-lying eigenvalues of the Hamiltonian of a many fermion system.
The Hamiltonian is assumed to be rotationally invariant and to conserve parity, 
and is limited to two-body (and three-body, in progress) forces. Otherwise
few assumptions are made.  

BIGSTICK allows for two species of fermions, such as protons and 
neutrons.  BIGSTICK is flexible, able to work with 
"no-core" systems and phenomenological valence systems alike, and can compute the electronic 
structure of single atoms or cold fermionic gases (in which cases the 
two species are interpreted as "spin-up" and "spin-down").  BIGSTICK 
has a flexible many-body truncation scheme that covers many common 
truncations. For nuclei it can assume isospin symmetry or break isospin 
conservation.  Interaction matrix elements must be pre-computed by 
a third-party program and stored as a file, but BIGSTICK accepts a variety of 
matrix element formats.

\section{What can {\tt BIGSTICK} do?}

 {\tt BIGSTICK}  can:

\begin{itemize}

\item  compute the ground state energies and low-lying excitation spectra, including 
angular momentum and, if relevant, isospin, of many-body systems with a 
rotationally invariant Hamiltonian; wave functions are also generated;


\item compute expectation values of scalar one- and two-body 
operators; 

\item  compute one-body densities, including transition densities, among the low-lying 
levels, which allows one to calculate transition rates, life times, moments, etc.; 

\item compute transition strength probabilities or strength functions for one-body 
transition operators, useful when one needs to model transitions to 
many excited states;

\item use the strength function capability to decompose the wave function by 
the eigenvalues of an operator, such as the Casimir of some group.

\end{itemize}

Along with this, one can ask, what are {\tt BIGSTICK}'s limitations?  This largely 
depends upon the computer used and the many-body system.  In low-energy 
nuclear structure physics, which is the main focus of our research, one can easily 
run on a laptop any nuclide in the phenomenological $sd$ space, and on a workstation 
reach most nuclides in the phenomenological $pf$ space. Although dimensionality 
is not the most important determination of computational burden, one can generally 
run cases of dimension up to a few million or even tens of millions, if one is patient, on a 
laptop, a few hundred million on a workstation, and a few billion on a parallel supercomputer.

As always, of course, much of the limitations depend upon the user. Although we provide a 
few example input files, it is generally up to the user to provide files for the model space, 
the interaction, and codes to postprocess density matrices into transitions. (We do provide 
some tools for this.)

\section{Downloading and compiling the code}

{\tt BIGSTICK} was developed for UNIX/Linux/MacOSX systems. We made no 
effort to adapt it to running under Microsoft Windows.

To get {\tt BIGSTICK}, download it from GitHub:

\begin{verbatim}
git clone https://www.github.com/cwjsdsu/BigstickPublick/
\end{verbatim}

\subsection{Directory structure}

With more recent versions of {\tt BIGSTICK}, the distribution includes the following folders/subdirectories:
\begin{verbatim}
bin             doc             make            src    
\end{verbatim}
plus a {\tt README.dat} file.  The {\tt src} directory contains all the source code. The {\tt doc} directory 
contains this manual and the {\tt WHATSNEW.txt} file, plus others. The makefile is in the {\tt make} directory. 
The generated executables are put in the {\tt bin} directory.

\subsection{Compilation}

Your distribution includes a makefile. To access it, go into the {\tt make} directory.
We have developed {\tt BIGSTICK} to compile and run successfully with 
Intel's ifort compiler and GNU gfortran.  You may need to edit the makefile to 
put in the correct compiler and/or if you wish to use for example LAPACK libraries.  
We have written the code to require minimal special compile flags.

For example, 

\noindent {\tt PROMPT> make serial}

\noindent makes a serial version of the code with the Intel ifort compiler by default.
Several other options are:

\bigskip

\noindent {\tt PROMPT> make openmp} \hspace{1.8cm}  $\rightarrow$ an OpenMP parallel version using ifort

\noindent {\tt PROMPT> make gfortran} \hspace{1.4cm}  $\rightarrow$ a serial  version using gfortran

\noindent {\tt PROMPT> make gfortran-openmp} \hspace{.15cm}  $\rightarrow$ an OpenMP   version using gfortran

\noindent and so on. To see all the options encoded into the makefile,

\noindent {\tt PROMPT> make help}

\bigskip 

Each of these generates an executable with the nonstandard extension {\tt .x}, chosen to make deletion easy: {\tt bigstick.x,  bigstick-openmp.x, bigstick\--mpi.x}, and 
{\tt bigstick-mpi-omp.x}.
These executables are created in the {\tt bin} directory.
There are options for compiler on a number of supercomputers. Please keep in mind, however, that 
compilers and compile flags on supercomputers are a Red Queen's Race, and it is up to the user to tune the makefile for any given configuration.
 
\bigskip

\textit{Libraries}.  In routine operations, {\tt BIGSTICK} uses the Lanczos algorithm to reduce the Hamiltonian matrix to a truncated 
tridiagonal matrix whose eigenvalues approximate the extremal eigenvalues of the full matrix. This requires an eigensolver 
for the tridiagonal. For modest cases, one can also choose to fully diagonalize the Hamiltonian, using a Householder algorithm. 
(In practice we find this can be done quickly for basis dimensions up to a few thousand, and with patience can be done up to a 
basis dimension $\sim 10^4$.)  For both cases we use the LAPACK routine {\tt DSYEV}, which solves the real-valued, double-precision 
symmetric eigenvalue problem.  The actual matrix elements are given in single-precision, but we found when the density of eigenvalues 
is high, double-precision gives us better values for observables, including angular momentum $J$ and isospin $T$.    

Although in principle one could link to a library containing {\tt DSYEV}, in practice this is highly platform dependent.  Also, except for 
special cases where one is fully diagonalizing very large matrices and are impatient, the call to {\tt DSYEV} is a tiny fraction of the time. 
Hence we supply an unmodified copy of {\tt DSYEV} and required LAPACK routines, and there is no need to call any libraries.

\section{Required input files}

In order to solve the many-body Schr\"odinger equation, {\tt BIGSTICK} requires at least two inputs:

\noindent (a) A description of the single-particle space, usually through a file with extension {\tt .sps} 
(although if one is running a no-core shell model calculation, there is an option to generate this 
automatically); and

\noindent (b) A file containing the matrix elements of the interaction, in the form of single-particle 
energies and two-body matrix elements (and, optionally, three-body matrix elements). 

We supply several example cases for both inputs, including some commonly used spaces and 
interactions. But in general it is the user's duty to supply these input files and, importantly, to 
make sure they are consistent with each other, i.e., to make sure the ordering of single-particle 
orbits in the {\tt .sps} file is consistent with those in the interaction file.  We describe the file 
formats in detail in  Chapter \ref{detailruns}

\section{Running the code}

BIGSTICK has a simple interactive input.  It can also be run by pipelining the 
input into the code. 

To run:

\noindent{ \tt PROMPT>bigstick.x}

(we recommend you keep the source code, the executable, and the input data files 
in separate directories, and make sure the executable is in your path). We use the 
nonstandard extension {\tt .x} to denote executables.

First up is a preamble, with the version number, information on parallel processes, and 
a reminder for citations:
\begin{verbatim}
  BIGSTICK: a CI shell-model code, version 7.11.4Jun 2025 
  
  Please cite: C. W. Johnson, W. E. Ormand, and P. G. Krastev 
  Comp. Phys. Comm. 184, 2761-2774 (2013);  
  and C. W. Johnson, W. E. Ormand, K. S. McElvain, H.-Z. Shan 
   arXiv:1801.08432 and report UCRL LLNL-SM-739926 
  
  This code distributed under the MIT Open Source License 

 Running on NERSC_HOST: none, scratch_dir (*.wfn,...): .
 Number of MPI processors =           1 , NUM_THREADS =            6
\end{verbatim}

Next and most important is the main menu (see Sec.~\ref{mainmenu} and Appendix \ref{menu} for further explication)

\begin{verbatim}
  * * * * * * * * * * * * * * * * * * * * * * * * * * * * * * * * * * * * * * 
  *                                                                         * 
  *               OPTIONS (choose one)                                      * 
  * (i) Input automatically read from "autoinput.bigstick" file             * 
  *  (note: autoinput.bigstick file created with each nonauto run)          * 
  * (n) Compute spectrum (default); (ns) to suppress eigenvector write up   * 
  * (d) Densities: Compute spectrum + all one-body densities (isospin fmt)  * 
  * (2) Two-body density from previous wfn (default p-n format)             * 
  * (x) eXpectation value of a scalar Hamiltonian (from previous wfn)       * 
  * (o) Apply a one-body (transition) operator to previous wfn and write out* 
  * (s) Strength function (using starting pivot )                           * 
  * (g) Apply the resolvent 1/(E-H) to a previous wfn and write out         * 
  * (m) print information for Modeling parallel distribution                * 
  * (l) print license and copyright information                             * 
  * (?) Print out all options                                               * 
  *                                                                         * 
  * * * * * * * * * * * * * * * * * * * * * * * * * * * * * * * * * * * * * * 

  Enter choice
\end{verbatim}

The most common choice is `{\tt(n)}' for a normal run. For a guide to the various options, see Section~\ref{mainmenu} and Appendix~\ref{menu}. 

To facilitate batch runs or multiple runs with similar inputs, each time
{\tt BIGSTICK} runs it creates a file {\tt autoinput.bigstick}.  This file can be 
edited; choosing `{\tt(i)}' from the initial menu will direct {\tt BIGSTICK} to read all 
subsequent commands from that file. 

Next up:
\begin{verbatim}
  Enter output name (enter "none" if none)
\end{verbatim}

If you want your results stored to files, enter something like {\tt Si28run1}.

The code will then create the following files:

\noindent {\tt Si28run1.res} :          text file of eigenenergies and timing information.

\noindent {\tt Si28run1.wfn}:     a binary file (not human readable) file of the wavefunctions 
                      for post-processing or for other runs, e.g. ``x''
                      expectation values, etc.
                      
\noindent {\tt Si28run1.log}: a logfile of the run, useful for tracking the exact conditions under which 
the run happened, as well as diagnosing problems.              

Other files generated but not need by most users:


\medskip

\noindent {\tt Si28run1.lcoef }:       text file of Lanczos coefficients;

\smallskip

\noindent {\tt timinginfo.bigstick} and {\tt timingdata.bigstick}: files on internal timing;

\smallskip

\noindent {\tt distodata.bigstick}: a file contain information on distribution of work across MPI processes;

\smallskip

\noindent and others used primarily by the authors for diagonsing behavior.

\medskip

If you enter ``{\tt none},'' 
the {\tt .bigstick} files will be created but
no results file ({\tt .res}) and no wavefunction file ({\tt .wfn}). 

\begin{verbatim}
  Enter file with s.p. orbit information (.sps)
  (Enter "auto" to autofill s.p. orbit info ) 
\end{verbatim}
This provides information about the single-particle space. 
A typical answer might be {\tt sd}, which tells {\tt BIGSTICK} to open the file {\tt sd.sps}, and read in information about the 
$sd$ valence space.  (Please be aware that in most cases one does not enter the extension, such as {\tt .sps} or {\tt .int}.)
The {\tt auto} option can only 
be used for ``no-core'' nuclear shell-model calculations.  

\begin{verbatim}
  Enter # of protons, neutrons 
\end{verbatim}
These are the valence protons and neutrons.  So, for example, if one wants to 
compute $^{24}$Mg, which has 12 protons and 12 neutrons, but the $sd$ single-particle 
space assumes a closed $^{16}$O core, so one has 4 valence protons and 4 valence neutrons.
For other kinds of fermions, 
see the appendix. 

\begin{verbatim}
  Enter 2 x Jz of system 
\end{verbatim}
{\tt BIGSTICK}  is a ``M-scheme'' code, meaning the many-body basis states have fixed
total $M = J_z$ (as opposed to J-scheme codes such as  {\tt NuShell }
which the basis has fixed total $J$).   You must enter an integer which is twice the 
desired value of $M$.  If there are an even number of particles, this is usually 0.  
For an odd number of nucleons, you must enter an odd integer, typically $\pm 1$. 
Because the Hamiltonian is rotationally invariant, the results should not change 
for a value $\pm M$.  One can choose a non-minimal $M$ if, for example, you are 
interested in high-spin states.

\begin{verbatim}
Enter parity +/- :
\end{verbatim}
In addition to fixed $M$, {\tt BIGSTICK} has fixed parity. {\tt  BIGSTICK } automatically 
determines if more than one kind of parity is allowed and asks for the parity. 
The $sd$ space, for example, has only positive parity states, and so this input is 
automatically skipped.

If you would like to compute both parities, enter `{\tt 0}'. (At the current time, this is necessary 
if you want to compute parity-changing transitions, as for any transition calculations 
{\tt BIGSTICK} must work in the same basis.)

\begin{verbatim}
  Would you like to truncate ? (y/n)
\end{verbatim}

In some cases it is possible to truncate the many-body space, discussed in detail in section \ref{truncation}.


{\tt  BIGSTICK} will then generate the basis; in most cases this takes only a fraction of a 
second.  {\tt  BIGSTICK } will print out some information about the basis, which you can 
generally ignore.

The next item is to read in the matrix elements of the Hamiltonian.
\begin{verbatim}
  Enter interaction file name (.int)
  (Enter END to stop ) 
\end{verbatim}
You can enter in a number of interaction files.  The format for the interaction files 
will be discussed below.  

\begin{verbatim}
  Enter scaling for spes, A,B,X ( (A/B)^X ) for TBMEs 
  (If B or X = 0, then scale by A ) 
\end{verbatim}
\textbf{Important:} You \textit{must} enter {\tt end} to finish reading in interaction files.

After the interactions files have been read in, {\tt BIGSTICK} sets up the jump arrays for reconstructing the matrix elements on 
the fly. After that, the eigensolver menu comes up:
\begin{verbatim}
  / ------------------------------------------------------------------------\ 
  |                                                                         | 
  |    DIAGONALIZATION OPTIONS (choose one)          | 
  |                                                                         | 
  | (ex) Exact and full diagonalization (use for small dimensions only)     | 
  |                                                                         | 
  | (ld) Lanczos with default convergence (STANDARD)                        | 
  | (lf) Lanczos with fixed (user-chosen) iterations                        | 
  | (lc) Lanczos with user-defined convergence                              | 
  |                                                                         | 
  | (td) Thick-restart Lanczos with default convergence                     | 
  | (tf) Thick-restart Lanczos with fixed iterations                        | 
  | (tc) Thick-restart Lanczos with user-defined convergence                | 
  | (tx) Thick-restart Lanczos targeting states near specified energy       | 
  |                                                                         | 
  | (sk) Skip Lanczos (only used for timing set up)                         | 
  |                                                                         | 
  \ ------------------------------------------------------------------------/ 
\end{verbatim}
As noted, the standard choice is `{\tt ld}' for default Lanczos. Other options are discussed later. 

\begin{verbatim}
ld
  Enter nkeep, max # iterations for lanczos 
  (nkeep = # of states printed out )
\end{verbatim}
Except for very small cases, {\tt BIGSTICK}  does not find all the eigenvalues. Instead it 
uses the Lanczos algorithm (introduced by Whitehead et al to nuclear physics) to 
find the low-lying eigenstates.  The variable {\it nkeep} is the number of targeted 
eigenpairs; typical values are 5-10.  One can either set a fixed number of iterations, 
typically 100-300, or set a maximal number of iterations and allow {\tt BIGSTICK} to stop 
sooner using a test for convergence (discussed in detail below). 

{\tt BIGSTICK} will then carry out the Lanczos iterations, printing out intermediate eigenvalues. 
The final result, which if a output file name was chose is also written to the 
{\tt .res} file, looks like
\begin{verbatim}
 State      E        Ex         J       T      par
    1    -99.44646   0.00000     0.000  -0.000    1
    2    -98.64234   0.80413     2.000  -0.000    1
    3    -97.74552   1.70094     4.000  -0.000    1
    4    -96.26342   3.18305     6.000  -0.000    1
    5    -96.13075   3.31571     2.000  -0.000    1
\end{verbatim}
This is fairly self-explanatory.  $E$ is the absolute energy, $Ex$ the excitation energy 
relative to the first state, and $J$ and $T$ are the total angular momentum and isospin, 
respectively.   Even though only $M$ is fixed, because the Hamiltonian commutes with 
$\hat{J}^2$ the final states will have good $J$.  Lack of good $J$  most likely 
signals lack of convergence, or states  degenerate in energy but with different $J$).  
Lack of good $J$ can also signal an error in the input file (specifically, a disallowed 
$J$ for a particular set of orbits; we have written error traps to catch such a problem), or,
lastly and only infrequently, a bug in the code itself. 

If the input matrix elements respect isospin, then $T$ should also be a good quantum 
number. {\tt BIGSTICK} allows one to read in isospin-breaking matrix elements, discussed in 
more detail in section \ref{pnham}.

Note that parity (1 or -1) is also listed. \textbf{Note: if parity option `0' is chosen,} that is, both parities allowed, \textbf{the listed parity can be wrong for incompletely converged states.} This is because the code looks at the parity of the first nonzero amplitude.

{\tt BIGSTICK} can also compute one-body density matrix elements at the end of a run; 
choose option {\tt d} in the initial menu. The format and conventions for the density 
matrices are in section \ref{densitymatrix}. 

The wavefunctions are saved to a {\tt .wfn} file, unless you choose option {\tt ns} in the 
initial menu. {\tt  BIGSTICK }can then post-process the files, for example computing 
the expectation value of a scalar (Hamiltonian-like) operator, section \ref{expectation};  compute 
overlap between wavefunctions from two different runs, section \label{overlap}; or  apply a 
non-scalar transition operator to a wavefunction and then compute the strength 
distribution of that transition, sections \ref{apply1body}, \ref{apply1bodybasic}, and \ref{apply1bodygoodJ}.

\section{Some sample runs}

In the directory {\tt examples} that should be found in your distribution, you will find 
various examples of runs, along with sample outputs to check the code is working 
correctly. 

\section{Typical run times}

In this section we survey `typical' run times for calculations using {\tt BIGSTICK}. Of course, these depend upon
the clock-speed of your chip as well as the compiler, as well as how much parallelism you are exploiting.  As we 
show below, {\tt BIGSTICK} does scale well in parallel mode.

Table \ref{timetable} gives, for a variety of nuclides, the dimensionality 
of the space, the number of operations (which is approximately though not exactly the number of nonzero 
matrix elements), the minimal storage which would be required to store the nonzero matrix elements, and 
finally an approximate run time, assuming 150 Lanczos iterations on a serial machine. The actual time may vary a lot, 
depending on clock speed and how efficiently the operations are actually processed.  Parallelism, of course, can 
speed up the wall clock times considerably. 

Empirically, one finds that the number of nonzero matrix element (here, operations) generally scales like 
$(\mathrm{dim})^{1.25}$ for two-body interactions, and $\approx (\mathrm{dim})^{1.5}$ for three-body forces.

Of course, running in parallel will speed up the code. To estimate the run time, divide by the number of cores used. 
These estimates are crude, but usually within about a factor of two. 

\begin{table}
\begin{tabular}{rrrrrr}
\hline 
Nuclide & space &  dim & $\#$ ops & min. store & ~ run time \\
\hline\\
$^{24}$Mg & $sd$ & 28,503 & 8.6M & 34 Mb & 5 s \\
$^{48}$Cr & $pf$ & 2 M & 1.5 B &  6 Gb & 15 min \\
$^{51}$Mn & $pf$ & 44M & 41B & 160 Gb & 9 hr \\
$^{56}$Fe & $pf$ & 500M & 0.9 T & 3.6 Tb & 6 d \\
$^{60}$Zn & $pf$ & 2.3B & 5T & 20 Tb & 35 d \\
$^{12}$C & $N_\mathrm{max}$ 6 & 32M  & 41 B & 160 Gb & 7 hr \\
$^6$Li & $N_\mathrm{max}$ 12 & 49M  & 180 B & 700 Gb & 30 hr \\
$^{12}$C & $N_\mathrm{max}$ 8 & 594M  & 1.2T  &  5 Tb &  8 d\\
$^{16}$O & $N_\mathrm{max}$ 8 & 1B  & 2 T & 8 Tb & 14 d \\
$^{10}$B & $N_\mathrm{max}$ 10 & 1.7B  & 5 T & 20 Tb & 35 d \\
$^6$Li & $N_\mathrm{max}$ 16 & 800M  & 7T & 27 Tb & 46 d \\
\hline
\end{tabular}
\caption{`Typical' run times for various nuclides, running \textit{in serial}
for 150 Lanczos iterations. To get approximate speed-up in parallel modes, divide by the number of cores. 
Here `min. store' is an estimate of the 
minimal storage required for nonzero matrix elements. \label{timetable}}
\end{table}

\chapter{Using {\tt BIGSTICK}, in detail}
\label{detailruns}

{\tt BIGSTICK} has two basic modes. It can calculate many-body spectra and wave functions, and it can process those wave functions in several ways. 
In order to generate the low-lying spectrum and wave functions, you need to, first, define the model space, and second, provide an interaction. 

\section{Overview of input files}

{\tt BIGSTICK} uses three classes of externally generated files.  Mandatory  are:  files which define the single-particle space, and files for interaction matrix elements. Optionally, {\tt BIGSTICK} can also use files for one-body transition matrix elements. Here we briefly summarize those files, and in later 
sections give more details. 

Files which define the single-particle space have the extension either {\tt .sps} (preferred) or {\tt .sp} (`legacy' from {\tt NuShellX} inputs). When prompted, the user only supplies the name, not the 
extension, i.e., if the file is {\tt sd.sps} only enter {\tt sd}. {\tt BIGSTICK} will automatically search for both {\tt sd.sps} and, if not found, then {\tt sd.sp}.
These files can assume isospin symmetry or separate proton-neutron orbits, but at this time, {\tt BIGSTICK} requires that the proton and neutron 
single-particle spaces initially be the same. {\tt BIGSTICK} can however truncate the proton and neutron spaces differently. 

If the user is carrying out a `no-core shell-model' calculation where the single-particle orbits are assumed to occur in a default order, {\tt BIGSTICK} 
has an `{\tt auto}' option for defining the single-particle space and no input file is required.

{\tt BIGSTICK} accepts two classes of files for interaction matrix elements. The default format is derived from {\tt OXBASH/NuShell}. It can be in 
isospin-conserving format or in explicit proton-neutron format. Be aware that the latter has two possibilities for normalization of the proton-neutron 
states.  These files are used primarily though not exclusively for phenomenological spaces and interactions.  All files with this format must end in 
the extension {\tt .int}, and as with the single-particle files, one enters only the name, i.e., if the file is {\tt usda.int} one enters in only {\tt usda}.
If the file is in isospin-conserving format, you only need to enter the name of the file. If the file is in proton-neutron format, you must first tell 
{\tt BIGSTICK} the normalization convention, see section \ref{pnham}.  These files have broad options for scaling the magnitudes of matrix elements, 
see section \ref{scaling}.

{\tt BIGSTICK} also accepts files in a format readable by the {\tt MFDn} code, which can be generated by the {\tt NuHamil} code (\ref{nuhamil})..  Here one must enter in the \textbf{full} name of the file, even if it 
has the extension {\tt .int}, so that if the file is {\tt TBME.int} you enter {\tt TBME.int} not {\tt TBME}; this signals to {\tt BIGSTICK} to expect 
the {\tt MFDn} format.  Go to 
section \ref{MFDinput} for more details.

Finally, {\tt BIGSTICK} can apply a one-body operator to a wave function in order to generate a transition strength function. These have extension 
{\tt .opme}. These are defined in section \ref{apply1body}, with advanced instruction and examples in sections \ref{apply1bodybasic} and 
\ref{apply1bodygoodJ}.

While we supply sample files of these various formats, in general it is \textbf{the responsibility of the user to generate or obtain input files}. 

All other files {\tt BIGSTICK} needs, such as wave function files with extension {\tt .wfn}, must been generated by a run of {\tt BIGSTICK} itself.

\section{Defining the model space}

\label{sps}

A many-body model space is defined by a single-particle space, the valence $Z$ and $N$,  a total $M$ value, a total parity (if applicable), and, optionally, truncations on 
that model space. Note that if you are carrying out what we call a secondary option, which starts from an existing wave function as stored in a 
{\tt .wfn} file, {\tt BIGSTICK} will automatically read from that file the information on the basis.  You only need to define the model space when 
carrying out a `primary' option. 

The single-particle space is defined one or two ways. Either  read in a file defining the single-particle space, or, for so-called \textit{no-core 
shell model} calculations, automatically generate the basis in a pre-defined form, using the autofill or `{\tt auto}' option.

 For consistency, we generally refer to 
\textit{orbits} as single-particle spaces labeled by angular momentum $j$ but not 
$j_z$, while \textit{states} are labled by both $j$ and $j_z$. 

Our default format for defining the single-particle space are derived from the format for {\tt OXBASH/NuShell/NuShellX} 
files.  A typical file is the {\tt sd.sps} file:
\begin{verbatim}
! sd-shell
iso
3
       0.0  2.0  1.5  2
       0.0  2.0  2.5  2 
       1.0  0.0  0.5  2
\end{verbatim}
There is no particular formatting (spacing) to this file.  Any header lines starting with an exclamation point ! or a hash mark $\#$ are skipped over. 
The first non-header line denotes about the isospin symmetry or lack thereof. 
{\tt iso} denotes the single-particle space for both species is the same; one can still 
read in isospin breaking interactions. 
The second line (3 in the example above) is the number of single-particle orbits. 
The quantum numbers for the single-particle orbits as listed are: $n, l, j, w$; the first three numbers 
are real or integers, $j$ is a real number.
 $n$ is the radial quantum number, which play no role in {\tt BIGSTICK} except to 
distinguish between different states. $l$ is the orbit angular momentum and $j$ is 
the total angular momentum; for the case of nucleons $j = l \pm 1/2$.  In {\tt BIGSTICK }
the most important quantum number is $j$; $l$ is used internally only to derive the 
parity of each state.  

While for most applications $j$ is a half-integer, i.e., 0.5, 1.5, 2.5, etc., it can also be integer. In that 
case $l = j$ and one should intepret `protons' and `neutrons' as `spin-up' and `spin-down.' One can 
compute the electronic structure of isolated atoms, for example. 

While $n$ and $l$ are not internally significant for {\tt BIGSTICK}, they aid the human-readability 
of the {\tt .sps} files; in addition, they can be invaluable as input to other code 
computing desired matrix elements. 

{\tt BIGSTICK } automatically unpacks each orbit to arrive at the $2j+1$ single-particle 
states with different $j_z$. 

The last `quantum number,' $w$, is the \textit{weight} factor, used for many-body 
truncations,  described  in Section \ref{truncation}. It must be a nonnegative 
integer. 

{\tt BIGSTICK} can handle any set of single-particle orbits; the only requirement is that 
each one have a unique set of $n,l,j$. (Although $n$ and $l$ are written above as 
real numbers, for historical reasons, they must have integer values.  $j$ can take either 
half-integer values or integer values with $l=j$; this latter we refer to as LS-coupling 
and is discussed in detail later on. All the $j$-values in a {\tt .sps} file must be 
consistent, that is, all half-integer or all integer.)

For example, one could have a set of 
$l= 0, j = 1/2$ states:
\begin{verbatim}
iso
4
       0  0  0.5  0
       1  0  0.5  0
       2  0  0.5  0
       3  0  0.5  0
\end{verbatim}

As of the current version of {\tt BIGSTICK}, one cannot define completely independent 
proton and neutron spaces.  One can however specify two variations where protons and neutrons can have different weights.
The preferred format is {\tt pnw}, where one lists the quantum numbers as well as the proton and neutron weights in two columns:
\begin{verbatim}
pnw
3 
       0.0  2.0  1.5  3   3
       0.0  2.0  2.5  2   3
       1.0  0.0  0.5  2   3
\end{verbatim}
For some more details on using the {\tt pnw} format, especially in delineating different proton and neutron valence spaces, see Section~\ref{protonneutron} below.

An alternate, older (and no longer recommended) format is,
{\tt wpn}, where first proton, then neutron orbits are listed in order.  
\begin{verbatim}
wpn
3 
       0.0  2.0  1.5  3
       0.0  2.0  2.5  2
       1.0  0.0  0.5  2
       0.0  2.0  1.5  3
       0.0  2.0  2.5  2
       1.0  0.0  0.5  3
\end{verbatim}
While the proton and neutron orbits can have different weights, at this time the sets of quantum numbers 
must be the same and they must be listed in the same order. In the example above, we have 
proton $0d_{3/2}$, $0d_{5/2}$, and $1s_{1/2}$, and then the same for neutrons. Only the $w$ values can be different.
(In older versions one had to list the number of both proton and neutron orbitals, but by default these now must be the same.)

The ordering of the single particle orbits is important and must be consistent with 
the input interaction files.  If one uses our default-format interaction files, one 
must supply a {\tt .sps} file. 

It is possible to set environmental variables so that {\tt BIGSTICK} automatically 
searches for {\tt .sps} files in a different directory:
\begin{verbatim}
  You can set a path to a standard repository of .sps/.sp files 
  by using the environmental variable BIG_SPS_DIR. 
  Just do : 
  export BIG_SPS_DIR = (directory name) 
  export BIG_SPS_DIR=/Users/myname/sps_repo 
  Currently BIG_SPS_DIR is not set 
\end{verbatim}

While we recommend the default {\tt .sps} format, we  also allow for {\tt NuShell/NuShellX}-compatible {\tt .sp} files, which have a similar 
format. Like our default format, they also come in isospin-symmetric and proton-neutron format. An annotated example of the former is
\begin{verbatim}
! fp.sp
t             ! isospin-symmetric
40 20         ! A, Z of core
4             ! number of orbits
1 4           ! number of species, orbits per species
1 1 3 7       ! index, n, l, 2 x j
2 2 1 3
3 1 3 5
4 2 1 1
\end{verbatim}
As with the default format, {\tt BIGSTICK} will skip over any header lines starting with ! or $\#$.  The next line, {\tt t}, denotes 
 isospin symmetry. (Note that, however, because {\tt BIGSTICK} requires the proton and neutron spaces to be the same, 
 one does not need this option, and independent of the form of the single-particle space file one can read in interaction matrix 
 elements in either isospin-conserving or -breaking format.) The next line, {\tt 40 20} are the $A$ and $Z$ of the core; these are not actually needed but 
are inherited. 

The third non-header line, here {\tt 4} denotes the number of indexed orbits.  The fourth non-header line, {\tt 1  4}, tells us 
there is just one `kind' of particle with 4 orbits. The next four lines are the orbits themselves, with the orbital index, radial quantum 
number $n$, orbital angular momentum $l$, and twice the total angular momentum $j$.  Here $n$ distinguishes between different 
orbits which otherwise have the same $l$ and $j$.  In this example, $n$ starts at 1, while in our other example $n$ starts at 0. 
This makes no difference for {\tt BIGSTICK}'s workings.

This can be contrasted with the {\tt pn} option for the same space, which has separate indices for protons and neutrons.
\begin{verbatim}
! fppn.sp
pn
40 20
8
2 4 4
1 1 3 7
2 2 1 3
3 1 3 5
4 2 1 1
5 1 3 7
6 2 1 3
7 1 3 5
8 2 1 1
\end{verbatim}
The main differentce are in the third and fourth lines.  There are a total of 8 orbits labeled, among two kinds or `species' of particles, 
each with 4 orbits. The first 4 orbits are attributed to protons and the the next 4 to neutrons. 
While {\tt BIGSTICK} accepts both formats, in practical terms it does not make a difference. 
At this time {\tt BIGSTICK} does not allow for fully independent proton and neutron spaces, and the ordering of proton and neutron orbits must 
be the same. (We hope to install the capability for more flexible spaces in the future.)

Notice that the {\tt NuShell}-compatible {\tt .sp} format does not include the weighting number $w$, which is assumed to be zero. Hence no 
many-body truncations are possible with these files. 

If, instead, one uses an {\tt MFDn}-formatted interaction file, one can use the \textit{autofill} option  for defining the single-particle states, by entering {\tt auto} in place of the name of the {\tt .sps} file:
\begin{verbatim}
  Enter file with s.p. orbit information (.sps)
  (Enter "auto" to autofill s.p. orbit info ) 
auto
  Enter maximum principle quantum number N 
  (starting with 0s = 0, 0p = 1, 1s0d = 2, etc. ) 
\end{verbatim}
The autofill option creates a set of single-particle orbits assuming a harmonic oscillator, 
in the following order: $0s_{1/2}, 0p_{1/2}, 0p_{3/2}, 1s_{1/2}, 0d_{3/2}, 0d_{5/2}$, etc.,  that is, for given $N$, in 
order of increasing $j$, 
up to the maximal value $N$.
 It also associates a value $w$ equal to the principal 
quantum number of that orbit, e.g., $2n+l$, so that $N$ above is the maximal principal 
quantum number.   So, for example, if one choose the principle quantum number $N= 5$ this includes up to the 
$2p$-$1f$-$0h$ shells, which will looks like
\begin{verbatim}
iso
21
0.0  0.0  0.5  0
0.0  1.0  0.5  1
0.0  1.0  1.5  1
1.0  0.0  0.5  2
0.0  2.0  1.5  2
0.0  2.0  2.5  2
...
1.0  3.0  3.5  5
0.0  5.0  4.5  5
0.0  5.0  5.5  5
\end{verbatim}

\subsection{Particle-hole conjugation}

\label{particlehole}

{\tt BIGSTICK} constructs the many-body basis states by listing the occupied particle states.
Because the available single-particle space is finite, one can alternately list the 
unoccupied hole states.  Such a representation can be advantageous if the single-particle 
space is more than half-filled, which only happens in phenomenological spaces: while the dimension of the Lanczos basis is unchanged, 
because of our jump technology the matrix elements can take much more space and 
memory. To understand this, , consider diagonal matrix elements, $\langle \alpha | \hat{V} | \alpha 
\rangle$ which are a sum over occupied states:
$$
\langle \alpha | \hat{V} | \alpha \rangle 
= \sum_{a,b \in \alpha} V(ab,ab).
$$
The number of terms in the sum is quadratic in the number of `particles' in the system. 
Switching to holes can dramatically decrease the terms in this sum: if one has 12 single-particle 
states, for example, having two holes rather than ten particles makes a difference of 
a factor of 25! The overall scaling is not so simple, of course, for off-diagonal matrix elements 
(quickly: matrix elements of the form $\sum_b V(ab,cb), a\neq c$, that is, between two 
states which differ only by one particle,  go linearly in the number of 
particles, while those $V(ab,cd), a\neq c, b\neq d$, that is, between two states 
which differ by two particles, are independent of the number of 
particles), in large model spaces one can see a big difference. In particular cases with 
a large excess of neutrons, so that we have a small number of protons but nearly fill the 
neutron space, can lead to enormous slow downs, as well as requiring many more jumps. 
Here transformation from particles to holes make for much greater efficiency.  In order 
to obtain the same spectra and observables (density matrices), the 
matrix elements must be transformed via a \textit{Pandya} transformation.

How to invoke particle-hole conjugation: When you are asked to enter the 
number of particles, you are told the maximum number of particles:
\begin{verbatim}
  Enter # of valence protons (max  12 ), neutrons (max  12)
\end{verbatim}
Simply enter the number of holes as a negative number, i.e.,
\begin{verbatim}
-2 -5
\end{verbatim}
{\tt BIGSTICK} will automatically carry out the Pandya transformation:
\begin{verbatim}
           2  proton holes =           10  protons 
           5 neutron holes =           7  neutrons
\end{verbatim}
You can conjugate protons, or neutrons, or both.  If you enter the maximum 
number of particles in a space, {\tt BIGSTICK} will automatically regard it as 
zero holes.    Calculation of density matrices works correctly with particle-hole 
conjugation. 

When written to file, hole numbers are also written as negative integers as a flag, and 
when  post-processing, {\tt BIGSTICK} will correctly interpret them. 

We find there is little significant performance difference in spaces with up to about 20 single particle states, 
i.e. the $pf$ shell, but beyond 20 the timing difference can become quite dramatic.

Note that if you want to completely fill a space (a \textit{plenum} rather than a vacuum), for example, to have all the neutron orbits filled, you should enter in the maximum valence number; as long as the flag {\tt iffulluseph} in module {\tt bmodule\_flags.f90} is set to {\tt .TRUE}, the code will automatically convert it to 
particle-hole. This will work with if you have a truncation and set $W=0$; this can be used if you want to force protons and neutrons to be in different spaces.  I would advice against trying this with a nontrivial truncation ($W > 0$).

\subsection{Truncation of the many-body space}
\label{truncation}

Given a defined single-particle space, the basis states have fixed total $M$ and fixed parity. If we 
allow all such states, we have a \textit{full configuration} many-body space. 
Sometimes, motivated  either by physics or computational tractability, one wants to further truncate this 
many-body space. 
{\tt BIGSTICK} allows a flexible scheme for truncating the many-body space which 
encompasses many, though not all,  truncations schemes.
We truncate the many-body space based upon
 single particle occupations. One could truncate based upon many-body quantum numbers, such as from 
 non-Abelian groups (e.g., SU(2) for the $J$-scheme, or the symmetry-adapted SU(3) scheme),  but that is beyond the scope our algorithms.

Each single-particle orbit is assigned a weight factor $w$. This is read in from the 
{\tt .sps} file or if the autofill option is used, is equal to the harmonic oscillator 
principal quantum number. $w$ must be a nonnegative integer.  If all orbits have the 
same $w$ then no truncation is possible and {\tt BIGSTICK} does not query about truncations. 

$w$ is treated as an addititive quantum number: each basis state has a total $W$ 
which is the sum of the individual $w$s of the occupied states.  Because $w$ is assigned to 
an orbit, it does not violate angular momentum or parity, and the total $W$ is the same for 
all many-body basis states that are members of the same configuration, e.g., 
$(0d_{5/2})^2 (1s_{1/2})^1 (0d_{3/2})^1$.  Typically one assigns the same $w$ to 
equivalent proton and neutron orbits (in principle one could assign different $w$s, 
which would break isospin, but 
we haven't explored this in depth).

Given the basis parameters, the single-particle orbits and their assigned $w$s and the 
number of protons and neutrons, {\tt BIGSTICK} computes the minimum and maximum total $W$ 
possible. The difference between these two is the maximal excitation:
\begin{verbatim}
  Would you like to truncate ? (y/n)
y
  Max excite =           20
  Max excite you allow 
\end{verbatim}
The user chooses any integer between 0 and ``{\tt Max excite}.''   {\tt BIGSTICK} then creates 
all states with total $W$ up to this excitation. 

\begin{figure}
\includegraphics [width = 7.5cm]{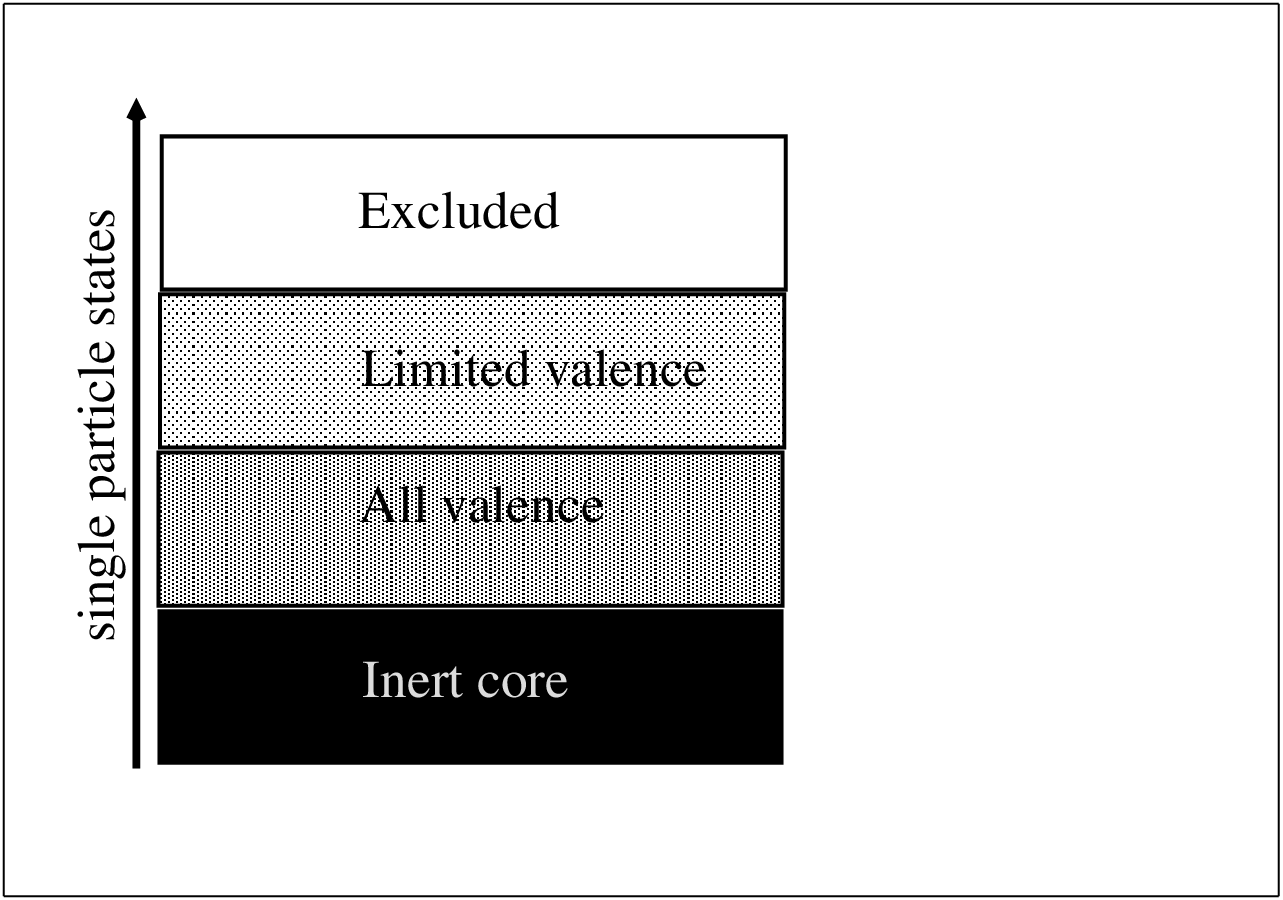}
\caption{\label{valence} Segregation of single-particle space. 'Inert core'
has all states filled. `Excluded' disallows any occupied states.  `All valence' can 
have states up to the number of valence particles filled, while `Limited valence' 
can only have fewer states filled (e.g. one, two, three...). See  text for discussion.
Figure taken from \citet{BIGSTICK}.
}
\end{figure}

This scheme encompasses two major trunction schemes.   
The first kind of truncation is  called a particle-hole
truncation in nuclear physics, or sometimes $n$-particle, $n$-hole; in atomic physics (and occasionally
in nuclear physics), one uses the notation `singles,'   `doubles,'
`triples,' etc. To understand this truncation scheme, begin by
considering a space of single-particle states, illustrated in Figure
\ref{valence}.  Any single-particle space can be partitioned into
four parts. In the first part, labeled `inert core', the states are
all filled and remain filled. In the fourth and final part, labeled
`excluded,' no particles are allowed. Both the core and excluded
parts of the single-particle space need not be considered
explicitly, only implicitly. In some cases there is no core.

More important are the second and third sections, labeled `all
valence' and `limited valence', respectively.  The total number of
particles in these combined sections is fixed at $N_v$, and this is
the valence or active space.

The difference between the `limited valence' and the `all valence'
spaces is that only some maximal number $N_l < N_v$ of particles are
allowed in the 'limited valence' space. So, for example, suppose we
have four valence particles, but only allow at most two particles
into the 'limited valence' space. In this case the `all valence'
might contain four, three, or two particles, while the 'limited
valence' space might have zero, one, or two particles. In more
standard language, $N_l = 1$ is called `one-particle, one-hole' or
`singles',  while $N_l = 2$ is called `two-particle, two-hole' or
'doubles', and so on. There are no other restrictions aside from
global restrictions on quantum numbers such as parity and $M$.

The second truncation is commonly used in no-core shell model calculations, 
 where center-of-mass considerations
weigh heavily. For all but the lightest systems, one must work in the
laboratory frame, that is, the wavefunction is a function of
laboratory coordinates, $\Psi = \Psi(r_1, r_2, r_3,\ldots)$. It is
only the relative degrees of freedom that are relevant, however, so
ideally one would like to be able to factorize this into relative
and center-of-mass motion:
\begin{equation}
\Psi(r_1,r_2, r_3,\ldots) = \Psi_\mathrm{rel}(\vec{r}_1 -\vec{r}_2,
\vec{r}_1 - \vec{r}_3, \ldots ) \times \Psi_\mathrm{CM} (\vec{R}_\mathrm{CM} )
\end{equation}
(note that we have only sketched this factorization).
In a harmonic oscillator basis and with a translationally
invariant interaction, one can achieve this factorization exactly, \textit{if}
the many-body basis is truncated as follows (see \citep{palumbo1967intrinsic,palumbo1968effects,gloeckner1974spurious}):

$\bullet$ In the non-interacting harmonic oscillator, each
single-particle state has an energy $e_i = \hbar \Omega (N_i +
3/2)$. Here $N_i$ is the principal quantum number, which is 0 for
the $0s$ shell, 1 for the $0p$ shell, 2 for the $1s$-$0d$ shell, and
so on. The frequency $\Omega$ of the harmonic oscillator is a
parameter but its numerical value plays no role in the basis truncation.

$\bullet$ We can then assign to each many-body state a
non-interacting energy $E_{NI} = \sum_i e_i$, the sum of the
individual non-interacting energies of each particle.   There will
be some minimum $E_\mathrm{min}$ and all subsequent non-interacting
energies will come in steps of $\hbar \Omega$--in fact for states of
the same parity, in steps of $2\hbar \Omega$.

$\bullet$ Now choose some $N_\mathrm{max}$, and allow only states with non-interacting energy
$E_{NI} \leq E_\mathrm{min} + N_\mathrm{max} \hbar \Omega$. In practice, restricting
states to the same parity means that the `normal' parity will have
$E_{NI} = E_\mathrm{min}$,  $E_\mathrm{min}+ 2 \hbar \Omega$,
$E_\mathrm{min}+ 4 \hbar \Omega, \ldots$, $E_\mathrm{min}+  N_\mathrm{max} \hbar \Omega$, while `abnormal' parity will have
$E_{NI} = E_\mathrm{min}+  \hbar \Omega$,
$E_\mathrm{min}+ 3 \hbar \Omega, \ldots$,
$E_\mathrm{min}+  N_\mathrm{max} \hbar \Omega$.

This is sometimes called the $N_\mathrm{max}$ truncation, the $N\hbar\Omega$ truncation, or simply the
energy truncation. It is more complicated than the previous
`particle-hole' truncation. We identify with each principal quantum
number $N_i$ a major shell; for a $4\hbar\Omega$ we can excite four
particles each up one shell, one particle up four shells, two
particles each up two shells, one particle up one shell and another
up three shells, and so on. While complicated, such a truncation
allows us to guarantee the center-of-mass wavefunction is a simple
Gaussian.

More generally, one can adjust the truncation scheme further, based upon skillful 
choice of single-particle $w$s.  The assigned $w$s need not be contiguous; the 
only requirement is that they be nonnegative. 

\subsection{Advanced truncation options}

All truncation is based upon the $w$ weight factors. In most applications, both protons and neutron orbits have the same 
weights, and one typically truncates equally.  A more general truncation scheme is possible.

First, as discussed in section \ref{sps}, it is possible for proton and neutron orbits to have different values of {\tt w}, if the 
{\tt .sps} file has the `{\tt pnw}' format:
\begin{verbatim}
pnw
3 
       0.0  2.0  1.5  3    2
       0.0  2.0  2.5  2    3
       1.0  0.0  0.5  2     3
\end{verbatim}
The dimensions of the proton and neutron orbits must be the same, as the order of all the quantum numbers besides $w$. 
The values of the $w$s can be different for proton and neutron orbits, however, as above.

It is possible to get a more fine-grained truncation.  When asked,
\begin{verbatim}
  Would you like to truncate ? (y/n/?=more information)
\end{verbatim}
choosing `{\tt p}'   allows different truncation on protons and neutrons:
\begin{verbatim}
  Max excite for sum, protons, neutrons? 
  (must be less than or equal to   8   4    4, respectively )
\end{verbatim}
That is, the maximum values of $W_p +W_n$, $W_p$, and $W_n$, respectively.  If you do not choose this option, then the limits are the 
same for all three. Please note, however, this truncation may not be robust for post-processing options such as 
expectation values (option `{\tt x}') and strength functions (`{\tt s}'), so we recommend avoiding this option.

\subsection{How to handle `different' proton-neutron spaces}

\label{protonneutron}

As of the current version, {\tt BIGSTICK} cannot directly handle independently defined proton and neutron spaces. You can, however trick it into behaving that way, with a 
small cost.   Both involve deft usage of the truncation and, in many cases, of particle-hole truncation.

Let's consider two toy cases.  First, suppose the proton and neutron spaces are entirely separate. For example, let's suppose valence  
protons occupy only the $0f_{7/2}$ space and valence neutrons only the $1p_{3/2}$. The {\tt .sps} file can look like:
\begin{verbatim}
iso
   2
0.0  3.0  3.5  0
1.0  1.0  1.5  1
\end{verbatim}
By choosing a {\tt Max excite} of zero, you will assure no particles are excited out of the $0f_{7/2}$ into the $1p_{3/2}$. (It is your responsiblity to 
set up the correct interaction file. You do not have to include cross-shell matrix elements if they are not needed; however if they are included, 
they will induce an effective single-particle energy so choose wisely.)

A more general, and \textbf{recommended}, approach is to use the {\tt pnw} format: suppose you want protons active in $0f_{7/2}$, $1p_{3/2}$ and $1p_{1/2}$, and neutrons in 
$1p_{3/2}$, $1p_{1/2}$, and $0f_{5/2}$.  Set up the {.sps} file
\begin{verbatim}
pnw
4  
       0.0  3.0  3.5  0    0
       1.0  1.0  1.5  0   99
       1.0  1.0  0.5  0   99
       0.0  3.0  2.5 99   99
\end{verbatim}
It is required that the proton and neutron orbits be the same, though the weight factors $w$ is the last column can differ. 
A weight of 99 signals that the orbital is `sterile' for either protons or neutrons, which means it will not be used.
Again, choosing {\tt Max excite} of zero will keep the protons and neutrons in their respective valence spaces.  If the valence spaces are 
significantly different, we strongly recommend utilizing particle-hole conjugation for the neutrons.  

One can make the truncations even more complex, for example allow a few protons to be excited but no neutrons, by careful usage of the 
options provided.   For example, setting
\begin{verbatim}
pnw
4  
       0.0  3.0  3.5  0     99
       1.0  1.0  1.5  1     99
       1.0  1.0  0.5  1     99
       0.0  3.0  2.5  99     0
\end{verbatim}
and setting the maximum truncation to 2, you can excite up to 2 protons out of the $0f_{7/2}$ into the 
$1p_{3/2}$ and $1p_{1/2}$ orbits, but none into the $0f_{5/2}$, while you will have only neutrons 
in the $0f_{5/2}$ but none in the $0f_{7/2}$-$1p_{3/2}$ -$1p_{1/2}$ orbits.

Here you must carefully consider the nature of the proton-neutron interaction.  Suppose you wanted 
four valence protons in the $0f_{7/2}$ -$1p_{3/2}$ - $1p_{1/2}$ space and 2 neutrons in the 
$0f_{5/2}$.  You could also set 
\begin{verbatim}
pnw
4  
       0.0  3.0  3.5  0    0
       1.0  1.0  1.5  1    0
       1.0  1.0  0.5  1    0
       0.0  3.0  2.5  99  99
\end{verbatim}
Because the $0f_{7/2}$ -$1p_{3/2}$ - $1p_{1/2}$ space has a total of 14 states, you have have instead 
set valence $N=14+2=16$.  With {\tt max excite} = 2, the neutrons in the $0f_{7/2}$ -$1p_{3/2}$ - $1p_{1/2}$ space
will be fixed.  In such cases it is often more efficient to use particle-hole conjugation ( section \ref{particlehole}).
In the valence neutron $0f_{5/2}$ space one wants 2 valence neutrons or 4 neutron holes, one then 
sets the number of valence neutrons to -4.  {\tt BIGSTICK} will confirm this corresponds to 16 neutrons all together, 
although it is not clever enough to tell you that 14 of them are fixed in a closed core. 

In this example, while the neutrons in $0f_{7/2}$ -$1p_{3/2}$ - $1p_{1/2}$ are fixed, they can have matrix 
elements with other particles, producing a change in single-particle energies. You should therefore understand 
carefully both your model space and your interactions. 

\textbf{Important}: Be careful in how you read in your interaction file. Although you are treating the proton and neutron spaces separately, in many cases the supplied interaction file, at least for empirical valence spaces, 
will still be in {\tt iso} format (see the next section for detail). You can test this by trying a small cases in your space, for example, just two protons and two neutrons. If you have set up correctly, you will get integer values of $J$. Alternately, if you get irrational values of $J$, the most likely culprit is that you have put in the wrong format for the interaction file.

\section{Interaction files}
\label{detailinput}

After the model space is defined, {\tt BIGSTICK} needs interaction matrix elements. All matrix elements are defined in the one-, two-, or possibly 
three-body-space.   {\tt BIGSTICK}'s job is to embed these matrix elements into a many-body space and solve the eigenvalue problem. 
(Because three-body interaction files are highly specialized, we do not discuss their format.)

The default format for two-body interaction file is derived from {\tt OXBASH/NuShell} and always ends in the extension {\tt .int}.  When entering the 
name of the file, only enter the name, not the extension, i.e., {\tt usdb} not {\tt usdb.int}; otherwise {\tt BIGSTICK} will misinterpret the file. 
\begin{verbatim}
!  Brown-Richter USDB interaction
63     2.1117   -3.9257   -3.2079  
  2  2  2  2    1  0  -1.3796
  2  2  2  1    1  0   3.4987
  2  2  1  1    1  0   1.6647
  2  2  1  3    1  0   0.0272
  2  2  3  3    1  0  -0.5344
  2  1  2  1    1  0  -6.0099
  2  1  1  1    1  0   0.1922
  2  1  1  3    1  0   1.6231
  2  1  3  3    1  0   2.0226
  1  1  1  1    1  0  -1.6582
  1  1  1  3    1  0  -0.8493
  1  1  3  3    1  0   0.1574
. . . 
\end{verbatim}
There is no specific spacing for this file. {\tt BIGSTICK} will skip any header lines starting with ! or $\#$.
The first line is

\smallskip

\textit{ Ntbme \qquad   spe(1) \qquad  spe(2) \qquad   spe(3) ... }

\smallskip

where \textit{Ntbme} is the number of \textit{two-body matrix elements} (TBMEs)
 in the rest of the 
file, and \textit{spe(i)} is the \textit{single-particle energy} of the $i$th orbit.  
(Note: older version required only 10 single particle energies are on each line. This has been changed and is no longer 
required.)  As a check, however, you should confirm that the code is reading in the correct first two-body matrix element, 
which is written both to screen and to the log file:
\begin{verbatim}
  As a check, first two-body matrix element is   -1.37960005    
\end{verbatim}

The rest of the file are the two-body matrix elements. This is defined as 
\begin{equation}
V_{JT}(ab,cd) = \langle ab; JT | V | cd; JT \rangle,
\label{tbmedef}
\end{equation}
where $a,b,c,d$ label orbits, as ordered in the {\tt .sps} file or as created by the autofill option; 
$J$ and $T$ are the total angular momentum and total isospin of the two-body states 
$| ab; JT \rangle$, which are normalized. This follows the convention of Brussaard and
Glaudemans.  Each matrix element is read in as

\smallskip

\textit{ a \qquad b \qquad  c \qquad d \qquad J \quad T } \qquad $V_{JT}(ab,cd)$ 

\smallskip

For input purposes, the order of $a,b,c,d$ is not important (as long as one has the 
correct phase), nor is the ordering of the TBMEs themselves.  When reading in the file, 
{\tt BIGSTICK} automatically stores the matrix element according to internal protocols, 
appropriately taking care of any relevant phases. 

Matrix elements that are zero can be left out, as long as $Ntbme$ correctly gives the 
number of TBMEs in the file. \textbf{More than one file can be read in; enter} {\tt end} 
\textbf{to  tell} {\tt BIGSTICK} \textbf{you are finished reading interaction files. }

\medskip

\textbf{Important}: {\tt Ntbme} cannot be 0. If it is zero, then {\tt BIGSTICK} will assume there are no matrix elements.  Some {\tt NuShell} input files 
have a zero here, but that will cause a problem. {\tt BIGSTICK} will give a warning:
\begin{verbatim}
  NO TWO-BODY MATRIX ELEMENTS FOUND 
\end{verbatim}
Note, however, that sometimes you might want to have no two-body matrix elements, for example, to add in single-particle energies only.

You can, however, set {\tt Ntbme} larger than the actual number of two-body matrix-elements, and {\tt BIGSTICK} will recover gracefully when it reaches the end 
of the file.

\medskip

\subsection{Scaling and autoscaling}

\label{scaling} 

Empirical studies with phenomenological interactions have found best agreement with experiment if one scales the two-body matrix elements 
with mass number $A$. (There is some justification based upon the scaling of harmonic oscillator wave functions with $A$). A standard scaling factor 
is 
\begin{equation}
\left ( \frac{ A_\mathrm{0} }{A} \right )^x
\end{equation}
where $A_\mathrm{0}$ is the reference mass number (typically $A$ of the frozen core +2, as it is fit to the interaction of two particles above
the frozen core), $A$ is the mass of the desired nucleus, and $x$ is an exponent, typically 
around $1/3$.  To accomodate this scaling, when reading in the default format, {\tt BIGSTICK} requests
\begin{verbatim}
  Enter scaling: spescale, A0,A,X 
 ( spescale scales single particle energies, 
  while TBMEs are scaled by (A0/A)^X ) for TBMEs 
  (If A or X = 0, then TBMES scaled by A0 ) 
\end{verbatim}
Typically the single particle energies are unscaled, but we allow for it. A typical entry, for example for the USDA/B interactions (\citep{PhysRevC.74.034315}), would 
be 
\begin{verbatim}
1   18   24   0.3
\end{verbatim}
Here the single particle energies are unscaled, the core has mass number 16 and hence the reference mass $A_0$ is 18, the target mass in this case has mass number $A=$ 24, and the 
exponent is 0.3.    Whoever provides the interaction has to provide the exponent.  If unsure, just enter
\begin{verbatim}
1, 1, 1, 1
\end{verbatim}

Many files used with {\tt NuShell} have autoscaling. For example, for the USDA/B file, the first lines are 
\begin{verbatim}
! 1=d3/2 2=d5/2 3=s1/2
! the first line has the three single-particle energies
! the - sign tells oxbash that the tbme have a mass dependence of the form
! [18/(16+n)]^0.3 where n is the number of valence particles
-63     1.9798   -3.9436   -3.0612   16.0000   18.0000 0.30000
\end{verbatim}
 A negative integer for the number of two-body matrix elements  (here, -63) initiates autoscaling.  
 The next three numbers are the single-particle energies, and the next numbers are $A_\mathrm{core}$, the reference mass, 
 and the exponent. If {\tt BIGSTICK} encounters a negative integer for the number of two-body matrix elements, it autoscale the 
 two-body matrix elements as described above. To turn off autoscaling, change -63 to 63.
 
 Keep in mind that not all interactions will be scaled.  \textit{Ab initio} interactions are almost never scaled, and `phenomenological'
 interactions depend on how they were derived and fit. See your interaction provider for more information.
 
 If you enable autoscaling (by setting  the number of matrix elements negative) and set the three parameters ($A_\mathrm{core},$ reference mass, and exponent) to zero, 
 i.e., so it looks like
 \begin{verbatim}
 -63     1.9798   -3.9436   -3.0612   0    0     0
 \end{verbatim} then all parameters will be left unchanged, that is, autoscaled by one; furthermore,  you will not be asked to enter in scaling factors. Autoscaling in both forms may be useful for impatient users and and users not comfortable with scaling.

\subsection{Proton-neutron and other isospin-breaking formats}

\label{pnham}

Often one needs to break isospin.  There are three modifications of the default format which break isospin. In addition, \textit{ab initio} inputs in 
the {\tt MFDn} format, described in section \ref{MFDinput}, also generally break isospin. 

The most robust format, which we recommend, is the explicit proton-neutron formalism.  Here one has separate labels for proton and neutron 
orbits; however, at this time \textbf{the proton and neutron orbits must have the same quantum numbers and be listed in the same order.}
For example, one might label the proton orbits $1= 0d_{3/2}$, $2=0d_{5/2}$, and $3=1s_{1/2}$. Then the neutron orbits must be 
$4= 0d_{3/2}$, $5=0d_{5/2}$, and $6=1s_{1/2}$.

While {\tt BIGSTICK} generally allows for arbitrary order, for the proton-neutron matrix elements the proton labels must be in the first and third columns and neutron labels in the second and fourth columns, 
that is, for $V_J(ab,cd)$, $a$ and $c$ \textit{must} be proton labels and $b,d$ must be neutron labels. 
With twice as many defined orbits, one must also provide separate 
proton and neutron single particle energies.  As an example, here is part of the file of the $p$-shell Cohen-Kurath matrix elements with good isospin:
\begin{verbatim}
! ORDER IS:  1 = 1P1/2    2 = 1P3/2 
   15      2.419      1.129
  1   1   1   1        0   1      0.2440000
  1   1   1   1        1   0     -4.2921500
  2   1   1   1        1   0      1.2047000
  2   1   2   1        1   0     -6.5627000
  2   1   2   1        1   1      0.7344000
  2   1   2   1        2   0     -4.0579000
  2   1   2   1        2   1     -1.1443000
  2   2   1   1        0   1     -5.0526000
\end{verbatim}
and here is an excerpt in proton-neutron formalism
\begin{verbatim}
    34    2.4190    1.1290    2.4190    1.1290
   1   3   1   3     0   1   0.24400
   1   1   1   1     0   1   0.24400
   3   3   3   3     0   1   0.24400
   1   3   1   3     1   1  -4.29215
   1   3   1   4     1   1  -0.85185
   1   3   2   3     1   1   0.85185
   1   3   2   4     0   1  -5.05260
   1   1   2   2     0   1  -5.05260
   3   3   4   4     0   1  -5.05260
   1   3   2   4     1   1   1.76980
   1   4   1   4     1   1  -2.91415
   1   2   1   2     1   1   0.73440
   3   4   3   4     1   1   0.73440
\end{verbatim}
In no case are headers required, but they do help as a check for the definition of the orbits. {\tt BIGSTICK} automatically checks 
that angular momentum and parity selections are not violated.  In the explicit proton neutron format $T$ is given in the sixth column but not actually used. 

\bigskip

There is one more question of convention one must deal with: the normalization of the two-body states in the definition of 
matrix elements.  All formats assume two-proton and two-neutron states are normalized, and states with good isospin are normalized. 
Files set up for {\tt NuShellX}, however, have \textit{unnormalized} proton-neutron states.  

{\tt BIGSTICK} can read in default-format proton-neutron interactions with either normalized (`{\tt xpn}' or explicit proton-neutron) 
or unnormalized (`{\tt upn}' or unnormalized proton-neutron) conventions. In both cases the files also include proton-proton and neutron-neutron
matrix elements, with normalized states. 

The relationship between the two is
\begin{equation}
V_J^{xpn}(a_\pi b_\nu, c_\pi d_\nu) = \frac{\sqrt{(1+ \delta_{ab})(1+ \delta_{cd})}}{2} V_J^{upn}(a_\pi b_\nu, c_\pi d_\nu) 
\label{upn2xpn}
\end{equation}
(\textit{Note: In older versions of this manual, the ratio in (\ref{upn2xpn}) was erroneously reversed. Eq.~(\ref{upn2xpn}) is now 
consistent with (\ref{xpndef}) and (\ref{upndef})})
Here we have marked the orbits $a,c$ as proton and $b,d$ as neutron, but the Kronecker-$\delta$s refer only to the quantum numbers $n,l, j$. 
For example, in the $sd$ shell, with the labels mentioned above, 
$$
V_J^{xpn}(16,25) = \frac{1}{\sqrt{2}} V_J^{upn}(16,25)
$$
because proton orbit 1 $(0d_{3/2})$ and neutron orbit 6 $(1s_{1/2})$are different, but proton orbit 2 and neutron orbit 5 are both $d_{5/2}$

(Another wrinkle:  {\tt NuShellX}-style files \textit{occasionally}, albeit rarely, include Hermitian conjugates of certain \textbf{proton-neutron} elements.
Specifically, for matrix elements of the form $V_J^{upn}(a_\pi b_\nu, b_\pi a_\nu) $, where $a\neq b$,
the matrix elements $V_J^{upn}(b_\pi a_\nu, a_\pi b_\nu) $ is also included in the file. 
Because {\tt BIGSTICK} automatically fills in matrix elements using Hermiticity, and allows for any ordering of $a,b,c,d$, (so that if $V(ab,cd)$ is read in, then $V(cd,ab)$ is \textit{always} automatically included); this means the code  
struggles to account for these extra elements. 
\textbf{Use such input files with caution}.) 

It is up the user to know whether or not the file uses normalized or unnormalized proton-neutron states. If the file was originally produced 
for use with {\tt NuShellX}, it is almost certainly the latter. 

(This arises out of the conversion of normalized isospin wave function to normalized proton-neutron wave functions and the result matrix elements.
One finds
\begin{equation}
V_J^\mathrm{pn}(ab,cd) = \frac{ \sqrt{(1+\delta_{ab})(1+\delta_{cd})}}{2} \left [ V_{J,T=0}^\mathrm{iso} (ab,cd) + 
V_{J,T=1}^\mathrm{iso} (ab,cd) \right ],
\label{xpndef}
\end{equation}
but the unnormalized convention yields the simpler
\begin{equation}
V_J^\mathrm{upn}(ab,cd) = V_{J,T=0}^\mathrm{iso} (ab,cd) + 
V_{J,T=1}^\mathrm{iso} (ab,cd). \label{upndef}
\end{equation}
While our preference is for the former, given the prominence  of the latter  through {\tt NuShellX} we include it as an option.)

\bigskip

In order to read in proton-neutron matrix elements in the default format, you must first tell {\tt BIGSTICK} to expect it.
\begin{verbatim}
  For xpn/upn formats, you MUST specify the format. 
  In this format proton and neutron orbits are sequential and do not overlap, 
  E.g., proton orbits are 1,2,3 and neutron orbits are 4,5,6. 
  FOR NOW despite the distinct numbering the proton and neutron orbits 
  must encompass the same space. 
  NOTE:  upn format is typical for TBME files distributed with NuShell; 
  xpn/upn files must have the name XXX.int, but enter XXX when requested. 
\end{verbatim}
That is, you must \textit{first} enter either the code {\tt xpn} or {\tt upn}, and then the filename:
\begin{verbatim}
  Enter two-body interaction file name OR file format code (e.g., XPN) 
  (Enter "end" to finish; "opt" for file format options; "?" for  general info ) 
xpn
  Enter name of two-body interaction file in explicit proton-neutron format 
usdbpn
\end{verbatim}

As with default-format isospin-conserving files, the file name must be {\tt xxxx.int}, but 
the user enters in just `{\tt xxxx}'. 

Also as with default-format isospin-conserving files, after entering the name of the file, the user 
is prompted for scaling. For maximal flexibility, there are two layers of possible scaling. The first 
is the standard phenomenological scaling:
\begin{verbatim}
  Enter global scaling for spes, A,B,X ( (A/B)^X ) for TBMEs 
  (If B or X = 0, then scale by A ) 
1 18. 24. 0.3
\end{verbatim}
These scalings are applied to all single particle energies and to all two-body matrix  elements.
In addition, one can enter in separate scaling factors for protons single-particle energies, neutron 
single-particle energies, proton-proton two-body matrix elements, neutron-neutron two-body matrix elements,  
and finally proton-neutron two-body matrix elements:
\begin{verbatim}
  Enter individual scaling for: proton spes, neutron spes, pp TBMEs, nn TBMEs, p
 n TBMES 
  (If not sure, just enter 1 1 1 1 1 )
\end{verbatim}

\bigskip

There are two alternate formats for isospin-breaking files which build upon the default format. These involve reading in separate files 
for proton-proton, neutron-neutron, and proton-neutron, or for isoscalar, isovector, and isotensor components.  There are some tricky issues 
of definition, however. Thus we do 
not actively support these alternative formats, instead recommending the explicit proton-neutron format, whether normalized or unnormalized

\bigskip

One can mix all of these different formats. You can read in an isospin-conserving file, a proton-neutron format file, and so on, in
any order. To stop reading in interaction files, enter `{\tt end}' at the prompt. 

\subsection{General one-body interactions}
\label{onebodypotential}

Interactions generally include one-body and two-body contributions, with three-body used for advanced applications.
For historical reasons, the standard format adopted for {\tt BIGSTICK} only reads in the the diagonal part of 
the one-body Hamiltonian, usually referred to as \textit{single-particle energies.} In phenomenological spaces, 
such as the $sd$ and $pf$ spaces, this is all that is possible. In  multi-shell spaces, however, one can have off-diagonal matrix elements
of a one-body part of the Hamiltonian, between orbits with the same $l$ and $j$ but different $n$.

 If you need to use off-diagonal one-body matrix elements, i.e. for some kind of potential which is not diagonal in the basis, there is an extension 
 for both the {\tt iso} (default) and the {\tt xpn} interaction formats. In addition to the mandatory list of single-particle energies at the beginning 
 of the file, one can add 
  one-body matrix elements \textit{after} reading in the two-body matrix elements 
( \textit{be sure} that the number of two-body matrix elements are correctly specified at the 
tope of the file). Immediately after the last two-body matrix element, specify the number of one-body matrix elements.
\begin{verbatim}
N1me  ! = # of one-body potential matrix elements
\end{verbatim}
followed by a list of the orbital indices and the one-body matrix elements.
\begin{verbatim}
   a      b    U(a,b)
\end{verbatim}
where $a,b$ are the orbital labels in the {\tt xpn} ordering--hence proton and neutron labels are different--and $U(a,b)$ is the matrix element. (Note that, 
like the two-body matrix elements, these are \textit{not} reduced via the Wigner-Eckart theorem.) 
The order of $a,b$ is not important, and only nonzero matrix elements need to be added.  You only should list $a,b$ and not also $b,a$--but again, 
the order does not matter.  Diagonal ($a=b$) are okay if they were not already read in as ``single-particle energies.''  Hence these will look like
\begin{verbatim}
    4   !  # of one-body matrix elements.
    1     2    -0.0144
    1     3     1.0888
    2     3    -2.2220
    4     5     0.8877
\end{verbatim}
{\tt BIGSTICK} will automatically check for these matrix elements.  If they are not present {\tt BIGSTICK} will skip over them gracefully. 
 If you are working in the {\tt xpn} format, be sure that you have 
included the matrix elements for both protons and neutrons; even if identical.  The diagonal elements can either appear at the beginning of the file, 
as `single-particle energies,' or here. If you specify the diagonal one-body matrix elements at the end of the file, then you must set the single-particle energies
at the beginning of the file to zero; not listing single-particle energies, even zeroes, will confuse {\tt BIGSTICK}. 

Note: at this time, {\tt BIGSTICK} cannot handle off-diagonal one-body Hamiltonians for a single particle \textit{or} a single hole.
The reason is, {\tt BIGSTICK} generally converts the one-body part of the Hamiltonian to an effective, number-dependent two-body operator, but this does not work 
if there is only one particle (or, if one invokes particle-hole conjugation, as described in section \ref{particlehole}, a single hole).  For diagonal single-particle energies {\tt BIGSTICK} can 
handle a single particle or hole with a specialized routine. In principle this could be generalized, but to date has not been. 

We emphasize all of the above regards the \textit{Hamiltonian}. One-body operators read in for option `{\tt (o)}' can be completely general, i.e., non-scalar or having 
angular momentum rank  $K > 0$: see section \ref{apply1body}.



\subsection{{\tt MFDn } format input}
\label{MFDinput}

Another major configuration-interaction code is {\tt MFDn} (Many-Fermion-Dynamics, nuclear 
version) out of Iowa State University (\citep{MFDN}). While within {\tt MFDn} there are several variations on 
conventions, we describe here the most common conventions. 

Unlike the default format, to read in an {\tt MFDn}-format file, you must enter the \textit{entire} name, including any extensions.
This signals to {\tt BIGSTICK} to prepare to read in an an {\tt MFDn}-format file. {\tt BIGSTICK} will treat a file {\tt TBME.int} 
very differently if you answer `{\tt TBME}' versus `{\tt TBME.int}' for the file name. 
{\tt MFDn}-format files are almost always for \textit{ab initio} or so-called \textit{no-core shell model} calculations, and almost 
always assume a harmonic oscillator basis. 

The input file first line is

\smallskip

\noindent {\tt nTBME} (other stuff which are not needed)

\smallskip

where {\tt nTBME} is the number of TBMEs in the file.  For example
\begin{verbatim}
   2056271     13   14    20.0000     2.0000
\end{verbatim}
The only number {\tt BIGSTICK} requires is the first one.  The fourth number, 20.000, is $\hbar \Omega$ in MeV, but 
not all codes generate this information.

The {\tt MFDn} format does not include explicit single-particle energies. 
Subsquent lines are of the form

\smallskip

\noindent {\tt a b c d J T Trel  Hrel  Vcoul  V } 

\smallskip 

or, more commonly, 

\smallskip

\noindent {\tt a b c d J T Trel  Hrel  Vcoul  Vpn Vpp Vnn } 

\smallskip 

Here all matrix elements are of the form $\langle a b; JT | V | c d ; JT \rangle$, that 
is the matrix element between \textit{normalized} two-body states with 
{\tt a,b,c,d} labels of single particle orbits, {\tt J} (and, optionally, {\tt T}) are 
total angular momentum and isospin of the coupled two-body states. The isospin $T$ is not really used.

Now for the matrix element. {\tt Trel} is the relative kinetic energy, that is 
\begin{equation}
\hat{T}_\mathrm{rel} = \sum_{i < j} \frac{ (\vec{p}_i - \vec{p}_j)^2}{2 M},
\end{equation}
These matrix elements are  computed in a harmonic oscillator basis for $\hbar \Omega =1 MeV$, 
and $A = 2$, and thus must be rescaled correctly for the $A$-body system, that is, must 
be multiplied by $2 \hbar \Omega/A$.

Now to the final matrix elements. The actual Hamiltonian one wants is 
\begin{equation}
\hat{H} = \hat{T}_\mathrm{rel} + \hat{V}_\mathrm{rel} 
+ \beta_{c.m.} (\hat{H}_\mathrm{cm}-\frac{3}{2} \hbar \Omega).
\end{equation}
Here $\hat{H}_\mathrm{cm}$ is the center-of-mass Hamiltonian, used to push up spurious states via the Palumbo-Lawson-Glocke method (\citep{palumbo1967intrinsic,palumbo1968effects,gloeckner1974spurious}):
\begin{equation}
\hat{H}_\mathrm{cm} = \frac{P^2_\mathrm{cm} }{2A m_N} 
+ \frac{1}{2} A m_N \Omega^2 R^2_\mathrm{cm}.
\end{equation}
where $m_N$ is the nucleon mass and 
\begin{equation}
\vec{R}_\mathrm{cm} = \frac{1}{A} \sum_i \vec{r}_i, \,\,\,
\vec{P}_\mathrm{cm} = \frac{\hbar}{i} \sum_i \vec{\nabla}_i.
\end{equation}
These have the correct commutation relation, that is, 
$[ \vec{R}_\mathrm{cm}, \vec{P}_\mathrm{cm} ] = i \hbar$,
so that $\vec{P}_\mathrm{cm}$ is the conjugate momentum to 
$\vec{R}_\mathrm{cm}$. 

It is useful to separate $\hat{H}_\mathrm{cm}$  into one- and two-body parts:
\begin{eqnarray}
\hat{H}_\mathrm{cm} =\frac{1 }{2Am_N} \sum_i p^2_i + \frac{1}{2A} m_N \Omega^2
\sum_i r^2_i + \nonumber \\
\frac{1 }{2A m_N} \sum_{i \neq j} \vec{p}_i \cdot \vec{p}_j 
+ \frac{1}{2A} m_N \Omega^2 \sum_{i \neq j} \vec{r}_i \cdot \vec{r}_j 
\end{eqnarray}
The first two terms are the single particle energies, with values $ \hbar \Omega (N+3/2)/A$, with $N$ the 
principal quantum number, and the second two terms is $\hat{H}_{rel} \times \hbar \Omega/A$. {\tt BIGSTICK} 
automatically accounts for all the factors, as long as you provide the correct $ \hbar \Omega$ as shown below:

When you select an {\tt MFDn}-format file, you will be prompted for the following:
\begin{verbatim}
  For MFD-formatted input choose one of the following :
  (I) No isospin breaking 
  (P) Explicit proton-neutron formalism 
  (C) Isospin breaking only through adding Coulomb 
\end{verbatim}
Almost always you should select option `{\tt p}'.

In order to use an \textit{ab initio} file, you need to enter in the value of $\hbar\Omega$ for both the kinetic energy 
term and the center-of-mass Hamiltonian to push up spurious states:
\begin{verbatim}
  Enter oscillator frequency (in MeV) and center-of-mass strength 
\end{verbatim}
You should know the frequency at which the file was created. The second term is $\beta_{c.m.}$. Typical values of $\beta_{c.m.}$ are 1-10. 

 \subsection{Using {\tt NuHamil}}

\label{nuhamil}

 Takayuki Miyagi's open source code {\tt NuHamil}, (arXiv:2302.07962, 
 github.com/Takayuki-Miyagi/NuHamil-public)
 can generate a 
 number of interactions from chiral effective field theory. One can 
 get {\tt NuHamil} to generate an file in MFDn format by including the 
 line
 \begin{verbatim}
 ext_2bme = ".MFDn”    
 \end{verbatim}
in the Python script {\tt exe/NuHamil\_2BME.py} included with the 
NuHamil distribution.  We take no responsibility for installing or running {\tt NuHamil}.

\subsection{Three-body forces}

While {\tt BIGSTICK} has a validated capability for  three-body forces, it is not optimized for large calculations; the 
main issue is storage of the large number of matrix elements. If you have the capability to generate three-body forces, 
please contact us, {\tt cjohnson@sdsu.edu}. We do not have the codes or capability to generate three-body forces for users.

In your distribution three-body forces are likely disabled. They can be re-enabled by setting the logical flag {\tt threebodycheck = .true.}
in module {\tt flags3body} in the file {\tt bmodules\_3body.f90}.  If this flag is enabled, {\tt BIGSTICK} will query if you want to 
use three-body forces:
\begin{verbatim}
  Do you want 3-body forces (y/n) ?
\end{verbatim}
If you answer `{\tt n},' {\tt BIGSTICK} will proceed with just 2-body forces.   If you answer `{\tt y},' {\tt BIGSTICK} will ask for the name of the 
file. Actually using three-body forces is complicated and beyond the scope of this current manual. 

You can, however, use two-body forces  in three-body mode (the matrix elements are multiplied internally by $(\hat{N}-2)/(A-1)$ to turn them into 
genuine three-body forces), by answering `{\tt none}' to the question of the name of the file of three-body forces. Most users will not be interested in this.

\section{Primary runtime options}

\label{mainmenu}

Here we outline the major run time options, although some issues are discussed in more detail elsewhere. The 
main menu can be divided into two categories, primary and secondary. We discuss secondary options, which require results of a 
previous run, in 
section \ref{postprocess}.  In most primary runs one solves the matrix eigenvalue problem, which invokes 
the diagonalization options menu, discussed in Appendix \ref{diagonalmenu}.

\subsection{Autoinput}

\label{autoinput}

\begin{verbatim}
  * (i) Input automatically read from "autoinput.bigstick" file             * 
  *  (note: autoinput.bigstick file created with each nonauto run)          *
\end{verbatim}
Each time {\tt BIGSTICK} runs, it writes the user's responses to a file {\tt autoinput.bigstick}. This file can be edited or used as the 
basis of a batch file. 
The \textit{autoinput} option, `{\tt i},' will read in the {\tt autoinput.bigstick} instead of taking responses from the user. 

\subsection{Standard or normal runs}

\label{normal}

\begin{verbatim}
  * (n) Compute spectrum (default); (ns) to suppress eigenvector write up   * 
\end{verbatim}
The \textit{normal} run, `{\tt n},' will generate the low-lying eigenspectrum and wave functions. This is the most common option. 
Two variations on it are {\tt ns} which will compute the eigenspectrum and the $J$ and $T$ values, but not write the wavefunctions to a 
file, and {\tt ne}, which will only compute eigenenergies. These latter options can save on time and file storage, but in most cases are not 
necessary. 

\subsection{One-body density matrices and occupations}
\label{densitymatrix}

One of the most important options for {\tt BIGSTICK} is to generate the one-body density matrices, defined as 
\begin{equation}
\rho^{fi}_K(a^\dagger b) =  [K]^{-1}  \langle J_f || (a^\dagger b)_K || J_i \rangle
\end{equation}
where we  use the choice of reduced matrix elements from  \citet{edmonds1996angular}, 
\begin{equation}
\langle J_f || O_K || J_i \rangle = [J_f] ( J_f M_f, KM | J_i M_i)^{-1} \langle J_f M_f | O_KM | J_i M_i\rangle
\end{equation}
The advantage of  this definition of the density matrix is that the reduced matrix element  of a general, non-scalar one-body operator is just 
the density matrix $\times$ the reduced matrix elements of that operator, that is
\begin{equation}
\langle \Psi_f, J_f || \hat{\cal O}_K || \Psi_i, J_i \rangle = \sum_{ab} \langle a || \hat{\cal O}_K || b \rangle \rho^{fi}_K(a^\dagger b) 
\end{equation}
where $a,b$ are labels for single-particle orbits, and $\langle a || \hat{\cal O}_K || b \rangle$ are the reduced one-body matrix 
matrix elements for the operator $ \hat{\cal O}$ with angular momentum $K$. 

{\tt BIGSTICK} has a number of options to generate density matrices. 
\begin{verbatim}
  * (d) Densities: Compute spectrum + all one-body densities                * 
  * (dx[m]) Densities: Compute one-body densities from previous run (.wfn)  * 
  *     optional m enables mathematica output                               * 
  * (dxp) Compute one-body densities from prior run (.wfn) in p-n format.   * 
\end{verbatim}
The density matrix option `{\tt d}' runs just like the normal option, except at the end of the run it generates the one-body density matrices, which 
we describe more fully in section \label{densities}.   
If the interaction file has good isospin, then the one-body density matrices will be coupled up to good isospin. 
If the interaction file breaks isospin, the density matrices will be in proton-neutron format.  If you use an interaction with good isospin but 
want the density matrices in proton-neutron format, use the option `{\tt dp}.'

Three variations are the option `{\tt dx},' which reads in a previously computed wave function file and from it computes
the one-body density matrices in isospin format; `{\tt dxm}', which does the same but  generates the density matrices in a format readable by Mathematica;
and {\tt dxp} which computes from a prior wave function file the one-body densities in proton-neutron format.  The output files have the extension {\tt .dres}.  
(At this time, there is not an option to write out one-body densities in a proton-neutron format readable by Mathematica.)  

When invoked, these options will ask for a range of initial and final states, .e.g,
\begin{verbatim}
           5  states 
  Enter start, stop for initial states 
  (Enter 0,0  to read all )
0 5
  Enter start, stop for final states 
  (Enter 0,0  to read all )
2 3
\end{verbatim}
 This is useful for cases where one has generated many states but only wants to extract 
densities for a few states.

\medskip

\textbf{Note}: When running in MPI mode, the amount of work need to calculate one-body densities is far less than 
for computing the two-body Hamiltonian. {\tt BIGSTICK} can get hung up when trying to distribute work for densities over a large number of MPI ranks. In that case, it is better to simply generate the wave functions using a large number of MPI ranks, and 
then re-run with option {\tt dx/dxp} with few MPI ranks.

\subsubsection{One-body densities: binary format}

\label{density_binary}

This is still in progess and will be updated in future versions. See 
\cite{gorton2024shell}

\subsection{Single-particle occupations}

\label{occ1b}

\begin{verbatim}
  * (p) Compute spectrum + single-particle occupations; (ps) to suppress wfn* 
  * (occ) single-particle occupations (from previous wfn)                   * 
\end{verbatim}
A restricted version of the one-body densities are the single-particle occupations. In principle given the former one can compute the latter, as 
described in section \ref{occfromden}, but for convenience we give an option to do this directly. Option `{\tt p}' does this as in the normal option, but also writes the 
single-particle occupations to the {\tt .res} file.  Option '{\tt ps}' does the same but does not write the wave functions to a file on disk.  Finally, the secondary 
option `{\tt occ}' reads in an existing wave function file and generates the single-particle occupations.
\begin{verbatim}
  Single particle state quantum numbers
ORBIT :      1     2     3  
   N :      0     0     1  
   J :      3     5     1  
   L :      2     2     0  
  
  State      E        Ex         J       T 
    1    -62.78960   0.00000     2.500   1.500
     p occ:  0.136    1.590    0.275
     n_occ:  0.353    4.299    0.347
\end{verbatim}

A detailed discussion of using one-body density matrices to get transition probabilities can be found in Chapter \ref{densities}.

When you run a density matrix option, such as `{\tt (d)}', etc., 
in addition to the {\tt .dres} file the occupations will be automatically 
written to a {\tt .occres} file. The format here is more machine-readable:
\begin{verbatim}
  Single particle state quantum numbers
ORBIT      N     L   2 x J  
     1     0     2     3
     2     0     2     5
     3     1     0     1
  
 State #    1 E =  -87.10445 2xJ, 2xT =    0   0
   1    0.5245    0.5245
   2    3.0420    3.0420
   3    0.4335    0.4335    
\end{verbatim}

Experience suggests this is a better route in 
most cases than our somewhat clumsy format

\section{Other primary options}

\label{other}

Here we briefly discuss a number of other options from the primary menu.

\subsection{Modeling}

\label{model}

\begin{verbatim}
  * (m) print information for Modeling parallel distribution                * 
\end{verbatim}
The modeling option `{\tt m}' is useful for seeing if enough nodes and memory can be allocated for a large parallel run. See also
section \ref{modeling}.  No interaction files are read in and no diagonalization is carried out. 

The option `{\tt(m0)}' will compute only the basis dimension, while the option `{\tt(md)}' will compute the memory requirements for computing one-body densities in MPI.

\subsection{Traces}

\label{trace}

There is an additional option most users are unlikely to use but which we mention nontheless. 
\begin{verbatim}
  * (c) Compute traces                                                     
\end{verbatim}
{\tt BIGSTICK} can  compute the trace of the Hamiltonian and the trace of the Hamiltonian squared, using the option `{\tt c}.'  Specifically, 
it computes the centroid, which is the trace divided by the dimension, and the width, which is the square root of the variance. 
\begin{verbatim}
 for dimension    28503, centroid =   -43.544457, width =    11.298695
  (saved to file trace.bigstick )
\end{verbatim}
As shown above, the results are written to the file {\tt trace.bigstick}. We caution users that in particular computing the width can be time consuming.
We have not fully tested this option in parallel. 

\subsection{Configurations and configuration occupations}

\label{config}

Finally, the menu option `{\tt (co)}' will, after a normal run, compute the \textit{configuation occupation}. (The option `{\tt (cx)}' will compute the configuration occupation from a previously computed wave function.)  
Here a `configuration' is sometimes also called a `partition,' and means a 
subspace defined by the occupation of orbits. That is, if we have proton orbits $0s_{1/2, \pi}$, $0p_{3/2, \pi}, 0p_{1/2, \pi}$, etc, and similarly for neutrons, we label a subspace by the number of protons and neutrons in 
each orbit, e.g. $(0s_{1/2, \pi})^2 (0p_{3.2, \pi}^3 (0p_{1/2, \pi}^1,  0s_{1/2, \nu})^1 (0p_{3.2, \nu}^2 (0p_{1/2, \nu}^2$.  The `{\tt (co)}' option produces what fraction of the 
wave function is in each configuration subspace/partition.

In addition to producing the usual {\tt .res} and {\tt .wfn} files, a file with extension {\tt .cfo } will contain the configuration occupations, which are in proton-neutron format. 
The  {\tt .cfo} begins with a description of the single-particle valence space and a list of the orbits:
\begin{verbatim}
 Z =    4 N =    4    ! valence # of protons and neutrons
  Proton orbits 
    0    2    3    0     ! n   l    2xj    w
    0    2    5    0
    1    0    1    0
  Neutron orbits 
    0    2    3    0
    0    2    5    0
    1    0    1    0
  Parity =            1  ! note: if all particles have same parity, then parity=1
                              ! otherwise parity = 1 is +, parity = 2 is -
  No W truncation 
\end{verbatim}
It then lists the proton and then neutron 'configurations':
\begin{verbatim}
  Proton configurations 
     1 :   4  0  0  
     2 :   3  1  0
     3 :   3  0  1
     4 :   2  2  0
...
     8 :   1  2  1
     9 :   1  1  2
    10 :   0  4  0
...
\end{verbatim}
where, e.g, 
\begin{verbatim}
     2 :   3  1  0
\end{verbatim}
means proton configuration $\#2$ has 3 protons in orbit 1 (which, from the list of orbits, is the $0d_{3/2}$ orbit), 
1 proton in orbit 2 ($0d_{5/2}$) and none in orbit 3 ($1s_{1/2}$) .

After the proton configurations the neutron configurations are listed in a similar manner, and then finally, 
for each eigenstate,  the combined configurations and the fraction of the wave function in each configuration:
\begin{verbatim}
  State            1
config   p config   n config      fraction
       1       1       1        0.0000039
       2       1       2        0.0000000
       3       1       3        0.0000000
       4       1       4        0.0001510
...
     116      10       8        0.0216867
     117      10       9        0.0024511
     118      10      10        0.1866053
     119      10      11        0.0367340
     120      10      12        0.0250714
....
\end{verbatim}
Note that for this state, $18.7\%$ of the wave function is in proton configuration 10 and neutron configuration 10. Looking up at the 
list of proton configurations above, that configuration has four protons in the $0d_{5/2}$ orbit; although not shown, it is the same here for 
the neutron configuration.

\section{Diagonalization options}
\label{diagonalmenu}

After the interactions files have been read in, {\tt BIGSTICK} sets up the jump arrays for reconstructing the matrix elements on 
the fly. After that, the eigensolver menu comes up:
\begin{verbatim}
  / ------------------------------------------------------------------------\ 
  |                                                                         | 
  |    DIAGONALIZATION OPTIONS (choose one)          | 
  |                                                                         | 
  | (ex) Exact and full diagonalization (use for small dimensions only)     | 
  |                                                                         | 
  | (ld) Lanczos with default convergence (STANDARD)                        | 
  | (lf) Lanczos with fixed (user-chosen) iterations                        | 
  | (lc) Lanczos with user-defined convergence                              | 
 |                                                                         | 
  | (bd) Block Lanczos with default convergence (STANDARD)                  | 
  | (bf) Block Lanczos with fixed (user-chosen) iterations                  | 
  | (bc) Block Lanczos with user-defined convergence                        | 
  |                                                                         | 
  | (td) Thick-restart Lanczos with default convergence                     | 
  | (tf) Thick-restart Lanczos with fixed iterations                        | 
  | (tc) Thick-restart Lanczos with user-defined convergence                | 
  | (tx) Thick-restart Lanczos targeting states near specified energy       | 
  | (tb) Thick-restart block Lanczos with default convergence  
  |                                                                         | 
  | (sk) Skip Lanczos (only used for timing set up)                         | 
  | (li) Lanczos iterations only, no further eigensolutions                 | 
  |                                                                         | 
  \ ------------------------------------------------------------------------/ 
\end{verbatim}

The full diagonalization option, `{\tt ex},' creates the entire Hamiltonian matrix, stores it in memory, and solves it using the Householder algorithm 
as implemented in the LAPACK routine {\tt DSYEV}. As such, it should not be used except for relatively small dimensions.  On workstations 
one can solve up to $\sim 10^3$ in a few or tens of minutes. We have solved up to $\sim 10^4$, but that can take hours. In principle MPI versions 
of Householder exist, but we have not installed one, as one seldom has need for all eigensolutions of a very large matrix. (If you do not need wave 
functions or the angular momenta, choosing option `{\tt ne}' will speed this up dramatically, as {\tt DSYEV} will run faster if one wants \textit{only} 
eigenvalues.) 

Under this option you can choose how many low-lying states, or all of them if you wish, to keep. These get written to file.

\bigskip

The primary eigensolver is the Lanczos algorithm, described in Chapter \ref{lanczos}.  Most of the time you will use option `{\tt ld}', the default 
Lanczos choice. Here you get asked
\begin{verbatim}
  Enter nkeep, max # iterations for lanczos 
  (nkeep = # of states printed out )
\end{verbatim}
{\tt BIGSTICK} will run until the standard convergence criterion, see section \ref{converge}, is satisfied, or until the maximum number of 
iterations is exceeded. The latter must be specified to reserve memory for the Lanczos vectors. Although the Lanczos and related Arnoldi algorithms are 
among the most studied in applied mathematics, there is no simple, robust rule for the number of iterations needed. For phenomenological 
spaces, the ground state will often 
converge in under 50 iterations, the first 5 states in 100 to 150 iterations, and so on. For no-core shell model calculations, the time to convergence is 
usualy longer. 

The default convergence check is discussed in the next section, \ref{converge}.  If you want a fixed number of iterations without checking
convergence, choose `{\tt lf}.'  If you want finer control over convergence, choose `{\tt lc},' discussed in \ref{converge}.

\bigskip

Block Lanczos is discussed below in section \ref{block}, with additional information in section \ref{blocklanczos}.

\bigskip

The Lanczos vectors are stored in memory. For large-dimension cases, especially on a laptop or desktop, one can run out of memory just 
storing these vectors. Alternately, if one needs a large number of converged states, after a number of iterations reorthogonalization actually starts to 
take more time than matvec. A robust alternative is the thick-restart Lanczos algorithm \citep{wu2000thick}, 
which requires fewer vectors stored in memory 
but requires more iterations.  While standard Lanczos finds the lowest $N_\mathrm{keep}$ eigensolutions with $N_\mathrm{iter}$ iterations, 
thick-restart has three numbers: $N_\mathrm{keep} < N_\mathrm{thick} < N_\mathrm{iter}$.  As described more fully in section \ref{thickrestart}, 
after  $N_\mathrm{iter}$ iterations the approximate Hamiltonian is diagonalized, and $N_\mathrm{thick}$ of these eigenvectors are kept for 
restarting. This process is repeated  until convergence or until a maximum number of restarts has been exhausted.
\begin{verbatim}
td
  Enter # of states to keep, number of iterations before restarting 
5 50 
  Enter max # of restarts 
10
\end{verbatim}
As with standard Lanczos, the values chosen usually come with experience. 
 We find we usually want $N_\mathrm{thick}  > \sim 3 \times N_\mathrm{keep}$, 
and we take $ N_\mathrm{iter}$ as large as practical.  Specifically, {\tt BIGSTICK} chooses 
$$
N_\mathrm{thick} = \max( 3 N_\mathrm{keep}, N_\mathrm{keep} + 5),
$$ 
as long as this is not larger than $ N_\mathrm{iter}$.

If you want a fixed number of iterations, choose `{\tt tf}', and you will be prompted for $N_\mathrm{keep}$, 
$N_\mathrm{iter}$, and then $N_\mathrm{thick}$:
\begin{verbatim}
tf
  Enter # of states to keep, # of iterations before thick-restart 
5 50
  Enter # of vectors to keep after thick-restart 
  (Typical value would be           20 )
  ( Must be between            5  and           50 )
\end{verbatim}

If you want to control the convergence, choose `{\tt tc}' and you will be prompted for convergence choices much as for 
standard Lanczos. 

Finally, `{\tt tx}' is an experimental mode attempting to find highly excited states. It   modifies thick-restart by choosing eigenpairs in the vincinity of
a selected absolute energy.  In our experience the convergence is not very good, but it does yield an eigenvector with very strong overlaps with 
the true eigenvectors in the vicinity of the target energy. 

Option `{\tt sk}' is only for testing timing of set-up to this point.  Option `{\tt li}' carries out the Lanczos iterations but does not solve the 
matrices; it can be used for example, if one wants a very large number of Lanczos $\alpha$ and $\beta$ coefficients.

\subsection{Convergence}
\label{converge}

As {\tt BIGSTICK} iterates, it checks for convergence. Every ten iterations it prints to screen the current $N_\mathrm{keep}$ lowest eigenvalues 
and the convergence criterion:
\begin{verbatim}
    1          -135.86073
    2          -133.92904
    3          -131.25354
    4          -131.02439
    5          -129.53058
  
          80  iterations 
      (energy convergence   0.70356 > criterion  0.00100)
  
    1          -135.86073
    2          -133.92904
    3          -131.25354
    4          -131.02439
    5          -129.53059
  
          90  iterations 
      (energy convergence   0.30353 > criterion  0.00100)
\end{verbatim}
As far as we can tell from the literature, there is no robustly ideal convergence criterion for general Lanczos. 
Our default convergence is on energy: {\tt BIGSTICK} takes the sum of the absolute value of the differences in energy between the 
current iteration and one previous, not only for the $N_\mathrm{keep}$ lowest energies but also for the next 5, and divides by the 
square root of $N_\mathrm{keep}+5$, that is,
\begin{equation}
\delta_\mathrm{conv} \equiv \frac{ \sum_{i = 1}^{N_\mathrm{keep}+5}  \left | E_i^\mathrm{new} - E_i^\mathrm{old} \right |}{\sqrt{ N_\mathrm{keep}+5}}
\end{equation}

 The reason for testing additional eigenvalues is to avoid the problem of plunging eigenvalues, 
well known to occur in Lanzos (\cite{lanczos}); it happens when by accident a low-lying state has a tiny overlap with the pivot or initial 
vector. We divided not by the number of energies compared but by the square root, because for a large number of energies one outlier could 
get washed out by many small deviations.  For our purposes this has worked well enough, but it is not rigorously tuned. 

If you want a different criterion, choose `{\tt lc}' on the diagonalization menu. One then gets a series of questions with which to tune the 
convergence, including comparing eigenvectors rather than eigenvalues:
\begin{verbatim}
  Enter nkeep, max # iterations for lanczos  
5 150
  Enter how many ADDITIONAL states for convergence test 
  ( Default=   5 ; you may choose 0 ) 
10
  
     Enter one of the following choices for convergence control :
 (0) Average difference in energies between one iteration and the last; 
 (1) Max difference in energies between one iteration and the last; 
 (2) Average difference in wavefunctions between one iteration and the last; 
 (3) Min difference in wavefunctions between one iteration and the last; 
2
  Enter desired tolerance 
 (default tol =  0.100E-04 )
\end{verbatim}

Similar options are available for thick-restart Lanczos and block Lanczos

\subsection{Block Lanczos}
\label{block}

The standard Lanczos algorithm, discussed in detail in Chapter \ref{lanczos}, applies in each iteration the Hamiltonian to a vector to create a new vector.
In the block Lanczos algorithm, the Hamiltonian is applied to a block of vectors to create a block of new vectors. If used judiciously, this can improve the 
performance of the code. 

Choosing the default option, `{\tt bd}', will lead to the following questions
\begin{verbatim}
  Enter nkeep, dimension of blocks, max # of block iterations 
  (nkeep = # of states printed out; typically ~ dim block )
\end{verbatim}
As with standard, or vector Lanczos, {\tt nkeep} is the number of final states desired.  The dimension of the block is the number of vectors in a block, 
and the number of block iterations is exactly what it sounds like--the number of times one iterates to create a new block. Therefore, all things being equal, 
$\#$ of block iterations $\times$ dimension of block $\approx$ $\#$ of vector Lanczos iterations. For example, if one chose a block dimension of 10, that is, 
ten vectors in each block, and carried out 15 block iterations, that is roughly similar to $10 \times 15 = 150$ ordinary Lanczos iterations. 

In general we recommend that {\tt nkeep} $\leq$ {\tt dim block}. The exception is for the `block strength' option `{\tt(sb)}' in the main menu; see 
Section \ref{introstrength}.

As we discuss in detail in section \ref{blocklanczos}, however, the actual Hamiltonian $\times$ block multiplication runs much faster than ordinary Hamiltonian $\times$ 
vector multiplication--up to twice as fast!  The downside is that, if one just uses a random pivot, one requires significantly more iterations than the equivalent 
vector Lanczos run.   If one uses a block of pivot vectors which are good approximations, however, the number of required iterations can be dramatically reduced.
For example, one could construct states in a truncated model space.  We have written a tool to project vectors from a smaller space to a a larger space, although it is 
not yet ready for release.  

To read in one or more pre-calculated pivot vectors, use option `{\tt (np)}' in the main menu.   You will be asked to enter choices from a list of  pre-calculated states.
First you will be asked whether you want to read in a 
contiguous block (e.g., states 1 through 10 or 20 through 30), or if you want to specify all the states.
\begin{verbatim}
 Need to pick a set of states for the pivot block 
  Choose either (c) a contiguous list or (s) list of specified states
\end{verbatim}
If you enter `{\tt c}', then enter the start and stop 
of the list. This must be within the number of states 
available. If this list does not fill up the pivot block, 
the rest will be created as random vectors, suitably 
orthogonalized.  This is a good option if you have large 
block dimensions.

If you enter `{\tt s}', then you will be asked to specify each 
of the vectors, i.e., {\tt 1,2,5,8, 13, \ldots}.
if you enter `0' then a 
random vector will be substituted.  


A description of an implementation of block Lanczos with thick-restart can be found in \citep{shimizu2019thick}, which includes discussion of some performance issues.

Running block Lanczos on parallel machines using MPI has additional constraints. See section \ref{blockMPI} for information.

\section{Secondary runtime options}

\label{postprocess}

Once {\tt BIGSTICK} has generated wave function, it can further process  the wave functions in secondary options. We discuss those options in detail here. 
Some of these were already mentioned in section \ref{mainmenu}.
All of these options will ask for the name of a previously generated {\tt .wfn} file.  \begin{verbatim}
  Enter input name of .wfn file 
\end{verbatim}
You do \textit{not} have to read in a {\tt .sps} or similar file to define
the model space; from the information in the {\tt .wfn} file, {\tt BIGSTICK} reconstucts the basis. Depending upon the option, you may be asked 
to enter names of appropriate files, such as interaction files.

We list these in the order they are presented in the menu, but the most important and commonly used option are `{\tt x},' expectation value of 
a scalar operator (section \ref{expectation}), `{\tt o},' apply a one-body non-scalar transition operator (section \ref{apply1body} and 
Chapter \ref{applications}), and `{\tt s/sn/su},' the strength function option (section \ref{introstrength} and Chapter \ref{applications}). 

\bigskip

\begin{verbatim}
  * (np) Compute spectrum starting from prior pivot                         * 
\end{verbatim}
Option `{\tt np}' allows you to choose a pivot, or initial starting vector, from a previously generated wave function. This might be useful, for 
example, if you wanted to try to get states of a particular quantum number such as $J$, although we do not currently have the capability to enforce 
this condition, and even if it is an exact quantum number numerical noise will allow states with other quantum numbers to creep in. 
A related and more widely useful option is the strength function 
option `{\tt s}' discussed below and in section \label{strength}.

The downside is that using an initial pivot will only accelerate the convergence of 
a single state.
Option `{\tt (np)}' is more useful for block Lanczos, where one can read in multiple vectors; see section~\ref{bootstrap}.

\bigskip

\begin{verbatim}
  * (dx[m]) Densities: Compute one-body densities from previous run (.wfn)  * 
  *     optional m enables mathematica output 
  * (dxp) Compute one-body densities from prior run (.wfn) in p-n format.   * 
\end{verbatim}
Options `{\tt dx}', `{\tt dxp},' and `{\tt dxm}' read in a previously generated wave function and compute the one-body density matrices.  The latter 
provides a Mathematica-friendly file format. The outputs for  `{\tt dx}' and `{\tt dxm}' are in isospin format while for `{\tt dxp}' it is in proton-neutron format.
The output files have the extension {\tt .dres}. 
More details about one-body density matrices are found in section \ref{densities}.

\bigskip
\begin{verbatim}
  * (occ) single-particle occupations (from previous wfn)                   * 
\end{verbatim}
This option computes the single-particle occupations from a previously generated wave function file.

\subsection{Expectation value}

\label{expectation}

\begin{verbatim}
  * (x) eXpectation value of a scalar Hamiltonian (from previous wfn)       * 
\end{verbatim}
The option `{\tt x}' allows you to compute the expectation value of a operator, which may have one-, two- (and in principle, three-) body components.
It must be an angular momentum scalar and thus is treated as a Hamiltonian, and is read in exactly as Hamiltonians, along with standard  requests for 
scaling information.  The results are written both to screen and, if an output name is given, to the {\tt .res} file.
\begin{verbatim}
   STATE      E            J           T^2           <H >      (norm)
    1      -92.7790     -0.0000      0.0000    499.287061    1.00000
    2      -91.1196      2.0000      0.0000    488.826115    1.00000
    3      -88.4779      2.0000      0.0000    509.922118    1.00000
    4      -87.9781      4.0000      0.0000    452.853182    1.00000
\end{verbatim}
The reason the norm of the vector is given is that after applying a one-body transition operator, as described in the next section (\ref{apply1body}), the wave function vector may no longer be normalized.

\subsection{Matrix elements of a scalar one+two-body operator}
\label{xme}

\begin{verbatim}
  * (h) Compute matrix elements of a scalar Hamiltonian (inputs as basis)  *     
\end{verbatim}

A generalization of computing the expectation value 
is to compute, for a set of wave functions, the 
matrix elements of an arbitrary (but angular momentum scalar) one+two-body operator, that is, something that looks like a Hamiltonian. This option works very similar to the expectation value option, except the outputs are written to a file with extension {\tt .xme}.  Zero matrix elements, including and especially those ruled out by angular momentu selection, are not written to file. The output looks like
\begin{verbatim}
  State      E        Ex         J       T      par
    1    -40.47233   0.00000     0.000   0.000    1
    2    -38.72564   1.74669     2.000   0.000    1
    3    -36.29706   4.17527     4.000   0.000    1
    4    -33.77415   6.69818    -0.000   0.000    1
    5    -32.92937   7.54296     2.000   0.000    1
 
           1           1  -6.62337160    
           2           2  -4.44876957    
           3           3  -3.29423618    
           4           1  -1.27810764    
           4           4  -4.65299320    
           5           2  0.628762603    
           5           5  -4.58469582    
          -1          -1   0.00000000    
\end{verbatim}
In the above, the integers refer to the labels of the states; the time reverse is not given but has the same value.  The values $-1, -1$ signal the end of the file.

(This option is useful for the {\tt PANASh} post-processing code.)

\subsection{Projection of states of good angular momentum}

\label{jproject}

\begin{verbatim}
  * (jp) Project states of good J from prior wfns and normalize             *     
\end{verbatim}

This option reads in a previous wave function with states that may be a mixture of different angular 
momentum $J$ and projects out and normalizes states with good $J$. 
In particular, this option assumes you have previously used option `{\tt (o)}' (see Sections ~\ref{apply1body}, \ref{apply1bodybasic}),
and applied a one-body operator with a definite angular momentum rank $K$. (Future 
work may make this more general.)

This option starts similar to many other options, first asking for the wave function file:
\begin{verbatim}
  Read in prior wfns, project good J and normalize 
  Enter input name of .wfn file     
\end{verbatim}
which will allow for automatic construction of the basis, and the for the output file
\begin{verbatim}
  Enter output name (enter "none" if none)    
\end{verbatim}
After the basis and jumps are constructed, you may select a range of vectors to project:
\begin{verbatim}
  There are           10  initial wavefunctions 
  
  Enter start, stop for initial states 
  (Enter 0,0  to read all )    
\end{verbatim}
You need to know the angular momentum rank (e.g., 1, 2, 3...) of the operator applied to the wave functions.
\begin{verbatim}
  What is the rank (i.e, angular momentum) of operator?     
\end{verbatim}
The reason is the code will then know how many Lanczos iterations to carry out for projection.
Applying an operator of rank $K$ can, on account of the triangle rule for addition of 
angular momentum, i.e., starting from a state with good angular momentum $J_i$
\begin{equation}
   | J_i - K| \leq J_f  \leq J_i + K 
\end{equation}
create at most $2K+1$ different final angular momenta $J_f$.

\subsection{Combining (and orthogonalizing) wave functions from several files}

\label{combine}

\begin{verbatim}
  * (ro) Read in multiple files of wfns and orthonormalize                  * 
\end{verbatim}

Section \ref{xme} describes how to generate the matrix 
elements of a Hamiltonian or some other Hermitian, 
scalar, one+two-body operator between a set of 
wave functions, which act as a basis of a subspace.  
In some cases, one may want to combine multiple files of wave functions,
which may not all be orthonormal, 
into a single wave function file (albeit all in the same basis).  This can 
be done with the {\tt `(ro)'} option. 

The workflow is somewhat clunky, because of  {\tt BIGSTICK}'s standard workflows. The 
basic steps are:
\begin{enumerate}

\item Enter name of first wave function file; this will cause the creation 
of the common basis.

\item Enter name of \textbf{output} wave function file to be created. There will be only one final wave function file.

\item Set the total number of wave functions, {\tt nkeep}, to be read in, usually from 
mutliple files.

\item Select the range of wave functions to be read, e.g., 1-5 or 15-20. 
Choosing {\tt 0,0} will select all in the file.  After being read in, 
the current {\tt .wfn} file will be closed.

\item  If {\tt nkeep} wave functions have not yet been read in, open a 
new {\tt .wfn} file. \textbf{These must be in the same basis as the first 
wave function file, and must have good angular momentum $J$}. 

 \item Select the range of wave functions to be read, e.g., 1-5 or 15-20. 
 Repeat until {\tt nkeep} wave functions read in.
\end{enumerate}

Here's how this looks when running:
\begin{verbatim}
  Enter choice 
ro
  Read multiple files of wfns and orthnormalize 
  Enter input name of .wfn file 
file1
 dimbasischeck=           ...
 ...
  Enter output name (enter "none" if none)
fileout

  .... Building basis ... 

  How many vectors to read in? 
...
  The first file has            5  wave vectors 
  Enter start, stop for initial states 
  (Enter 0,0  to read all )
...
  - - -  NEXT FILE - - - 
  Enter input name of .wfn file 
file2
...
  Enter start, stop for initial states 
  (Enter 0,0  to read all )
...
...
  All vectors read in 
  Next: orthonormalize!
\end{verbatim}

After all the wave function have been read in, {\tt BIGSTICK} will orthonormalize 
and write to the output {\tt .wfn} file.

(This option is useful for the {\tt PANASh} post-processing code.)

You can also combine wave functions from several files into one file, but without orthonormalization: 

\begin{verbatim}
  * (ru) Read in multiple files of wfns but DO NOT orthonormalize                  * 
\end{verbatim}

\subsection{Applying a one-body transition operator}
\label{apply1body}

One of {\tt BIGSTICK}'s important capabilities is to take a set of previously generated wave functions and apply a non-scalar one-body operator
to them:
\begin{verbatim}
  * (o) Apply a one-body (transition) operator to previous wfn and write out* 
\end{verbatim}
If you choose this option, you will be asked for the name of the input  {\tt .wfn} file as well as the name of an output {\tt .wfn} file. Then you will be 
asked for a file with the reduced matrix elements of the operator, which must have the extension {\tt .opme}:
\begin{verbatim}
  Enter name of .opme  file 
\end{verbatim}
Here is an annotated example {\tt .opme} file:
\begin{verbatim}
!  header: Gamow-Teller-like
iso                     ! assumes isospin
           3             ! # of single particle orbits
  1    0    2  1.5      ! index, n, l, j of orbits
  2    0    2  2.5
  3    1    0  0.5
           1           1     ! J, T of transition 
    1    1  -2.68328      ! a, b   < a |||  O  ||| b >
    1    2   5.36656
    2    1  -5.36656
    2    2   5.01996
    3    3   4.24264
\end{verbatim}
The only formatting is the the first non-header line, here {\tt iso}, must be flush against the left.
The file must contain the single-particle orbits, and {\tt BIGSTICK} checks against the orbits used to build the wave function.
After the list of orbits, the $J$ and $T$ of the operator come, and then the 
the non-zero reduced matrix elements. Here, assuming isospin is a good quantum number, we have doubly-reduced matrix elements. 
Although there is a symmetry $ \langle a ||| \hat{O} ||| b \rangle = (-1)^{j_a - j_b}   \langle b ||| \hat{O} ||| a \rangle $, at this time 
{\tt BIGSTICK} requires both elements. 

Many transition operators do not preserve isospin. Therefore {\tt BIGSTICK} can read in operators in an explicit proton-neutron 
symmetry:
\begin{verbatim}
!  M1 matrix elements in the sd shell
pns
           3
  1    0    2  1.5
  2    0    2  2.5
  3    1    0  0.5
           1           2
    1    1     0.1568000     1.4481392   ! a b  proton m.e.   neutron m.e.
    1    2     3.4710987    -2.8962784
    2    1    -3.4710987     2.8962784
    2    2     6.7871771    -2.7092204
    3    3     3.3425579    -2.2897091
\end{verbatim}
The code {\tt pns} in the first non-header line signals that the matrix elements are in proton-neutron formalism, with the same list of orbital 
quantum numbers for protons and neutrons.  
The 2 in the 5th line also signals that one is breaking isospin. 
Thus in the list of reduced (in $J$ only) matrix elements, the columns are for protons and neutrons, 
respectively. 

{\tt BIGSTICK} will read in all the wave functions $| \Psi_i \rangle$ from the initial wave function file, and write $\hat{O} | \Psi_i \rangle$ 
in the final wave function file.  These wave functions will generally not be normalized and will not have good angular momentum or isospin. 
More on this elsewhere. 

Currently, both the initial and final wave functions must be in the same basis. Thus, there are no explicit charge-changing transitions. 
To handle charge-changing transitions, one must use an interaction with good isospin and exploit isospin rotation, described in section \ref{isospinrotation}.
For transitions which change parity, one must use a basis with both parities, option {\tt 0} in the parity-selection menu.

Because transition operators are not in general unitary, the result wave function vectors are not normalized. This information is 
important, as it tells us about total transition strengths, also known as the non-energy-weighted sum rule. 

\subsection{Applying a two-body body scalar operator}

\label{apply2body}

\begin{verbatim}
  * (a) Apply a scalar Hamiltonian to a previous wfn and write out          * 
\end{verbatim}
Option `{\tt a}' works very similar to option `{\tt o}:' one reads in a previously-generated file of wave functions, applies an operator to each wave function, 
and writes the results to another file. The difference is here the operator must be a one-plus-two-body \textit{scalar} operator, that is, like a Hamiltonian. 
The files are read in the same as a Hamiltonian, along with scaling, and so on.

\subsection{Two-body transition densities}

\label{dens2body}

{\tt BIGSTICK} can now compute two-body densities.  One must carry out an ordinary run to create a {\tt .wfn} file, e.g., options such as `{\tt n}', `{\tt d}', etc..
Then run {\tt BIGSTICK} again, choosing `{\tt 2}' on the initial menu.   You will be asked for the name 
of the previously computed {\tt .wfn} file, as well as the mandatory name of the output file. The resulting file will be have a extension {\tt .den2b}. 
You will also be asked
\begin{verbatim}
  Enter start, stop for initial states 
  (This is because two-body densities are large)
  (Enter 0,0  to read all )
\end{verbatim}
with a similar choice for final states. 

The list of all two-body densities can be quite long and exhaustive. Two options are `{\tt 2d}' and `{\tt 2i}' which just computes ``diagonal'' two-body densities, where the initial and final 
states are the same, and only for scalar (two-body) densities.  While the two-body densities in proton-neutron formalism are still written to a {\tt .den2b} file, one can reinterpret these 
as expectation values of operators, written to the {\tt .res} file. 

More details are found in Section \ref{2bdens}.

\subsection{Generating strength function distributions}
\label{introstrength}

One of the most powerful and most useful capabilities is `{\tt s},' the strength function distribution option.  We give an overview here, with many 
more details of application in section \ref{strength}. 

Like all secondary options, the strength function option starts by prompting the user for a previously generated {\tt .wfn} file.  The user 
is then prompted for a Hamiltonian or Hamiltonian-like interaction file or files. Next the user must enter the number of iterations for Lanczos:
\begin{verbatim}
  Fixed iterations ONLY: 
  Enter nkeep, # iterations for lanczos  
  (nkeep = # of states printed out )
\end{verbatim}
The default way to make this option work is through standard Lanczos. The number of results to keep and the number of iteration 
depends upon the application; see section \ref{strength}.  Finally, the user must choose from the input file the pivot or starting vector, 
\textbf{a key decision}:
\begin{verbatim}
  There are           5  wavefunctions 
          5  states 
  
  STATE      E         J          <H > 
    1      -92.7790     -0.0000      0.0000
    2      -91.1196      2.0000      0.0000
    3      -88.4779      2.0000      0.0000
    4      -87.9781      4.0000      0.0000
    5      -87.4348      3.0000      0.0000
  Which do you want as pivot? 
 \end{verbatim}
 What happens next is that Lanczos runs normally, produces eigenvalues and eigenvectors and writes them to file. It also additionally computes 
 the \textit{overlap of the pivot with each of the eigenstates}, that is, $| \langle f | \mathrm{pivot} \rangle|^2$:
 \begin{verbatim}
   Energy   Strength 
  ______   ________ 
  17.30356   0.00007
  49.98777   0.00123
 110.94935   0.00956
 184.33815   0.01945
 249.75355   0.01641
 301.54676   0.08014
 383.34766   0.05833
 428.29023   0.14090
 498.95403   0.04282
 534.87775   0.17398
 588.87306   0.45712
  ______   ________ 
  \end{verbatim}
This is the \textit{strength function} or strength distribution. If the starting vector has a norm different from one, this is noted
\begin{verbatim}
  0.99999999895896363         = total input strength 
\end{verbatim}
and this is included in the strengths.  The usefulness of this capability cannot be overestimated, and is discussed in depth in section \ref{strength}.

By default, the output wave functions are normalized. This can be made explicit by using the option `{\tt(sn)}'. Alternately, the option `{\tt(su)}'  will instead normalize the output wave functions to the input normalization $\times$ the square root of the strength. This is useful 
when computing strength distributions, especially after projection, as discussed in sections 
\ref{apply1bodygoodJ} and \ref{GTstrength}.  
\bigskip

Alternately, if one chooses `{\tt ss}' then \textit{no} wave function will be written to file and the $J$, $T$ of the final wave functions will not be computed. 
This is recommended when working in large spaces and a large number of iterations have been carried out: the wave functions (and $J$, $T$) are generally 
not useful and take considerable time. 

\textit{Block strength}. 
Recently, we have added a block strength option, `{\tt(sb)}.'  Here one reads in a block of several vectors and carries out block Lanczos, thus getting 
strength functions for multiple starting vectors simultaneously.  

As with block Lanczos (see Section \ref{block}), you will be promped:
\begin{verbatim}
  Fixed block iterations ONLY: 
  Enter nkeep, dim of block,  # iterations for lanczos  
  (nkeep = # of states printed out )
\end{verbatim}
In standard block Lanczos, one generally wants {\tt nkeep } $\leq$ the dimension of the block.   For block strength, however, much as in standard strength option, 
one generally wants {\tt nkeep} $\sim$ the number of Lanczos \textit{vectors} generated = (dim block) $\times$ (\# of block iterations).

When reading in from a previously generated file (option `{\tt np}' in the main menu), , you can must, select the states for the pivot block. You will be prompted:
\begin{verbatim}
  Enter a list of  3  states for the pivot block
  (Please enter in order)
\end{verbatim}
If you enter `0' a random vector will be generated. 
After this it runs just as the regular strength function, albeit with producing strength distributions for multiple input vectors.  \textbf{Note}: the detailed strengths may 
differ from the single-state run.  However by carrying out a running sum you should be able to see the \textit{distribution} is the same.

When running large cases on parallel machines using MPI, there are additional constraints; see section \ref{blockMPI}.

Finally, by choosing `{\tt sbs}' the wave functions will not be written to file, nor $J$, $T$ computed. 

\subsection{Overlap or dot product of wave functions}
\label{overlap}

\begin{verbatim}
  * (v) Overlap of initial states with final states                         * 
\end{verbatim}
The output eigenstates are written as vectors.
Although most users are unlikely to need to use this, using the `{\tt v}' option {\tt BIGSTICK} can compute the dot product between two such 
wave functions, including from different files.  From the first file, you must choose a specific state:
\begin{verbatim}
  Which do you want as initial state? 
\end{verbatim}
A second file is then opened (you can reopen the first file), and the intial state is dotted against each of them. The results are written to the file
{\tt overlap.dat}:
\begin{verbatim}
  Initial state =            1
  state      E      J    T       <i|f>     |<i|f>|^2 
   1    -87.10445 -0.0  0.0     0.99885     0.99771
   2    -85.60214  2.0  0.0    -0.00000     0.00000
   3    -82.98830  2.0  0.0     0.00000     0.00000
   4    -82.73201  4.0  0.0     0.00000     0.00000
   5    -82.03407  3.0  0.0    -0.00000     0.00000
   6    -81.22187  4.0  0.0    -0.00000     0.00000
   7    -79.76617 -0.0  0.0    -0.02326     0.00054
\end{verbatim}
We found this option useful in validating other capabilities, such as the strength function capability.

\section{Output files}

{\tt BIGSTICK} generates a number of output files.  These fall into two broad categories. The most important output files have a name 
supplied by the user, e.g., {\tt mg24} followed by an extension, e.g. {\tt .res} or {\tt .wfn}.  Other files, which are not needed by most casual users,
have the same standard name upon each run, ending in {\tt .bigstick}.

\bigskip

\textbf{Results}. The most important file are the results files, which have an extension {\tt .res}. When you initiate {\tt BIGSTICK}, after the 
main menu choice, {\tt BIGSTICK} almost always asks you for the name of the output files:
\begin{verbatim}
  Enter output name (enter "none" if none)
\end{verbatim}
If you enter ``{\tt none}'' then several files are suppressed, in particular the results file.

The results file generally contains the output spectrum, e.g.
\begin{verbatim}
  State      E        Ex         J       T 
    1    -41.39657   0.00000     0.000   0.000
    2    -39.58581   1.81077     2.000   0.000
    3    -37.08646   4.31012     4.000   0.000
    4    -34.46430   6.93227     0.000   0.000
    5    -33.56871   7.82786     2.000  -0.000
\end{verbatim}
It also may contain one-body density matrices, or the results of strength function runs. 

\bigskip

\textbf{Wave functions}.  The {\tt .wfn} file contains wave function information, stored in binary (or ``unformatted.'') In addition to containing the 
wave function vectors, it has a header which contains enough information the basis can be recreated. 

\bigskip

\textbf{Autoinput}.  On each run {\tt BIGSTICK} generates a file {\tt autoinput.bigstick}. This saves the various input when run from the terminal. 
This is useful if one is making small tweaks to run, or to use as the basis for input directives. To use the autoinput file directly from terminal, 
choose `{\tt i}' at the opening menu.

\bigskip

\textbf{Log file}. The {\tt .log} file summarizes information about the run, such as the date, time, {\tt BIGSTICK} version number, dimensions, 
parallelization (number of MPI processes and OpenMP threads),
internal flag settings, and so on. While not needed by the casual user, they are useful to document the exact conditions under which a particular 
result ran and for debugging. If no output name is specified, this file is named {\tt logfile.bigstick}.

\subsection{Secondary files} 

{\tt BIGSTICK} generates some intermediate files which are not needed for ordinary runs but in some cases can be 
useful.  The most useful of these are the {\tt .lcoef} files, which in an ordinary Lanczos run contains the Lanczos coefficients $\alpha_i, \beta_i$. 
If no output name is specified, this file is called {\tt lanczosvec.lcoef}.


\subsection{Diagnostic files}  {\tt BIGSTICK} also generates a number of diagnostic files, primarily for development, 
tuning, and debugging. 

\bigskip

{\tt timingdata.bigstick} contains the time spent in different matvec modes (SPE, PP, etc) on each MPI process.

\bigskip

{\tt distrodata.bigstick} contains the type and size of jumps stored on each MPI process.

\bigskip

\section{Memory usage}

\label{memory}

The motivation for {\tt BIGSTICK}'s on-the-fly algorithm is to save memory over storing the nonzero many-body matrix elements.  Despite this, 
{\tt BIGSTICK} can still be quite memory-hungry.  The main sinks of memory are: the Lanczos vectors themselves, the jumps factorizing the 
many-body matrix elements, and the uncoupled two-body matrix elements.  Which dominates depends upon the system. For example, in 
large phenomenological calculations, the main memory usage is from Lanczos vectors. In no-core shell model calculations, it is typically the 
jumps, except for very light systems ($A=3,4$) where for large spaces the uncoupled matrix elements actually dominate.

{\tt BIGSTICK} gives a report on memory usage. It also has some default caps on memory and will halt if these are violated. The default caps can 
be changed by the user. 

Both in normal runs and in modeling runs, {\tt BIGSTICK} produces a report:
\begin{verbatim}
 RAM for 2 lanczos vector fragments (max)         :     3923.728 Mb 
 RAM for jumps in storage    (total)          :        2353.914 Mb 
 Max RAM for local storage of jumps           :         177.069 Mb 
 RAM for uncoupled two-body matrix elements   :        0.017 Mb 
\end{verbatim}
The RAM report above is for the initial and final Lanczos vectors in matvec. In order to  reorthogonalize, {\tt BIGSTICK} also stores all Lanczos 
vectors. When run in MPI, these Lanczos vectors are distributed across many MPI processes:
\begin{verbatim}
  Enter max number of Lanczos iterations 
150
  Assuming max memory per node to store Lanczos vectors    16.00000      Gb 
  Storage of Lanczos vectors distributed up across          128  nodes 
  Memory per node =    10.74658      Gb 
\end{verbatim}

The default memory caps can all be found in the module {\tt flagger} in the file {\tt bmodules\_flags.f90}.
The most important ones are 
\begin{verbatim}
   real    :: maxjumpmemory_default  = 16.0    ! in Gb 
   real    :: maxlanczosstorage1 = 16.000    ! in Gb
\end{verbatim}
These can be changed, though of course {\tt BIGSTICK} must be recompiled.

\chapter{Applications}

\label{applications}

In this chapter we discuss in more detail applications of {\tt BIGSTICK}, specifically one-body density matrices and one-body transition strengths. 

\section{One-body density matrices}

\label{densities}

{\tt BIGSTICK} can be directed to compute the reduced one-body density matrices, 
\begin{equation}
\rho_{K}^{fi}(ab) \equiv  \frac{1}{\sqrt{(2K+1)}}    \left \langle \Psi_f \left |  \left | \left  [ \hat{c_a}^\dagger \otimes \tilde{c}_b \right ]_{K} 
\right | \right |  \Psi_i \right \rangle    \label{onebodydensity}
\end{equation}
where we use reduced matrix elements as defined in Appendix \ref{reduced}. We use this particular definition because the reduced matrix element 
of a generic one-body operator is the sum of products of 
the density matrix  elements and the reduced matrix elements, namely,
\begin{equation}
\langle \Psi_f || \hat{O}_K || \Psi_i \rangle = \sum_{ab} \rho_{K}^{fi}(ab) \,\, \langle a || \hat{O}_K || b \rangle .
\end{equation}
It's important to note that $\langle a || \hat{O}_K || b \rangle$ are \textit{ matrix elements between single-particle states}, while the density matrices 
are matrix elements between \textit{many-body states}.  While some many-body codes compute the many-body matrix elements for specific 
operators, such as E2, M1, and so on, we chose for {\tt BIGSTICK} to produce one-body density matrices, allowing the user to compute the 
transition matrix elements for \textit{any} one-body operator.

For systems with good isospin one can also define ``doubly-reduced'' matrix elements, that is, reduced in both angular momentum and isospin: 
\begin{equation}
\rho_{K,T}^{fi}(ab) \equiv  \frac{1}{\sqrt{(2K+1)(2T+1)}}   \left . \left . \left \langle \Psi_f \right |  \right | \right | \left  [ \hat{c_a}^\dagger \otimes \tilde{c}_b \right ]_{K,T} \left | 
\left | \left |  \Psi_i \right \rangle \right . \right .  \label{onebodydensityJT}
\end{equation}

When you choose the density matrix option, {\tt BIGSTICK} will write to the {\tt .dres} file (but not to screen, and no longer to the {\tt .res} file)
the density matrices, e.g.:
\begin{verbatim}
 Initial state #    2 E =  -85.60214 2xJ, 2xT =    4   0
 Final state   #    1 E =  -87.10445 2xJ, 2xT =    0   0
 Jt =   2, Tt = 0        1 
    1    1  -0.01957   0.00000
    1    2   0.18184   0.00000
    1    3   0.09721   0.00000
    2    1  -0.18184   0.00000
    2    2  -0.35744   0.00000
    2    3  -0.26323   0.00000
    3    1  -0.09721   0.00000
    3    2  -0.26323   0.00000
\end{verbatim}
The first two lines are the labels and energies of the initial and final wavefunctions; {\tt Jt} and {\tt Tt} are the angular momentum and isospin of the 
one-body operator, and, for example, 
\begin{verbatim}
    1    3   0.09721   0.00000
\end{verbatim}
1 is the label of the first single-particle orbit, 3 the label of the second (as defined by the input {\tt .sps} file), 
and the two real numbers are the $T=0,1$ one-body density matrix elements, that is, 
$$
 \left \langle \Psi_2 \left | \left | \left |  \left [ \hat{a}^\dagger_1 \otimes \tilde{a}_3 \right ]_{J=2,T=0} \right | \right | \right | \Psi_1 \right \rangle = 0.09721
$$
while that for $T=1$ is, here, zero.

Between the listing of the spectra (i.e., energies, excitation energies, and angular momentum and isospin) and the density matrices, an ordered list of the single-particle orbitals is given, for convenience in post-processing, e.g.,
\begin{verbatim}
  Single particle state quantum numbers
ORBIT      N     L   2 x J  
     1     0     2     3
     2     0     2     5
     3     1     0     1
\end{verbatim}

{\tt BIGSTICK}  has two options for densities.  The option {\tt d } will compute one-body densities with good isospin,  where the output looks like 
(this example is $^{23}$Ne in the $sd$ shell with the USDB interaction):
\begin{verbatim}
 Initial state #    1 E =  -62.78960 2xJ, 2xT =    5   3
 Final state   #    1 E =  -62.78960 2xJ, 2xT =    5   3
 Jt =   0, Tt = 0        1 
    1    1   0.84730   0.28105
    2    2   8.32846   2.85633
    3    3   1.52286   0.13245
\end{verbatim}
Alternately, there is the option {\tt dp} which puts the density matrix elements into explicit proton-neutron form.
\begin{verbatim}
 Initial state #    1 E =  -62.78960 2xJ, 2xT =    5   3
 Final state   #    1 E =  -62.78960 2xJ, 2xT =    5   3
 Jt =   0, proton      neutron 
    1    1   0.16625   0.43288
    2    2   1.58968   4.29943
    3    3   0.47558   0.60124
\end{verbatim}
Option `{\tt dxp}' allows one to compute one-body densities in a proton-neutron format from a previously computed wave function.

\subsection{Symmetries of density matrix elements}

A useful symmetry relation is
\begin{equation}
\rho_{K,T}^{if}(ba)= (-1)^{j_a -j_b +J_i -J_f + T_i - T_f} \rho_{K,T}^{fi}(ab).
\end{equation}

\subsection{Particle occupations from densities}

\label{occfromden}

Particle occupations are the average number of particles in  single-particle orbit for a given wave function. 
Although there is an option, `{\tt p},' to compute the orbit occupation, you can also extract this information from the 
diagonal one-body density matrices. The total number of particles in orbit $a$ is 
\begin{equation}
n(a) = \frac{ [j_a]}{[J_i]} \rho^{ii}_{K=0}(a^\dagger a)
\end{equation}
If your densities are in proton-neutron format, you can extract the proton and neutron occupations separately. 
If you have your densities in isospin formalism, you can extract the total number of protons \textit{and} neutrons in an orbit
\begin{equation}
n_\pi(a)+ n_\nu(a) = \frac{ [j_a][1/2]}{[J_i][T_i]} \rho^{ii}_{K=0,T=0}(a^\dagger a)
\end{equation}

 To separately extract proton and neutron occupation one must take careful account of the Clebsch-Gordan coefficients.  
 One must have $f = i$, so that $J_f = J_i $ and $T_f = T_i$, as well as considering 
only $a=b$. Furthermore, the answer depends upon $T_z = (Z-N)/2$ (using the notation $[x] = \sqrt{2x+1}$)
\begin{eqnarray}
n_\pi(a) =\frac{\left [\frac{1}{2} \right  ]   \left [ j_a \right ] }{ \left [ J_i \right ]   \left [ T_i \right ]  }  \frac{1}{2} 
\left (  \rho^{ii}_{K=0,T=0} (a^\dagger a) + \frac{T_z \sqrt{3} }{\sqrt{T_i (T_i +1)}}  \rho^{ii}_{K,T=1} (a^\dagger a)  \right ), \label{npi}
 \\
n_\nu(a) = \frac{\left [\frac{1}{2} \right  ]   \left [ j_a \right ] }{ \left [ J_i \right ]   \left [ T_i \right ]  }  \frac{1}{2} 
\left (  \rho^{ii}_{K=0,T=0} (a^\dagger a) - \frac{T_z  \sqrt{3} }{\sqrt{T_i (T_i +1)}}  \rho^{ii}_{K,T=1} (a^\dagger a)  \right ).\  \label{nnu}
\end{eqnarray}

\subsection{Conversion from proton-neutron to isospin}

The one-body density matrices are internally in proton-neutron formalism, but can be converted to isospin formalism \textit{if} the initial and final states have good isospin. (This is 
done in the subroutine {\tt coupled\_densities} in the file {\tt bdensities.f90}.) We choose the convention that protons have $m_t = +1/2$, while neutrons have $m_t = -1/2$, hence 
$M_T= (Z-N)/2$.  If $\rho$ is a one-body density, with $\rho_T$ isospin densities with $T=0,1$ and $\rho_{m_t}$ proton/neutron densities, then 
\begin{equation}
\rho_T =  \frac{[T_f]}{ ( T_i \, T_z , T 0 | T_f \, T_z )} \frac{ 1}{ [T]}\sum_{m_t} \rho_{m_t}  \left (  \frac {1}{2} \, m_t, \frac{1}{2} \, -m_t  | T 0 \right ). 
\end{equation}
For the special case where $T_i = T_f $, and using analytic formulas for the Clebsch-Gordan coefficients, one gets
\begin{eqnarray}
\rho_{T=0} & =& \frac{[T_i]}{\sqrt{2}  } \left ( \rho^\pi + \rho^\nu \right ) , \\
\rho_{T=1} &=& \frac{[T_i]}{\sqrt{2}  } \left ( \rho^\pi - \rho^\nu \right )  \frac{ \sqrt{T_i (T_i+1)} }{\sqrt{3} \, T_z}.
\end{eqnarray}
Note that here for $T_z=0$, the $\rho_{T=1}$ density matrix must vanish. Finally, one can invert to get
\begin{equation}
\rho^{\pi,\nu} = \frac{ 1 }{\sqrt{2}  \, [ T_i]} \left ( \rho_{T=0} \pm \frac{T_z \sqrt{3} } { \sqrt{ T_i (T_i+1) }} \rho_{T=1} \right ) .
\end{equation}
This agrees with Eq.~(\ref{npi}), (\ref{nnu}). 

\subsection{Strengths from density matrix elements}

\label{transitionstrengthsfromdensities}

Given some transition operator $\hat{O}_K$ carrying definite angular momentum $K$, the 
transition strength between an initial and final state is just the square of the matrix element:
$$
\left | \langle J_f M_f | \hat{O}_{K M} | J_i M_i \rangle \right|^2.
$$
This is the matrix element that goes into Fermi's golden rule for decay and transition rates.

But in most experimental situations we cannot pick out specific values of $M_{i,f}$ 
(unless we are doing an experiment with polarization). The final result must then 
\textit{average} over initial states and \textit{sum} over final states, that is, 
\begin{equation}
\frac{1}{2J_i +1} \sum_{M_i} \sum_{M_f} 
\left | \langle J_f M_f | \hat{O}_{K M} | J_i M_i \rangle \right|^2.
\end{equation}
In most cases there is also implicitly a sum over $M$.  (If not, 
the final result will be different.) 
Now we can use the Wigner-Eckart theorem to rewrite the average/sum as:
\begin{equation}\frac{1}{2J_i+1}
\sum_{M_f}  \sum_{M_i}  \sum_M
\left | ( J_i M_i,  K M | J_f M_f )  \right |^2  \left | ( J_f || \hat{O}_K || J_i ) \right|^2.
\label{sumoverms}
\end{equation}
Now we can use the the selection rule $M_f = M_i + M_f$ to eliminate the sum over $M_f$ and the 
orthogonality of the Clebsch-Gordan coefficients to sum over 
$M_i$ and $M$
\begin{equation}
\sum_{M_i} \sum_{M} | ( J_i M_i,  K M | J_f M_i + M_f )|^2  = 1 
\end{equation}

Thus we get the result in terms of reduced matrix elements,
\begin{eqnarray}
\frac{1}{(2J_i +1)} \sum_{M_i} \sum_{M_f}  \sum_M
 \left | ( J_f M_f| \hat{\cal O}_{K M} | J_i M_i ) \right|^2 \nonumber \\
= \frac{1}{(2J_i +1)} \left | ( J_f || \hat{\cal O}_K || J_i ) \right|^2,
\end{eqnarray}
As one often calls
\begin{equation}
\frac{1}{2J_i +1} \left | ( J_f || \hat{\cal O}_K || J_i ) \right|^2, \label{Bvalue}
\end{equation}
the  $B$-value, written $B({\cal O})$ (for example, $B(GT)$ for 
Gamow-Teller, $B(E2)$ for electric quadrupole, etc.), this says the strength for 
an operator is $B({\cal O} )$.

In the {\tt BIGSTICK} code and most other shell-model codes, we  compute transition strengths using transition density matrix elements: 
the doubly reduced transition matrix element for a one-body operator $\hat{\cal O}_{K,T}$ of angular momentum rank $K$ and isospin rank $T$ is 
\begin{equation}
 \left . \left . \left \langle \Psi_f \right |  \right | \right |\hat{\cal O}_{K,T} \left | 
\left | \left |  \Psi_i \right \rangle \right . \right . = 
\sum_{ab} \rho_{K,T}^{fi}(ab)
 \left . \left .  \left \langle a \right | \right | \right | \hat{\cal O}_{K,T} 
\left | \left | \left | b \right \rangle \right . \right . .
\label{transME}
\end{equation}

Although the default output is doubly-reduced matrix elements, the definition of $B$-values do \textit{not} sum or average over `orientations' in isospin space, because $T_z = (Z-N)/2$ is fixed. Hence we have to account for that by undoing the Wigner-Eckart reduction in isospin, so that, for non-charge changing transitions (e.g., $\gamma$-transitions),  
\begin{eqnarray}
B({\cal O}: i\rightarrow f) =
\frac{1}{2J_i +1} \left | (\Psi_f:  J_f || \hat{\cal O}_J ||\Psi_i  J_i ) \right|^2 \nonumber \\
= \frac{1}{2J_i +1} \left | (\Psi_f:  J_f T_f ||| \hat{\cal O}_{J,T} |||\Psi_i  J_i T_i ) \right|^2 \times \frac{ \left | ( T_i T_{z} , T 0 | T_f T_z)\right|^2}{2T_f+1}.
\label{qconservetrans}
\end{eqnarray}
Note the last line uses the result of Eq.~(\ref{transME}).

\subsection{Sample case: spin-flip}

Let's consider a couple of simple cases, both in the \textit{sd} shell with the USDB
interaction \citep{PhysRevC.74.034315}.  Let's consider the spin-flip operator $\vec{\sigma} = 2 \vec{S}$, which has the following doubly-reduced matrix elements:

\medskip

\begin{tabular}{ll}
One-body matrix element  & value \\
\hline
$\langle 0d_{3/2} ||| \vec{\sigma} ||| 0d_{3/2} \rangle$ &  -2.19089  \\
$\langle 0d_{3/2} ||| \vec{\sigma} ||| 0d_{5/2} \rangle$ &   4.38178  \\
$\langle 0d_{5/2} ||| \vec{\sigma} ||| 0d_{3/2} \rangle$ &   -4.38178  \\
$\langle 0d_{5/2} ||| \vec{\sigma} ||| 0d_{5/2} \rangle$ &    4.09878 \\
$\langle 1s_{1/2} ||| \vec{\sigma} ||| 1s_{1/2} \rangle$ &   3.46410 \\
\end{tabular}

\medskip

The  nuclide $^{19}$F, which has only one valence proton and two valence neutrons, has, with appropriate scaling of the 
matrix elements, the low-lying spectrum:
\begin{verbatim}
  State      E        Ex         J       T 
    1    -23.86096   0.00000     0.500   0.500
    2    -23.78367   0.07729     2.500   0.500
    3    -22.09059   1.77037     1.500   0.500
    4    -21.26237   2.59858     4.500   0.500
    5    -19.25724   4.60371     6.500   0.500
\end{verbatim}
The density matrix from the second state ($J=5/2$) to the third ($J=3/2$) state is, up to some overall phases, 
\begin{verbatim}
 Initial state #    2 E =  -23.78367 2xJ, 2xT =    5   1
 Final state   #    3 E =  -22.09059 2xJ, 2xT =    3   1
 Jt =   1, Tt = 0        1 
    1    1  -0.08640  -0.01635
    1    2   0.44978  -0.36112
    1    3   0.01255  -0.09014
    2    1  -0.16826   0.08815
    2    2  -0.00280  -0.35352
    3    1  -0.03483   0.07521
    3    3   0.28978  -0.20874
\end{verbatim}
Because the vector of Pauli matrices $\vec{\sigma}$ carries one unit of angular momentum and no isospin, we only use the  ({\tt Tt= 0}) set of matrix elements
(column second from the right ). {\tt BIGSTICK} also generates the transition matrix elements for {\tt Jt = } 2, 3, and 4, not shown. 
 Applying (\ref{qconservetrans}), we get a B($\sigma: 2 \rightarrow 3$)=1.2609
 
 A second case is $^{20}$Ne.  The ground state is at -40.4723 MeV, which the first $J=1,T=0$ state, 
 state $\# 25$, is at -27.8364 MeV (or 12.636 MeV excitation energy).  The density matrix is
 \begin{verbatim}
  Initial state #    1 E =  -40.47233 2xJ, 2xT =    0   0
 Final state   #   25 E =  -27.83635 2xJ, 2xT =    2   0
 Jt =   1, Tt = 0        1 
    1    1   0.00069   0.00000
    1    2   0.14575   0.00000
    1    3  -0.10567   0.00000
    2    1   0.18722   0.00000
    2    2  -0.04822   0.00000
    3    1  -0.02309   0.00000
    3    3   0.28308   0.00000
\end{verbatim}
and B($\sigma:1\rightarrow 25$) = 0.3597.

\subsection{Charge-changing transitions}

\label{isospinrotation}

Charge-changing transition such as Gamow-Teller are a little more subtle. If we have isospin-conserving interactions, so that our initial and final states have good isospin,
we can use \textit{isospin rotation} so that we don't have to change basis.  If we want to have a transition 
$$
^A_ZX_N \rightarrow ^A_{Z\pm 1}Y_{N\mp 1},
$$
that is, from some initial $T_{z,i} = (Z-N)/2$ to some final $T_{z,f} = (Z-N)/2 \pm 1$, we must work in the basis with the smaller $T_z$; then both initial and final 
states will be somewhere in the spectrum.   What we want to calculate is 
$$
\left | \langle \Psi_f : J_f, T_f T_{z,f} || \hat{\cal O} || \Psi_i : J_i , T_i T_{z,i} \rangle \right |^2,
$$
but what we can actually calculate with {\tt BIGSTICK} is 
 $$
\left | \langle \Psi_f : J_f, T_f T_{z} || \hat{\cal O} || \Psi_i : J_i , T_i T_{z} \rangle \right |^2.
$$
Fortunately this can be accomplished with only a small modification of the above procedure:
\begin{eqnarray}
B({\cal O}: i\rightarrow f) =
\frac{1}{2J_i +1} \left | (\Psi_f:  J_f || \hat{\cal O}_J ||\Psi_i  J_i ) \right|^2 \nonumber \\
= \frac{1}{2J_i +1} \left | (\Psi_f:  J_f T_f ||| \hat{\cal O}_{J,T} |||\Psi_i  J_i T_i ) \right|^2 \times \frac{ \left | ( T_i T_{z,i} , T \pm 1 | T_f T_{z,f})\right|^2}{2T_f+1},
\label{qchangetrans}
\end{eqnarray}
where the difference between Eq.~(\ref{qconservetrans}) and (\ref{qchangetrans}) is in the isospin Clebsch-Gordan. There is, however, one more subtle point in 
treating the isospin raising/lowering operator, $\tau_{\pm}$.  If one treats $\tau$ as a rank-1 spherical tensor in isospin space, one can show that 
\begin{equation}
\tau_\pm = \frac{1}{\sqrt{2}} \tau_{1,\pm 1}.
\end{equation}
Therefore, formally, in the above calculations, we are actually using $2^{-1/2} \tau_{1,0}$ in our calculation, and then rotating to a charge-changing transition.

It's always good to have a way to check calculations, and in the case of Gamow-Teller it's the Ikeda sum rule, which says
\begin{equation}
\sum_f B(\vec{\sigma} \tau_+ : i \rightarrow f) - B(\vec{\sigma} \tau_- : i \rightarrow f) = 3(N-Z)
\end{equation}
independent of the  initial state $i$.
Here the isospin raising operator $\tau_+$ changes a neutron into a proton and hence is the operator for $\beta^-$ decay, while 
the isospin lowering operator $\tau_-$ changes a proton into a neutron and hence is the operator for $\beta_+$ decay. This assume 
our convention that protons are isospin `up' and neutrons isospin `down;' many authors have opposite conventions.

\subsection{Sample case: $^{19}$F}
Let's calculate the Gamow-Teller B-value. The matrix elements for Gamow-Teller are the same as for $\vec{\sigma}$ as shown above 
except multiplied by $\sqrt{3/2}$.
Then using the {\tt Tt= 1} one-body density matrix elements, we get B(GT: $2\rightarrow 3$) = 1.3990.

For $^{20}$Ne, there is at $J=1,T=1$ state at -29.3066 MeV (or 11.166 MeV excitation energy, state $\# 15$); the density matrix is 
\begin{verbatim}
 Initial state #    1 E =  -40.47233 2xJ, 2xT =    0   0
 Final state   #   15 E =  -29.30659 2xJ, 2xT =    2   2
 Jt =   1, Tt = 0        1 
    1    1   0.00000   0.05163
    1    2   0.00000   0.09951
    1    3   0.00000  -0.03397
    2    1   0.00000   0.18236
    2    2   0.00000   0.32717
    3    1   0.00000  -0.03311
    3    3   0.00000  -0.08363
\end{verbatim}
Here B(GT) = 0.1654, for either $\beta+$ or $\beta-$.

\subsection{Transitions utilities}

A set of codes for generating common one-body operator matrix elements and for post-processing of one-body density matrices into, e.g., transition $B$-values, 
is available to download from GitHub, in the {\tt /util/} directory of {\tt https:// github.com /cwjsdsu/BigstickPublick/}. A detailed manual and sample runs are included.

\section{Two-body densities}

\label{2bdens}

{\tt BIGSTICK} can now compute two-body densities.    (At this time, two-body densities have not yet been enables for MPI.)
The two-body density matrix elements are defined in parallel to one-body densities.
We want 
$$
\langle \Psi_f, J_f || \hat{\cal O}_K || \Psi_i, J_i \rangle = \sum_{ab} \langle a || \hat{\cal O}_K || b \rangle \rho^{fi}_K(a \tilde{b}) 
$$
so we define
 \begin{eqnarray}
\rho^{fi}_J(ab, J_{ab}; {c}{d},J_{cd}) \nonumber  \\
\equiv
[J]^{-1}  \left\langle J_f \left |\left | \left [ \hat{A}^\dagger_{J_{ab}}(ab)  \otimes \tilde{A}_{J_{cd}}(cd) \right ]_J \right | \right  | J_i  \right \rangle
\frac{1}{\sqrt{(1+\delta_{ab}) (1+\delta_{cd}) }}
\label{tbtddef}
\\
= - [J]^{-1}  \left\langle J_f \left |\left | \left [ \left [\hat{a}^\dagger \otimes \hat{b}^\dagger \right ]_{J_{ab}} \otimes
\left [ \tilde{c} \otimes \tilde{d} \right ]_{J_cd} \right ]_J \right | \right  | J_i  \right \rangle \zeta^{-1}_{ab} \zeta^{-1}_{cd}, \nonumber
\end{eqnarray}
where the factor $\zeta_{ab} \equiv \sqrt{1+\delta_{ab}}$ is needed for normalized two-body states; see Appendix \ref{ops2ndquant}.   
In proton-neutron format, 
$\zeta=1$ always, that is, proton and neutron orbitals are considered distinct.

Note also that we have defined
\begin{equation}
\tilde{A}_{J\, M}(ab) = - \left [ \tilde{a} \otimes \tilde{b} \right ]_{JM} = (-1)^{J+M} \hat{A}_{J\, -M}(ab).
\end{equation}
where the time-reversed operator is
\begin{equation}
\tilde{c}_{j_c, \, m_c} = (-1)^{j_c + m_c} \hat{c}_{j_c, \, - m_c}.
\end{equation}
so that
\begin{eqnarray}
 \hat{A}_{JM}(ab) \equiv
-\left [ \hat{a} \otimes \hat{b} \right ]_{JM} = \left( \hat{A}^\dagger_{JM}(ab) \right)^\dagger \\
= - (-1)^{J-M} \left [ \tilde{a} \otimes \tilde{b} \right ]_{J\, -M} .
\end{eqnarray}
The  two-body matrices so defined are reduced with respect to angular momentum but not with respect to isospin.

 One must carry out an ordinary run to create a {\tt .wfn} file, e.g., options such as `{\tt n}', `{\tt d}', etc..
Then run {\tt BIGSTICK} again, choosing `{\tt 2}' on the initial menu.   You will be asked for the name 
of the previously computed {\tt .wfn} file, as well as the mandatory name of the output file. The resulting file will be have a extension {\tt .den2b}. 
You will also be asked
\begin{verbatim}
  Enter start, stop for initial states 
  (This is because two-body densities are large)
  (Enter 0,0  to read all )
\end{verbatim}
with a similar choice for final states.

The output {\tt .den2b} file begins by defining the proton and neutron orbits:
\begin{verbatim}
 !#  Two-body densities from BIGSTICK version 7.9.8  Sept  2020
 !#  Run date: 2020-02-15
 !# Densities written in explicit proton-neutron formalism 
 !# Single-particle orbits information follows 
 !# Number of single-particle orbits (same for both protons, neutrons ) 
           3
 !# Proton, neutron index  N     L   2xJ 
         1    4           0      2      3
         2    5           0      2      5
         3    6           1      0      1
 !#   Z   N 
      4   4
\end{verbatim}
In this explicit proton-neutron format, orbits 1-3 refers to protons, and 4-6 to neutrons. 
Then for each set of initial and final states, we get the two-body densities:
\begin{verbatim}
 !# Ini state      Energy         J     
         1       -92.77905       0.0
 !# Fin state      Energy         J     
         1       -92.77905       0.0
 !# a   b    Jab    c   d    Jcd  Jmin Jmax   & 
  (< f || [[ab:Jab]^+[cd:Jcd]]_J || i >,/sqrt(2J+1) J=Jmin,Jmax)
    1   1     0     1   1     0     0   0       0.199810
    1   1     2     1   1     2     0   0       0.057047
    1   1     2     2   1     2     0   0       0.027265
    1   1     0     2   2     0     0   0       0.428295
    ...
!# Ini state      Energy         J     
         2       -91.11964       2.0
 !# Fin state      Energy         J     
         2       -91.11964       2.0
 !# a   b    Jab    c   d    Jcd  Jmin Jmax  & 
  (< f || [[ab:Jab]^+[cd:Jcd]]_J || i >, J=Jmin,Jmax)
    1   1     0     1   1     0     0   0       0.367200
    1   1     0     1   1     2     2   2       0.024823
    1   1     2     1   1     0     2   2       0.024823
    1   1     2     1   1     2     0   4       0.121627  -999.000000     0.009363  .....
    1   1     0     2   1     2     2   2       0.064846
    1   1     0     2   1     4     4   4      -0.023200
    1   1     2     2   1     1     1   3    -999.000000    -0.018076  -999.000000
    1   1     2     2   1     2     0   4       0.061892  -999.000000    -0.012186  .....
    1   1     2     2   1     3     1   4    -999.000000    -0.021697  -999.000000   ....
    1   1     2     2   1     4     2   4       0.036903  -999.000000     0.000209
    1   1     0     2   2     0     0   0       0.785513
    1   1     0     2   2     2     2   2       0.084297
...    
\end{verbatim}
What this means is as follows: we couple up two destruction operators, with orbit labels $c$ and $d$ to total angular momentum {\tt Jcd},  and 
two creation operators in orbits $a$ and $b$ coupled up to {\tt Jab}, and the total transition operator is coupled up to some $J$. The range of 
$J$ is from {\tt Jmin = |Jab-Jcd|} to {\tt Jmax = Jab+Jcd}; furthermore, we must have $|J_i - J_f | \leq J \leq J_i +J_f$.  For each allowed value of $J$, we have a value of the density matrix. 

A value of {\tt -999.0000} means that the matrix element could not be computed due to a vanishing Clebsch-Gordan coefficient; in that case, one should rerun 
with a different $M$ value.

\subsubsection{`Diagonal' two-body densities}

An alternate option is to compute only the ``diagonal'' densities, which we define here as the same initial and final states and only scalar densities, that is, 
only $J=0$ in the output. While the usual two-body densities in proton-neutron formalism are still written to a {\tt .den2b} file, one can reinterpret these 
as expectation values of operators, which have a different normalization, written to the {\tt .res} file. 
The options are `{\tt 2d},' which writes the expectation values in proton-neutron formalism, and `{\tt 2i}', which write the expectation values in isospin formalism. 

With these restrictions, less information is needed:
\begin{verbatim}
 !#   State      Energy         Jstate     
         1       -40.47233       0.0
 !# a   b   c   d   Jpair   <  psi || [[ab:Jpair]^+[cd:Jpair]]_0 || psi > 
    1   1   1   1     0       0.073358
    1   1   1   1     2       0.011630
    1   1   2   1     2       0.011295
    1   1   2   2     0       0.200507
    1   1   2   2     2       0.035984
    1   1   3   1     2       0.015691
\end{verbatim}
Note that {\tt Jpair}  here is the angular momentum of the pair. Because we have time-reversal symmetry, we have that 
$\rho(ab,J;cd,J) = \rho(cd,J; ab;J)$. For the diagonal case, only one value is printed. 

From this one can extract the expectation value of components of the Hamiltonian. 
Let
\begin{eqnarray}
\hat{O}_J(ab,cd) =  \zeta^{-1}_{ab} \zeta^{1}_{cd} \sum_M \hat{A}^\dagger_{JM} (ab) \hat{A}_{JM}(cd)  \label{hamopform}
\end{eqnarray}
a scalar operator.
Now define for a state $i$ the expectation value 
\begin{eqnarray}
\langle i, J_i M | \hat{O}_J(ab,cd) | i,J_i M \rangle = \frac{ [ J ] }{[J_i ] }
\rho^{ii}_0 (ab, J; cd, J ) .
\end{eqnarray}
Note that, because of time-reversal symmetry, 
$\langle i, J_i M | \hat{O}_J(ab,cd) | i,J_i M \rangle = \langle i, J_i M | \hat{O}_J(cd,ab) | i,J_i M \rangle$. In a Hamiltonian, 
which is time-reversal-symmetric, a matrix element $V_J(ab,cd)$ actually applies to both. Therefore we define an expectation value which 
is the sum of both,
\begin{eqnarray}
X_J^i(ab,cd) \equiv 
\frac{ \left ( 
\langle i, J_i M | \hat{O}_J(ab,cd) | i,J_i M \rangle +  \langle i, J_i M | \hat{O}_J(cd,ab) | i,J_i M \rangle \right )}{(1+ \delta_{ac} \delta_{bd} )
} \nonumber  \\
= \frac{2}{1+ \delta_{ac} \delta_{bd} } 
\frac{ [ J ] }{[J_i ] } \rho^{ii}_0 (ab, J; cd, J ) 
\end{eqnarray}
This is equivalent to taking the expectation value of a Hamiltonian with exactly one matrix element, $V_J(ab,cd) = 1$. 
It is important to note that proton and neutron orbits are defined to be different, so that $a_p \neq a_n$. 
Such expectation values are useful for, e.g., perturbative fitting of matrix elements.  It is also useful as a consistency check, as one should get the sum rule
\begin{eqnarray}
\sum_{ab} \frac{1+\delta_{ab}}{2}  X^i_J(ab,ab) =
\sum_{a \leq b} \sum_J 
 X^i_J(ab,ab) \nonumber \\
= 2 \sum_{a \leq b} \sum_J  
\frac{ [ J ] }{[J_i ] } \rho^{ii}_0 (ab, J; ab, J ) 
= A(A-1), 
\end{eqnarray}
where $A$ is the number of valence particles.  When running this option, {\tt BIGSTICK} automatically prints out this sum rule. 

If one choose `{\tt (2d)}' the expectation values are printed in proton-neutron formalism. 
One can also convert the densities to isospin format, using option `{\tt (2i)}',  though be aware, if the state has nonzero isospin, then one can have isospin tensors, i.e., non-isoscalar operators. 
Nonetheless we can generalize (\ref{hamopform}) 
\begin{eqnarray}
\hat{O}_{JT}(ab,cd) = \frac{\zeta^{-1}_{ab} \zeta^{-1}_{cd} }{1+\delta_{ac} \delta_{bd} } \sum_{M,M_T}  \hat{A}^\dagger_{JM,TM_T} (ab) \hat{A}_{JM, TM_T}(cd)  \nonumber \\
+ \hat{A}^\dagger_{JM,TM_T} (cd) \hat{A}_{JM,TM_T}(ab) ,
\end{eqnarray}
a scalar, isoscalar, time-symmetric operator, 
and extract expectation values
\begin{eqnarray}
X_{J,T=1}^i(ab,cd) \equiv 
\langle i, J_i M, T_i M_{T} | \hat{O}_{JT=1} (ab,cd) | i,J_i M, T_i M_T \rangle  \\
 = 
 X_J^i( a_p b_p, c_p d_p) +  X_J^i( a_n b_n, c_n d_n)  
 + \frac{1}{2\sqrt{(1+\delta_{ab})(1+ \delta_{cd})} }  
 \nonumber \\
\times   \left [ X_J^i( a_p b_n, c_p d_n)     + (-1)^{j_a + j_b + j_c + j_d} 
 X_J^i( b_p a_n, d_p c_n)   \right . 
 \nonumber \\
  \left . \left .
 - (-1)^{J+ j_a + j_b } 
 X_J^i( b_p a_n, c_p d_n) 
  - (-1)^{J+ j_c + j_d } 
 X_J^i( a_p b_n, d_p c_n)  \right ] \right . 
 \nonumber
\end{eqnarray}
and
\begin{eqnarray}
X_{J,T=0}^i(ab,cd) \equiv 
\langle i, J_i M, T_i M_{T} | \hat{O}_{JT=0} (ab,cd) | i,J_i M, T_i M_T \rangle  \\
 = 
\frac{1}{2\sqrt{(1+\delta_{ab})(1+ \delta_{cd})} }   \nonumber \\ 
\times \left [ X_J^i( a_p b_n, c_p d_n)  
  + (-1)^{j_a + j_b + j_c + j_d} 
 X_J^i( b_p a_n, d_p c_n)  \right .  \nonumber \\
 \left .
 + (-1)^{J+ j_a + j_b } 
 X_J^i( b_p a_n, c_p d_n)  + (-1)^{J+ j_c + j_d } 
 X_J^i( a_p b_n, d_p c_n)  \right ] \nonumber
\end{eqnarray}
Note in the mixed proton-neutron contributions, we keep fixed the order $pn,pn$, as that is how {\tt BIGSTICK} 
orders the labels. In the above, the labels $a$, $b$, etc. without suffixes are shared by both protons and neutrons.

It is important to remember: These expectation values are written to the output {\tt .res} file, with the reguarly-defined densities still written to the {\tt .den2b} file. Furthermore, 
the `densities' and the `expectation values' have different normalizations. 

If the state has isospin $T_i > 0$, however, there are expectation values for up to four additional non-isoscalar operators (one isotensor and 
three isovector). We leave those as an amusing exercise for the reader. 

\section{Strength function option}

\label{strength}

One important capability of {\tt BIGSTICK} is using the Lanczos algorithm to efficiently compute transition strength function distributions
and to decompose a wavefunction using a scalar operator, or option `{\tt s}' in the main menu. Note that this default strength function option, as well 
as '{\tt sn}' , will write \textit{normalized} wave functions to file. The option `{\tt su}' will write unnormalized wave functions to file. This is useful in 
some important applications, as discussed below in \ref{apply1bodygoodJ} and \ref{GTstrength}.

\subsection{Decomposition}

\label{decomposition}

We'll start with decomposition of a wavefunction using a scalar operator \citep{PhysRevC.91.034313}, because operationally it is the most straightfoward. 
Suppose you have a wavefunction, $| \Psi \rangle$, which you have previously computed using {\tt BIGSTICK} and have 
stored in a {\tt .wfn} file; further suppose you have some operator $\hat{\cal O}$ which is an angular momentum scalar, which in turn means its 
matrix elements can be stored in a file just like a Hamiltonian. This operator $\hat{\cal O}$ in turn has eigenpairs,
\begin{equation}
\hat{\cal O} | \Phi_\omega \rangle = \omega | \Phi_\omega \rangle.
\end{equation}
We can always expand $| \Psi \rangle$ into the eigenstates of $\hat{\cal O}$:
\begin{equation}
| \Psi \rangle = \sum_\omega c_\omega | \Phi_\omega \rangle
\end{equation}
and the fraction of the wavefunction $| \Psi \rangle$ labeled by $\omega$ is simply 
$$
\left | \langle \Phi_\omega  | \Psi \rangle \right |^2 = \left | c_\omega \right |^2.
$$
This is particularly useful when  $\hat{\cal O}$ is the Casimir of some group or subgroup, such as total orbital angular momentum $\hat{L}^2$ 
or total spin $\hat{S}^2$.  In that case we say we \textit{decompose} the wavefunction $| \Psi \rangle$ into its $L$- or $S$- components. 

{\tt BIGSTICK} can carry out this decomposition easily.  What you need is, first, a previously computed wavefunction in some 
{\tt .wfn} file, and a file or files which contain the matrix elements of the decomposing operator. 

\noindent \textit{To do this:}

1. From the initial menu choose the option `{\tt s}':

\begin{verbatim}
  * (s) Strength function (using starting pivot )                           * 
\end{verbatim}
$\vdots$
\begin{verbatim}
  Enter choice 
s
\end{verbatim}

Note: the \textit{pivot} is the starting vector; here it is the wavefunction you wish to decompose. {\tt BIGSTICK} can only decompose one wavefunction at a time.

\medskip

2. Enter name of file containing the wavefunction to be decomposed (i.e., the pivot):

\begin{verbatim}
  Compute strength function distribution using previous wfn 
  Enter input name of .wfn file 
mg24
\end{verbatim}
Here the choice of the wavefunction file is {\tt mg24.wfn}; you do not include the extension. At this point, {\tt BIGSTICK} reads in some information from the {\tt .wfn} 
file:

\begin{verbatim}
  testing magic number     31415926    31415926
 dimbasischeck=                 28503
  Valence Z, N =            4           4
  Single particle space : 
  N       L       2xJ 
           0           2           3
           0           2           5
           1           0           1
  
  Total # of orbits =            3
  2 x Jz =            0
\end{verbatim}

The `magic number' is a test of internal consistency to make sure, first, {\tt BIGSTICK} is correctly reading the file (in particular if the binary file was created on a 
different platform) and also between different versions of {\tt BIGSTICK} if the information protocol has changed.

From this information {\tt BIGSTICK} reconstructs the basis and checks the dimensions agree. It then asks for the output file:

\begin{verbatim}
  Enter output name (enter "none" if none)
\end{verbatim}

After this, {\tt BIGSTICK} will make the standard inquiries for the Hamiltonian. When decomposing a wavefunction, the `Hamiltonian' is actually an angular momentum 
scalar which is a Casimir of the group or sub-group; for example, it could be $\hat{S}^2$ or $\hat{L}^2$. Such files are in in the same format as any interaction file.
While we provide a sample operator, it is up to the user to generate these files. 

After the interaction file(s) have been read in, you must enter in the number of iterations and number of states to keep ({\tt BIGSTICK} automatically chooses a 
fixed-iteration run for Lanczos):
\begin{verbatim}
  Enter nkeep, # iterations for lanczos  
  (nkeep = # of states printed out )
\end{verbatim}
Exactly how many iterations to to choose requires some knowledge of the group, or, specifically, knowledge of the eigenvalues of the Casimir, and, in many cases, a few trials.   
Remember that the irreps of the group are labled by the eigenvalues of the Casimir, which means the eigenvalues are highly degenerate.  The number of iterations 
needed should be no greater than the number of distinct eigenvalues. So, for example, if one has 8 nucleons and is decomposing via spin, the values of $S$ can 
be $0, 1, 2,3,$ or 4. Therefore the number of iterations should be no more than 4 (because one wants a total dimension of 5). Often one can use fewer iterations. If you 
use too many iterations, you will get duplication of eigenvalues or, worse, unconverged duplicate eigenvalues.

Finally, {\tt BIGSTICK} will print out a list of the starting states in the pivot file, and their energies and $J$ and $T$ values, and ask you to choose a pivot:
\begin{verbatim}
  There are            5  wavefunctions 
           5  states 
  
  STATE      E         J          <H > 
    1      -92.7790     -0.0000      0.0000
    2      -91.1196      2.0000      0.0000
    3      -88.4779      2.0000      0.0000
    4      -87.9781      4.0000      0.0000
    5      -87.4348      3.0000      0.0000
  Which do you want as pivot? 
\end{verbatim}
Hence if you want to decompose the $J=3$ state, enter {\tt 5}. 

Immediately after reading in the pivot, {\tt BIGSTICK} will print out the norm of the input pivot
(that generally does play a role in this kind of decomposition, but will in transition strength functions):
\begin{verbatim}
  0.999999998379788         = total input strength 
\end{verbatim}
Often the norm or total input strength is far different from 1.

After carrying out the specified Lanczos iterations,
the result will look something like this, depending on how many iterations:
\begin{verbatim}
  Energy   Strength 
  ______   ________ 
   0.00000   0.63545
   2.00000   0.33880
   6.00000   0.02515
  12.00000   0.00059
  20.00000   0.00000
  ______   ________ 
\end{verbatim}
The `energies' on the left are the eigenvalues of the operator you are using to decompose the wavefunction, here $\hat{S}^2$.  Hence the $J=3$ state (or 
state 5 in the above example), is $63.5\% S=0$, $33.9\% S=1$, and so on. These results are written to the standard {\tt .res} file.

\subsection{Transition strength function distributions: the basics}

\label{apply1bodybasic}

Often we want the transition function between two states, that is $\left | \left \langle \Psi_f \right | \hat{\cal O } \left | \Psi_i \right \rangle \right |^2$
where $\hat{\cal O}$ is some one-body transition operator, for example the $E2$ or $M1$ transition operator. (As always, we assume the reader is familiar 
with these concepts.) 
 If one only wants one or two transitions, one can compute those using the one-body density matrices, which we describe above in \ref{transitionstrengthsfromdensities}.

 But sometimes we want many transitions to many final (or `daughter') states from a single initial (`parent') state, for example if we want to profile `giant' resonances. 
 We can do this using {\tt BIGSTICK} in three to four steps.  The first step is to generate and write to file an initial wavefunction. 
 
 The second step is to apply a one-body operator, $\hat{\cal O}$.  The matrix elements of the one-body operator must 
 be stored in file with extension {\tt .opme}, with the format defined in the next section. 
 To apply a one-body operator, choose option `{\tt o}' at the opening menu:
 \begin{verbatim}
   * (o) Apply a one-body (transition) operator to previous wfn.... 
 \end{verbatim}
 {\tt BIGSTICK} will then ask for the name of the input {\tt .wfn} file and an output name, required here. After reconstructing the basis from the information in the 
 input {\tt .wfn} file, it will ask:
 \begin{verbatim}
   Enter name of .opme  file 
\end{verbatim}
The matrix elements of the one-body operator are read in from a file with extension {\tt .opme} (`operator matrix element').  While we distribute some sample {\tt .opme} files with {\tt BIGSTICK}, in general it is 
up to the user to generate such files.   The format of such files are
\begin{verbatim}
 iso                    ! indicating good isospin 
           3            ! # of single-particle orbits
  1    0    2  1.5        ! index of orbit, n, l, and j
  2    0    2  2.5
  3    1    0  0.5
        1        0    ! J and T of operator
    1    1  -2.19089  !  a, b  < a ||| O ||| b >
    1    2   4.38178
    2    1  -4.38178
    2    2   4.09878
    3    3   3.46410
\end{verbatim}
{\tt BIGSTICK} first checks the list and order of single-particle orbits agrees with that of the read-in wavefunction. {\tt BIGSTICK} will then read in the matrix elements of the one-body operator and apply it to \textit{each} wavefunction stored in the input {\tt .wfn} file and write them to a new
output {\tt .wfn} file.

The final step is to run {\tt BIGSTICK} again, this time with the strength function option `{\tt s},' using the wavefunction generated in the second step as input.  
This time, when {\tt BIGSTICK} asks for the interaction file name, you should use the \textit{same} file(s) to generate the initial state, because you are diagonalizing 
the Hamiltonian.

\begin{figure}
\includegraphics [width = 10cm]{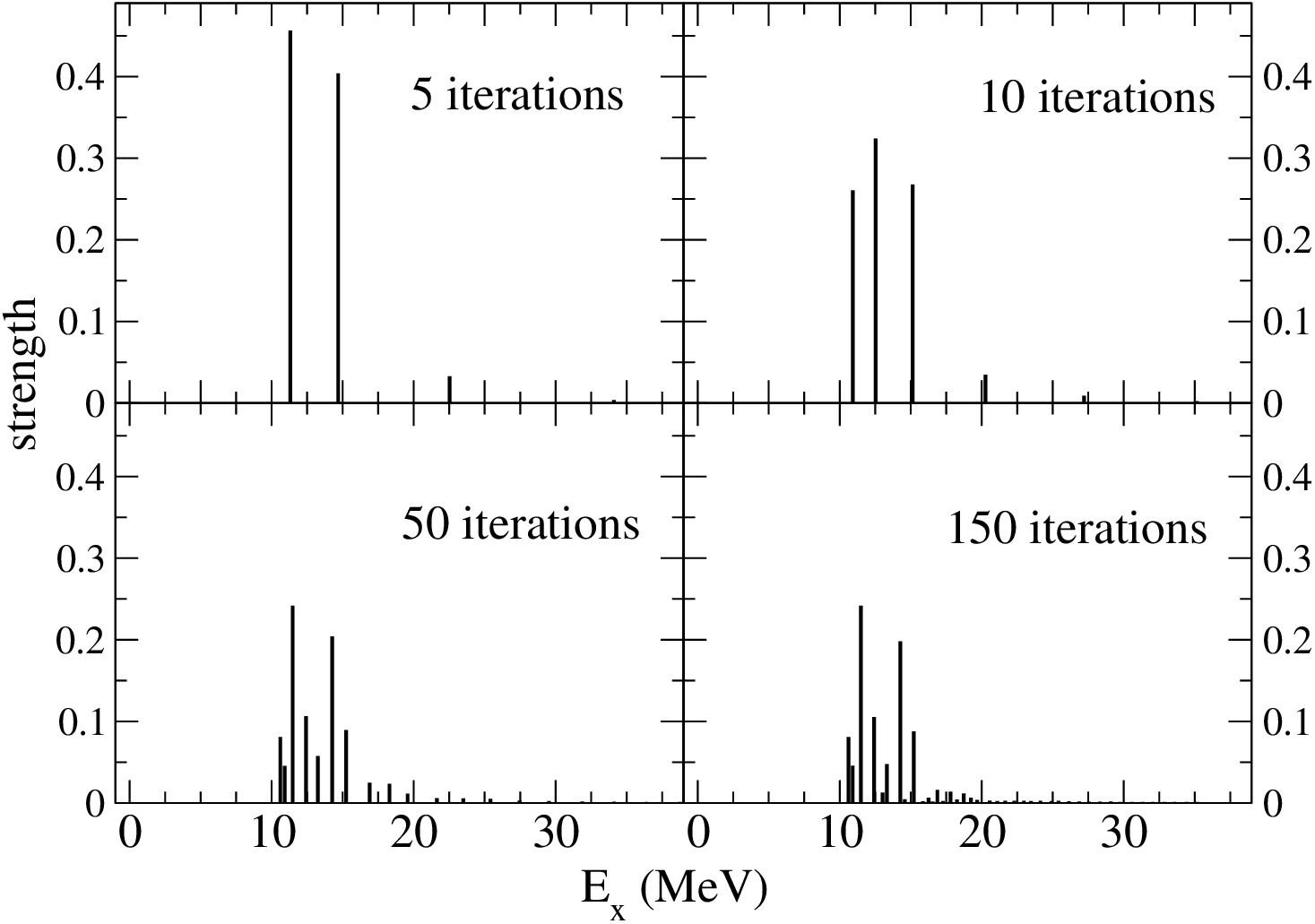}
\caption{\label{si28gamow} Illustration of how transition strengths evolve with increasing number of Lanczos iterations. 
In this example, the operator $\vec{\sigma} \tau_0$ was applied to the ground state of $^{28}$Si, calculated in the \textit{sd} shell 
with the USDB interaction.
}
\end{figure}

As with decomposition, {\tt BIGSTICK} will now carry out a fixed number of iterations and print out the transition strength. Because it includes the norm of the input 
pivot, these strengths can be greater than 1.

An important question is that of convergence.  As you probably know, in the Lanczos algorithm the extremal eigenpairs converge first, with interior eigenpairs converging 
later.  This is true as well for the strengths described above: the extremal strengths will converge quickly to strengths (and eigenenergies) of extremal levels, but 
interior strengths will often not be converged; instead they will be some sort of `local average' of strengths. 

\begin{figure}
\includegraphics [width = 10cm]{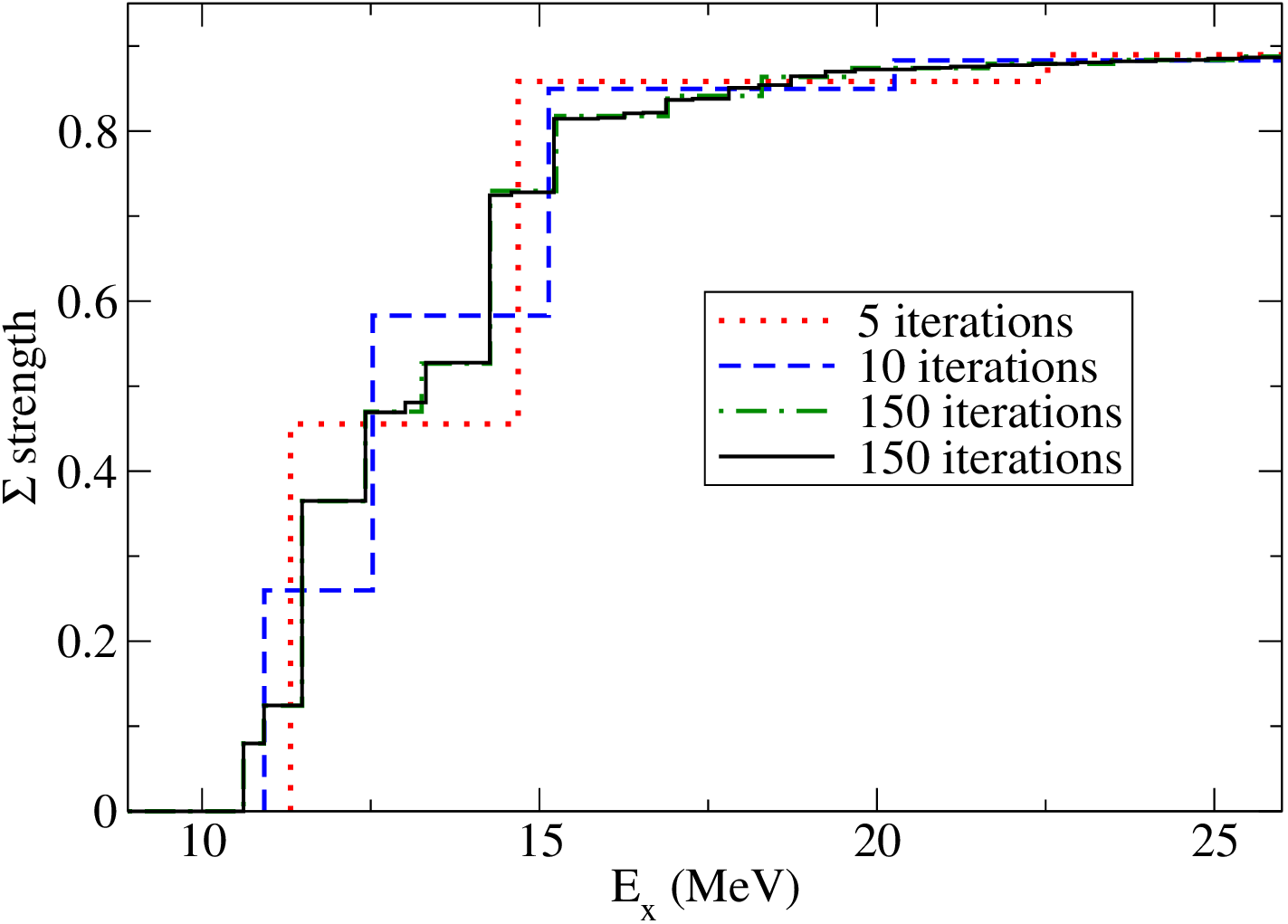}
\caption{\label{si28summed} Running sums of the strengths shown in Fig.~\ref{si28gamow}.
}
\end{figure}

This can be confusing at first, as illustrated in Fig.~\ref{si28gamow}. Here we computed in the $sd$ shell and using the USDB interaction \citep{PhysRevC.74.034315}m
$^{28}$Si (which has six valence protons and six valence neutrons on top of a frozen $^{16}$O core).  After generating the wave function in a `normal' run (option `{\tt (n)}', 
we applied the operator $\vec{\sigma} \tau_0$ (generated with additional tools now available in the {\tt BIGSTICK} GitHub repository) using the option `{\tt (o)}'.  Finally, 
we used the strength function option `{\tt (s)}', and applied 5, 10, 50, and 150 iterations.   When the individual strengths are plotted, it looks like the results are badly converged. 

What looks like a bug is actually a feature. In practice one often doesn't need each and every strength to be fully converged. Instead we only need integrals 
over the strengths to be converged, and this does happen. While we can only refer the reader to \citet{ca05} and references therein, we can state that the 
\textit{moments} of the distributions of strengths do converge. In fact, if one carries out $N$ Lanczos iterations, one has $\sim 2N$ moments of the 
distribution correctly. Hence often only thirty or fifty iterations suffice. This is illustrated in Fig.~\ref{si28summed}, which displays the running sums of the strengths in Fig.~\ref{si28gamow}.
Plotted thusly, one can see that how well they overlap, and indeed the 50 iteration case is nearly indistinguishable from the 150 iteration case.

As another example, consider $^{20}$Ne in the $sd$ shell with the USDB interaction.  If we apply the $\sigma$ operator, with the matrix elements given above,
and then apply the strength function option {\tt s}, the output will look something like:
\begin{verbatim}
  0.38353481218057317       = total strength 
          35  iterations 
  
  Energy   Strength 
  ______   ________ 
 -40.47233   0.00000
 -38.72564   0.00000
 -36.29706   0.00000
 -33.77148   0.00000
 -32.88892   0.00000
 -30.18655   0.00000
 -27.83635   0.11991
 -25.73419   0.00003
 -25.49045   0.17314
 -24.73655   0.01326
 -22.86589   0.00027
 -22.24522   0.01542
 \end{verbatim}
Notices the strength at -27.836 MeV (which is the $J;T=1;0$ state) is 0.11991; using the Clesbsch-Gordan coefficients gives a factor of 3, or a 
total strength of 0.3597, which agrees with our previous result.

Note: the option `{\tt ss}' will only compute the strength distribution, but will not write wave functions to file, nor compute $J$, $T$ of the final wave functions. 
This is useful for large dimension cases with many iterations, as the wavefunctions are often not used afterwards.

There is a small bug in this lovely ointment: it assumes we have treated angular momentum (and isospin) correctly, a topic we now turn to.

\subsection[Transitions with good angular momentum]{Transition strength functions with good angular momentum}

\label{apply1bodygoodJ}

In the prior subsection we glided over questions of angular momentum, which we treat more carefully here.  An important question is correct calculation of the 
$B$-values, as defined in Eq.~(\ref{Bvalue}) above, which assume an average over final states and a sum over final states. 
But what we computed in the previous section was 
$$
| \langle \Psi_f | \hat{\cal O} | \Psi_i \rangle |^2
$$
where the states have fixed $M$ and fixed $T_z$, because, as currently written, both initial and final wavefunctions must be in the same basis.  
(We plan at a later date to write a separate tool which will allow one to apply and operator from a wavefunction in one basis to a wavefunction in a different basis.)
If you want 
the applied operator to change parity, then both parities must be included in the basis (option 0 when entering parity in the initial calculation of wavefunctions). 

To extract the $B$-value, one has to invoke the Wigner-Eckart theorem:
\begin{eqnarray}
B({\cal O}: i\rightarrow f) =
\frac{1}{2J_i +1} \left | (\Psi_f:  J_f || \hat{\cal O}_J ||\Psi_i  J_i ) \right|^2
=  \nonumber \\ 
\frac{2J_f+1}{(2J_i +1) }
\left |
\frac{ \langle \Psi_f:  J_f M | \hat{\cal O}_{J 0} | \Psi_i: J_i M\rangle  }{( J_i M,  J 0 | J_f M ) } \right|^2.
\label{Bvalue_calc}
\end{eqnarray}

If one has $J_i = 0$ (which can only happen if $M=0$), then 
the $B$-value is straightforward to calculate:
$$
B({\cal O}: i\rightarrow f) = (2J+1)  | \langle  \Psi_f:  J 0| \hat{\cal O}_{J 0} |  \Psi_i: 0 0 \rangle|^2.
$$
It's more complicated with $J_i \neq 0$; in that case the triangle rule, $ | J_i -J | \leq J_f \leq J_i +J$ is in effect, and in fact the state produced by 
{\tt BIGSTICK}, 
$$
\hat{\cal O}_K | \Psi_i: J_i \rangle
$$
will be an admixture of states of different $J_f$.  Thus
one needs an additional step, of projecting out states of good angular momentum \textit{after} applying the transition operator 
but \textit{before} carrying out the strength function option via Lanczos. Fortunately we already know how to do this via decomposition as discussed in 
\ref{decomposition}.

Therefore, to properly carry out calculation of strength functions, you will need (a) files for the interaction, (b) a file 
for the one-body transition operator, and (c) files with matrix elements of $\hat{J}^2$ and, separately, ${T}^2$ (you only need the latter if your transition operator 
has isospin rank 1).  Then carry out the following steps:

\medskip

(1) With your interaction use option {\tt (n) } or similar option, generate a {\tt .wfn} file containing an initial state;

\medskip

(2) Use option {\tt (o)} to apply the one-body operator (note this will be applied to \textit{every} wavefunction in the file);

\medskip

(3) If your initial state has $J_i \neq 0$, you will need to filter out a state of good $J_f$ for every possible $J_f$; in fact what you will do will be to 
decompose the state from step (2) into its components $J_f$. Here you use option {\tt (s)}. If there are $N$ possible values of $J_f$ you only need to 
do $N-1$ iterations.

If your transition operator has $T=1$ and your initial $T_i \neq 0$, you will need to further decompose into possible $T_f$ states. 

\medskip

(4) Finally, use option {\tt (s)} again, but this time with the original interaction, to get the strength function distribution. You will have to apply the 
Wigner-Eckart theorem as in Eq.~(\ref{Bvalue_calc}), but now that step (3) guarantees a definite value of $J_f$ (and, if needed, $T_f$), you can carry this out.

You will have to repeat for each possible final value of $J_f$.  (An efficient way to do this is using the block strength function option.)


Here is an example using $^{19}$F: if we choose the $J=5/2$ state (state $\# 2$) as the pivot and apply $\vec{\sigma}$, and then use the strength function 
option with $\hat{J}^2$,
\begin{verbatim}
   1.3726179231928042       = total strength 
           3  iterations 
  
  Energy   Strength 
  ______   ________ 
   3.75000   1.13861
   8.75000   0.05630
  15.75000   0.17771
  ______   ________ 
  \end{verbatim}
  This means 83$\%$ of the pivot has $J=3/2$  and only 4.08$\%$ has $J=5/2$.  Next we run the strength function again  on the second state using the 
  original ({\tt usdb}) interaction:
  \begin{verbatim}
   1.0000000433422094       = total strength 
          35  iterations 
  
  Energy   Strength 
  ______   ________ 
 -23.86096   0.00000
 -23.78367   0.00000
 -22.09059   0.66440
 -21.26237   0.00000
 -19.25723   0.00000
\end{verbatim}
Note that the wavefunction is normalized when it is read in. 
(By the way, the zeroes show up because although there is no strength to them, or very little, roundoff error allows them to grow during Lanczos. This is the same phenomenon which forces us to orthogonalize new Lanczos vectors against old ones.)
We have to multiply $0.66440 \times 1.13861 = 0.756$  to get the `raw' strength, which here is $|\langle J_f M | \vec{\sigma} | J_i M \rangle |^2$, then 
we have to follow Eq.~(\ref{Bvalue_calc}):
$$
0.756 \times \frac{ 2(3/2)+1}{2(5/2)+1} \times \frac{1}{| ( 5/2 \, 1/2 , 1 \, 0 | 3/2 \, 1/2)|^2} = 0.756 \times \frac{4}{6} \times \frac{1}{2/5} =1.260
$$
which agrees with our previous result!  Note that you have to do each step with care; if don't scale the two-body matrix elements, you will get different results.

\subsection{ Gamow-Teller with strength function option}

\label{GTstrength}

Charge-changing transitions such as Gamow-Teller are a straightforward generalization but require even more care.  Here one transitions from a 
state with some initial $T_{z,i}$ to some final $T_{z,f} = T_{z,i} \pm 1$.  Because, as of the time of this writing, {\tt BIGSTICK} requires the same initial 
and final basis, we have to choose $T_{z,0} = \mathrm{min} \, ( \mathrm{abs} (T_{z,i}),  \mathrm{abs} (T_{z,f}) )$ and invoke isospin rotation. 
Typically you will have to filter both $J$ and $T$.  The B-value is given by
\begin{eqnarray}
B({\cal O}: i\rightarrow f) =
\frac{1}{2J_i +1} \left | \langle \Psi_f:  J_f, T_f \, T_{z,f}  || \hat{\cal O}_J ||\Psi_i:   J_i , T_i \, T_{z,i} \rangle \right|^2
=  \nonumber \\ 
\frac{2J_f+1}{(2J_i +1) }
\left |
\frac{ \langle \Psi_f:  J_f M ,  T_f \, T_{z,0}| \hat{\cal O}_{1\, 0,1 \,0} | \Psi_i: J_i M, T_i \, T_{z,0} \rangle  }{( J_i M,  J 0 | J_f M ) } \right|^2 
\label{Bvalue_calc_iso} \\
\times \left |
\frac{ ( T_i \,      T_{z,i} , 1 \, \pm 1 | T_f \, T_{z,f} )}
{ ( T_i \,      T_{z,0} , 1 \, 0 | T_f \, T_{z,0} )}
\right |^2
\nonumber
\end{eqnarray}
where the last line uses the isospin Wigner-Eckart theorem to transform from the isospin frame the calculation is carried out in, to the physically 
desired isospin frame. 

We use {\tt BIGSTICK}'s strength function option {\tt s} to compute  the matrix element
$ | \langle \Psi_f:  J_f M ,  T_f \, T_{z,0}| \hat{\cal O}_{1\, 0,1 \, 0} | \Psi_i: J_i M, T_i \, T_{z,0} \rangle|^2$.
This is slightly involved.  We give now a detailed example, again with $^{19}$F.  After applying the Gamow-Teller operator, we filter out first with $\hat{J}^2$,
using the second state as the pivot:
\begin{verbatim}
  0.72792934307990387       = total strength 
           3  iterations 
  
  Energy   Strength 
  ______   ________ 
   3.75000   0.61870
   8.75000   0.01998
  15.75000   0.08925
  ______   ________ 
\end{verbatim}
and then we filter this with $\hat{T}^2$ (applied to the first state, that is, the one with $J=3/2$
\begin{verbatim}
  0.99999997896552772       = total strength 
           3  iterations 
  
  Energy   Strength 
  ______   ________ 
   0.75000   0.77481
   1.10978   0.00000
   3.75000   0.22519
  ______   ________ 
\end{verbatim}
and then finally applying the strength function with the {\tt usdb} interaction:
\begin{verbatim}
   1.0000000877719881       = total strength 
          35  iterations 
  
  Energy   Strength 
  ______   ________ 
 -23.86096   0.00000
 -23.78367   0.00000
 -22.09059   0.87875
 -21.26237   0.00000
 -19.25722   0.00000
\end{verbatim}

Thus in this example the `raw' transition strength is that for the $J;T = 5/2;1/2 \rightarrow 3/2;1/2$ which is 
$$
0.61870 \times 0.77481  \times 0.87875 = 0.42125.
$$
This now has to be converted to a Gamow-Teller B-value by Eq.~(\ref{Bvalue_calc_iso}):
$$
B(GT) =\frac{2 \cdot \frac{3}{2} +1}{2 \cdot\frac{ 5}{2} +1} \times \frac{  0.42125}
{\left  | \left (\frac{ 5}{2} \,\, \frac{1}{2}, 1 \, 0 | \frac{3}{2} \,\, \frac{1}{2} \right )\right |^2} \times 
\left |
\frac{\left  ( \frac{1}{2} \,\, -\frac{1}{2}, 1 \, 1 | \frac{1}{2} \,\, \frac{1}{2} \right ) }
{\left (\frac{1}{2} \,\, +\frac{1}{2}, 1 \, 0 | \frac{1}{2} \,\, \frac{1}{2} \right )  }
\right |^2
$$

A good exercise is to compute low-lying transitions two ways, first with density matrices, and then via the strength function option, to confirm they agree with each other. 
For Gamow-Teller, one can and should verify results by using the Ikeda sum rule.

One can also use other operators for projection, for example, using center-of-mass to project out nonspurious states in no-core shell model calculations.  

We plan to later allow two-body transition operators, but as of version 7.8.1 these 
options have not yet been installed.

\section{Resolvent/Green's function}

\label{green}
One can   compute the action of the Green's function or resolvent, 
$(E_0 - \hat{H} )^{-1}$ on an initial state, the pivot $| \Psi_i \rangle$, which must be read in from a file.
The option  {\tt `(g)'} writes $(E_0 - \hat{H})^{-1} | \Psi_i \rangle $ to a {\tt .wfn} file.

In the related option, {\tt `(gv)'}, after calculating the resolvent on the pivot, one reads in a 
previously computed file of 
wave functions and computes the overlap. 

After reading in the interaction, you will be asked for 
the energy
\begin{verbatim}
  + + + + + + + + + + + + + + + + + + + + + + + + + + + 
  + Calculating resolvent/Green function on an initial vector 
  
  
  Enter energy E in resolvent/Green function 1/(E-H)
\end{verbatim}

If $E_0$ is extremal relative to the eigenvalues, Lanczos algorithm converges quickly, but if it is interior, the convergence can be slow. This can be understood through a 
spectral decomposition of the uncoverged eigenvalues, crossing $E_0$. 

The option {\tt `(gc)'} allows you to enter a complex energy 
$E_0$.
\begin{verbatim}
  Enter real, imaginary parts of E 
\end{verbatim}

The output is written to a file, as two vectors: the first 
vector is the real part, the second vector is the imaginary part.  You can then take overlaps or use as necessary.

The most general option is {\tt `(pv)'}. This option reads in a choice of pivot and carries out a fixed, user-defined number of iterations.  The resulting Lanczos vectors are then dotted against a second {\tt .wfn} input file, with the results 
written to the file {\tt overlap\_lanczos.dat}:
\begin{verbatim}
  Lanczvec     Finalstate       overlap    
        1         1         -0.01173
        2         1          0.04016
        3         1         -0.09216
        4         1          0.15468
. . . .    
\end{verbatim}
No energy is input, but from the Lanczos coefficients and overlaps with the Lanczos vectors, one can reconstruct the 
resolvent for an arbitrary energy.


\chapter{A peek behind the curtain}

\label{peek}

Although this manual  details how to use {\tt BIGSTICK}, it only outlines the algorithms and  program. 
The distribution includes an \textit{Inside Guide} which, while incomplete, contains many more details on the code, and the 
source code itself is heavily commented.  Even so, it is a complex program, with more than seventy Fortran files and on the order of 70,000 lines. 
Several  files are for specialized applications most users will not care about, as well as for features slated for obsolescence. 

There are ways to get {\tt BIGSTICK} to present more information about its inner workings, as well as ways to extert more control over the 
algorithm. A number of logical flags turn behaviors on and off. The most important flags, and some default settings, are found in 
{\tt bmodules\_flags.f90}. Some additional flags for output are in the module {\tt io} in {\tt bmodules\_main.f90}, and other flags can be found 
elsewhere.  To detail all the possibilities would expand this already long manual by a significant amount.

In this chapter we outline the major steps {\tt BIGSTICK} takes in carrying out a `normal' run, as well as telling the curious user how to 
print out an explicit representation of the basis and of the many-body Hamiltonian matrix.

\section{A normal run}

Here are the steps {\tt BIGSTICK} carries out in a `normal' run, that is, setting up a many-body Hamiltonian and finding the low-lying extremal eigensolutions.

\begin{itemize}

\item {\tt BIGSTICK} sets up the basis.

\item {\tt BIGSTICK} counts up the number of jumps (data needed for constructing the Hamiltonian on-the-fly).

\item {\tt BIGSTICK} gathers the interaction data.

\item If running in parallel, {\tt BIGSTICK} computes the distribution

\item {\tt BIGSTICK} generates the data needed (on a specific MPI process if running in parallel), specifically the jumps and the decoupled matrix elements.

\item {\tt BIGSTICK} sets up storage for the Lanczos vectors.

\item { \tt BIGSTICK} begins Lanczos iterations.

\item Upon completion of Lanczos, {\tt BIGSTICK} constructs the low-lying eigenvectors. It resets the jumps and computes angular momentum and isospin 
as expectation values. Eigenvalues and eigenvectors are written to file.  If density matrices requested, {\tt BIGSTICK} resets jumps for one-body 
operators and computes. 

\item Upon finishing, {\tt BIGSTICK} reports on timing and closes down.

\end{itemize}

\section{Writing out the basis}

\label{trdens}

Through factorization and other tricks, {\tt BIGSTICK} only implicitly stores the basis and the Hamiltonian. In actual operations, 
{\tt BIGSTICK} stores information on pieces of Slater determinants, which we call ``haikus.'' Haikus are organized by quantum numbers, 
as are the action of single-fermion creation and annihilation operators on the haikus. These latter we called ``hops'' and from them 
we construct jumps, and from jumps we construct many-body matrix elements, and so on.  Once the hops are created the haikus are not 
needed, and once the jumps are constructed the hops are not needed. 

It can be useful, however, to have explicit representations of both the basis and of the many-body Hamiltonian. In standard runs, 
the final eigenvectors are written to file with extension {\tt .wfn}. These files are unformatted to save space. Furthermore the 
detailed basis information is not saved; instead any basis {\tt BIGSTICK} is constructed in a standard order, and when reading 
a {\tt .wfn} file {\tt BIGSTICK} swiftly reconstructs the basis. 

From the main menu, however, the option `{\tt t}' will write out both the basis and the eigenvectors in explicit, human-readable form, 
to a file with extension {\tt .trwfn} (originally written as an input to Petr Navratil's density code {\tt TRDENS}). 

Here is an annotated example output from the $p$-shell, using the Cohen-Kurath interaction. First is a header describing the nucleus:
\begin{verbatim}
           4     ! valence Z
           4     ! valence N
 ckpot ! name of INTERACTION FILE 
   19.45492      ! HW (approx)           12
           1  ! # of majors shells 
   12    1 ! total p+n s.p.s, # shells core 
           0  ! Nmax (excitations) 
                    51  ! # of many-body configurations = basis dimension
           1  ! parity, + 
           0  ! 2 x Jz 
           5  ! # of eigenstates kept
\end{verbatim}
Next is a list of the eigenenergies, and numerical J and T values; the latter are written as real numbers.
Reading across we have $E$, $J$, and $T$.
\begin{verbatim}
  -71.04467      3.6173283E-06 -4.1723251E-07
  -66.39703       2.000000     -3.8743019E-07
  -58.59552       1.000001     -4.4703484E-07
  -57.57795      3.6235542E-06 -4.1723251E-07
  -57.54143       4.000000     -2.9802322E-07
\end{verbatim}
Then comes a list of the single particle state quantum numbers. Reading across we have
label, $n$ (number of radial nodes), $l$, $2\times j$, $2 \times j_z$, and $2 \times t_z$:
\begin{verbatim}
1           0           1           1          -1           1
2           0           1           3          -1           1
3           0           1           3          -3           1
4           0           1           1           1           1
5           0           1           3           1           1
6           0           1           3           3           1
7           0           1           1          -1          -1
8           0           1           3          -1          -1
9           0           1           3          -3          -1
10           0           1           1           1          -1
11           0           1           3           1          -1
12           0           1           3           3          -1
\end{verbatim}
So single-particle state 1 is a $0p_{1/2}$ with $j_z = -1/2$ and is a proton, single-particle state 2 is 
$0p_{3/2}$ with $j_z = -1/2$ and is a proton, etc.

Finally we have a listing of the 51 many-body basis states and their amplitudes for the first five eigenstates:
\begin{verbatim}
    1    2    3    5    8   10   11   12
  0.1989746094  0.1647298932 -0.0000000709 -0.0788139924 -0.1036236882
    1    2    3    5    7   10   11   12
  0.0805789307  0.0938614532  0.0855044648 -0.1246829554 -0.0399925224
    1    2    3    4    8   10   11   12
 -0.0805789307 -0.0938614309  0.0855044946  0.1246829703  0.0399925113
    1    2    3    4    7   10   11   12
 -0.0582427122 -0.0433882587  0.0000000068  0.0952372476  0.0186970048
    1    2    3    6    7    8   11   12
  0.0825415403 -0.0342168659 -0.0919694155 -0.0390651748  0.1794791222
    1    2    3    6    7    8   10   12
  0.0697834268 -0.0333072022  0.0155492499 -0.1079809889  0.1385308802
    1    2    3    6    9   10   11   12
 -0.1513192952  0.0582505427 -0.0000000329  0.0562667921 -0.3108672500
    2    3    4    5    7    8   11   12
  0.0560085103 -0.0361775197 -0.0000000029 -0.0111581217  0.1036224812
...
\end{verbatim}
So the first many-body basis states has occupied single particle states 1,2,3,5 (protons) and 8,10, 11, and 12 (neutrons). 
The five real numbers following are the amplitudes for this basis state for the five eigenstates whose eigenenergies are 
given above. You can see an example of factorization: the basis states 1 and 2 have the same proton occupancies loop 
over neutron occupancies; while basis states 3 and 4 have the same proton occupancies (but different from basis states 1 and 2) 
and loops over the \textit{same} neutron occupancies as basis states 1 and 2. By adding up the $j_z$ values, 
the proton ``Slater determinants'' have $M_p = -2$ and the neutron Slater determinants have $M_n= +2$. 
The next basis states, number 5 through 7, have $M_p = - M_n = -1$. In this way {\tt BIGSTICK} builds up the basis. 
As shown in the example, 
in constructing the basis via factorization, the innermost loop is over neutron Slater determinants while the outer is over protons. 

We recommend against using this option on a regular basis, because writing this information to a file is slow, and {\tt BIGSTICK} does 
not have postprocessing options for this format. Nonetheless it can be useful for understanding what is going on, and could be a basis 
for a user's own post-processing code. 

To facilitate user-written post-processing, we have added two additional options. Option {\tt `(tx)'} 
allows you to read in a standard format .wfn file and to write it out in the .trwfn format.
Option {\tt `(tw)'} goes in the opposite direction: with it one can read in a .trwfn file and 
write out as a .wfn file.  For this latter option, the user \textbf{must} enter in the information about 
the basis as for a normal {\tt BIGSTICK} run. Furthermore, the order of the coefficients in the .trwfn 
file cannot be changed; when reading in from a .trwfn file, {\tt BIGSTICK} assumes the order of 
the coefficients is standardized and does not check the actual occupations of the configurations.

\subsection{Alternate information on basis}

\label{altbasis}

To enable further post-processing, two new options for printing out basis information have been added. Option {\tt `(b)'} 
will produce a binary file, with extension {\tt .bas}, with basis information.  Alternately, and likely more useful for most readers, 
option {\tt `(ba)'} will produce a human-readable (ASCII) version of the same information, also with extension {\tt .bas}.
Unlike the option `{\tt (t) } ,' this option does not create the full basis. It does provide valuable information, however, on 
how the full basis is constructed from proton and neutron many-body states (Slater determinants--technically, the 
occupation representation of Slater determinants--or `SDs').  The basic idea is that each proton SD and each neutron SD is assigned
an index, {\tt ip} and {\tt jn}, respectively, and there are arrays {\tt pstart()} and {\tt nstart} which provide the index of the 
combined many-body state, that is, 

\medskip

\noindent {\tt index = pstart(ip) + nstart(jn)}.

\medskip

Furthermore, the occupied single particle states for each proton and neutron SD are provided.
Hence the  {\tt .bas} provides information on how to reconstruct the basis.

After writing the basis information to file, {\tt BIGSTICK} halts. These options do not create or solve the Hamiltonian matrix, and
do not write any wave function vectors to file; for that you will still need the {\tt (t)} option.  The many-body states and amplitudes 
written to the {\tt .trwfn} file will be in the same order as those defined

\medskip

(The example given is for $^{20}$Ne in the $sd$-shell.  Some of the tabs may be different in your output file. In addition, the 
comments do not appear in the actual {\tt .bas} file.)

\medskip

\begin{verbatim}
    31415926        ! 'magic number' 
                  1        ! wfn version number
\end{verbatim}
The `magic number' was introduced in case we made significant changes to  our wave function conventions. The version 
number serves a similar function
\begin{verbatim}
  2      2     ! valence Z and N
 T    ! isospin flag (hardly used any more)
  3       3 ! # of proton, neutron orbits
\end{verbatim}
The quantum numbers of the orbits (protons first, then neutrons, even if the same) are, in order:
\textit{nr} (radial quantum number), $2 \times j$, $l$, $\pi$ (parity) and $w$ (weighting for truncations):
\begin{verbatim} 
!  nr   2j     l   par  w 
    0    3    2    1    0
    0    5    2    1    0
    1    1    0    1    0
    0    3    2    1    0
    0    5    2    1    0
    1    1    0    1    0
   12     12  ! # of proton, neutron single particle states
\end{verbatim}
The quantum numbers of the single particle states (protons first, then neutrons, even if the same), are, 
in order: \textit{nr} (radial quantum number), $2 \times j$, $2 \times m$, $l$,  $w$ (weighting for truncations),
$\pi$ ( parity), orbit label corresponding to the above defined orbits, and finally the group number. Here the 
`group' is used to identify single particle states with the same parity, $w$, and $m$ value, used in creating arrays 
for interactions; however it is not 
set by the time the basis is written to file (and is not generally needed).
\begin{verbatim} 
!  nr   2j    m   l     w  par orb  group
    0    3   -1    2    0    1    1    0
    0    3    1    2    0    1    1    0
    0    5   -1    2    0    1    2    0
    0    5    1    2    0    1    2    0
    1    1   -1    0    0    1    3    0
    1    1    1    0    0    1    3    0
    0    3    3    2    0    1    1    0
    0    5    3    2    0    1    2    0
    0    5   -3    2    0    1    2    0
    0    3   -3    2    0    1    1    0
    0    5   -5    2    0    1    2    0
    0    5    5    2    0    1    2    0
...
\end{verbatim}
Because this information is mostly repeated later: {\tt BIGSTICK} initially reads in or creates a single-particle space, and later, 
depending upon truncations, may only use part of that single particle space. This is mostly in the context of \textit{ab initio} / 
`no-core' calculations. For example, one may define a single-particle space with 10 major oscillator shells, but then create
a $p$-shell nucleus allowing only 4$\hbar \Omega$ excitations, equivalent to choosing {\tt Max excite} (W) of 4; this will only 
use 6 major oscillator shells.  The reason for this is for the code to be more robust to use; also, for some interaction files the 
single-particle space in the file may be larger than that needed, in which case one must initially put in the single-particle space 
appropriate for the interaction file. If you are doing phenomenological calculations in a restricted space, this is generally not relevant 
for you.

\medskip

The following  quantum numbers are for the  full many-body space:
\begin{verbatim}
           0 +       0    ! Jz   parity  max W
 \end{verbatim}
The following are logical flags for single-particle space:
\begin{verbatim}          
 T T F   ! allsameparity, allsameW, spinless
              640  ! basis dimension
 \end{verbatim}
If there are no truncations, for example as in many phenomenological calculations, {\tt allsameW=.true.} and all values of $w$ 
are set =0.
 
 Now the single-particle states are written out again. As stated above, it may seem redundant, but it ensures that the many-body states 
 are correctly interpreted. The following single-particle states are the minimum needed. For phenomenological calculations, such as
 in the $sd$ or $pf$ shell, this is usually the same information as before.

 \begin{verbatim}
           6        6       0       6       6    ! nhsps(-2:2)
\end{verbatim}
These are dimensions of subspaces of the single-particle states. 
The number of proton single particle states used is {\tt nhsps(1)+nhsps(-1)}; 
the number of neutron single particle states used is {\tt nhsps(2)+nhsps(-2)};
{\tt nhsps(0)} contains no useful information.  This information is given above
but is repeated as a check

 Next come (again)
the single particle quantum numbers, here the single particle quantum states actually used.  These are needed to interpret the 
many-body states
\begin{verbatim}
! index   nr   l   2j   2m    w  par
     1    0    2    3   -1    0    1
     2    0    2    5   -1    0    1
     3    1    0    1   -1    0    1
     4    0    2    5   -3    0    1
     5    0    2    3   -3    0    1
     6    0    2    5   -5    0    1
     7    0    2    3    1    0    1
     8    0    2    5    1    0    1
...
\end{verbatim}
Finally the  many body states are written out, proton Slater determinants first, then neutron.
It is important to understand that the basis is organized by sectors. 
A sector is the set of proton or neutron Slater determinants defined by quantum numbers $J_z, \pi$, and $W$.
\begin{verbatim}
           9        66   ! # of proton sectors, proton Slater dets
\end{verbatim}
That is, in this example, there are 9 proton sectors, with a total of 66 proton Slater determinants distributed 
among them. 

Looping over the proton sectors,
\begin{verbatim}
     1      -8      1       0  ! index, 2Jz, par, and W for sector
     1       2      2   ! start_pSD, stop_pSD,  # proton SDs in sector
\end{verbatim}
One should have {\tt stop\_pSD - start\_pSD + 1} = number of proton Slater determinants in this sector.

Each proton sector has a \textit{conjugate} neutron sector; their quantum numbers $J_z, \pi, W$ combine 
correctly to the quantum numbers for the full system, i.e., $J_z(p)+J_z(n) = J_z(\mathrm{tot})$.
\begin{verbatim}     
           1           ! # of conjugate sectors for this sector 
           1           ! list of conjugate neutron sectors
\end{verbatim}
One only gets more than one conjugate sector when doing non-trivial truncations in $W$.  
\begin{verbatim}     
           0           0    !   obsolete information
\end{verbatim}
Finally, for each proton sector, the proton Slater determinants are listed in order, starting from {\tt start\_pSD} and 
ending with {\tt stop\_pSD}.  They are listed with: label; 
\begin{verbatim}          
!   label           pstart       occupied s.p. states...
           1              0           5           6
           2              2           4           6
\end{verbatim}
Here {\tt pstart} is a key array; neutron Slater determinants have a similar array {\tt nstart}.  
When combining a proton Slater determinant and a neutron Slater determinant, the sum
{\tt pstart+nstart} is the index in the combined basis.  

For each Slater determinant, the list of occupied single particle states (using the second indexing given above) describes 
the state.

After looping over the proton sectors and listing the occupied single particle states in each proton Slater determinant in the sector,
the process starts over again with neutron sectors.

\bigskip

The pseudocode for construction of the basis:

\noindent \textit{loop over proton sectors} 

\textit{loop over list of  neutron sectors conjugate to current proton sector}

\textit{   \hspace{0.5cm}  loop over proton Slater determinants {\tt ip} in proton sector }

\textit{     \hspace{1cm}      loop over neutron Slater determinants  {\tt jn} in neutron sector}

\textit{              \hspace{1.5cm}        basis index = }{\tt pstart(ip) + nstart(jn)}

\bigskip

The routines for writing to the binary {\tt .bas} file are {\tt write\_wfn\_header} in 
{\tt bwfnlib.f90} and {\tt basis\_out4postprocessing} in {\tt boutputlib.f90}; 
for the human-readable ASCII file  {\tt write\_wfn\_header\_ASCII} is used instead.

\section{Writing out the Hamiltonian and other operators}

\label{printham}

The many-body Hamiltonian can also be explicitly generated by choosing the option `{\tt (wh)}'. This will write the nonzero 
matrix elements to the file {\tt ham.dat}.The first line of the file is the basis dimension. The following lines are
$i,j, H_{ij}$, with $ i \geq j$, and only for $H_{ij}$ nonzero.   Choosing `{\tt wo}' will write out the matrix elements of a one-body transition operator in the same format. 
After writing the nonzero matrix elements to file, {\tt BIGSTICK} halts and does not solve the eigenvalue problem.

Alternately, if from the Lanczos menu you choose `{\tt ex}' for exact or full diagonalization, 
the entire Hamiltonian is created and explicitly stored and solve using the LAPACK routine {\tt DSYEV}.  This occurs in the 
routine {\tt exactdiag\_p} in the file {\tt blanczos\_main.f90}. If the basis dimension is less than 100, {\tt BIGSTICK} will automatically write out 
the Hamiltonian matrix elements to a file {\tt ham.dat}. This occurs around line 1246; the user can edit this part of the code to control 
the dimension cutoff for writing out (in general we do not encourage writing out for large dimensions) as well as the format.

\chapter{Lanczos algorithm}
\label{lanczos}

Today the  most common way to find all the eigenpairs of a Hermitian (here, real and symmetric) matrix 
is first reduce the matrix to tridiagonal form via a sequence of unitary transformations, the Householder algorithm\citep{parlett1980symmetric,numericalrecipesfortran}, and then solve the resulting 
tridiagonal matrix via QL decomposition with implicit shifts (or so we've been told). But for the very large dimensions of standard 
CI calculations, one neither can extract all eigenpairs nor would one want to. 

(We can understand why through the concept of intruder 
states, that is, a state outside the designated model space.  For example, in the $sd$-shell, one has only positive parity states, so any negative parity state is an intruder.  Yet, experimentally, at some 
point intruders--in our example, negative parity states, consisting of excitation from the $p$ shell into the $sd$ shell, or from the $sd$ shell into the $pf$ shell--will start appearing in the experimental spectrum, but are outside the calculation. 
Eventually intruder states dominate, simply because there are an infinite number of them, and any calculation in a finite model space is physically incomplete. 
There is no simple way, at least for the non-expert, to determine \textit{where} we expect intruders to dominate.)

Instead we turn to the Lanczos and related algorithms \citep{parlett1980symmetric,numericalrecipesfortran,lanczos}.  Lanczos is part of a family of so-called Arnoldi algorithms, which iteratively construct a new 
orthonormal basis, the Krylov subspace. In this new basis the Hamiltonian is tridiagonal, but unlike the Householder algorithm, one does not need to 
fully carry out the transformation.  The Lanczos algorithm is simple, beautiful, and powerful, though like all algorithms it is not without its own limitations.

\section{Standard Lanczos algorithm}

\label{vectorlanczos}

The Lanczos algorithm is exceedingly straightforward.  We will summarize it here, though we will not explicate it in detail. 
Starting from some initial vector $| v_1 \rangle$, called the \textit{pivot}, 
one iteratively generates a sequence of orthonormal vectors $\{ | v_i \rangle, i = 1, k  \}, \langle v_i | v_j \rangle = \delta_{ij}$:
\begin{equation}
\label{lanczos_matrix2}
\begin{array}{lllll}
\hat{H} |v_1 \rangle = & \alpha_1 | v_1 \rangle   &  + \beta_1 | v_2 \rangle       &                                    &                                    \\
\hat{H}| v_2 \rangle =  & \beta_1 |v_1 \rangle      &  + \alpha_2 | v_2 \rangle   &  +\beta_2 | v_3 \rangle   &  \\
\hat{H}| v_3 \rangle =  &                                       &    \beta_2 |v_2 \rangle      &  + \alpha_3 | v_3 \rangle   &  +\beta_3 | v_4 \rangle \\
                                    & \ldots                              &    & & \\
\hat{H}| v_i \rangle =   &                                         &  \beta_{i-1} | v_{i-1} \rangle & + \alpha_i | v_i \rangle  & +\beta_i | v_{i+1} \rangle \\
                                    & \ldots                              &    & & \\
\hat{H}| v_k \rangle =   &                                    &      &  \beta_{k-1} | v_{k-1} \rangle & +  \alpha_k | v_k \rangle    \\
\end{array}
\end{equation}
Each iteration generates a new Lanczos vector.  If we stop at the $k-1$th iteration, we have $k$ Lanczos vectors and a $k$-dimension Krylov 
subspace. Using orthonormality of the vectors, one can show that in this basis, the Hamiltonian is tridiagonal:
\begin{equation}
H_{i,i} = \langle v_i | \hat{H} | v_i \rangle = \alpha_i,
\end{equation}
\begin{equation}
H_{i,i+1} = H_{i+1,i} = \langle v_i | \hat{H} | v_{i+1} \rangle = \beta_i.
\end{equation}
and all other matrix elements are zero. 

The specific steps for creating the next Lanczos vectors are straightforward:

\smallskip

\noindent (1) $| w_i \rangle = \hat{H} |v_i \rangle$  \hspace{1cm} \textit{Initial matvec on vector} i;

\noindent (2) $\alpha_i = \langle  v_i | w_i \rangle $\hspace{1cm}  \textit{dot product to get} $\alpha_i$;

\noindent (3) $ | w_i \rangle \leftarrow | w_i \rangle - \alpha_i | v_i \rangle$ \hspace{1cm}  \textit{orthogonalize against initial vector} i;

\noindent (4) \textit{If} $ i > 1$ \hspace{.2cm} $ | w_i \rangle \leftarrow | w_i \rangle - \beta_{i-1} | v_{i-1} \rangle $
\hspace{.4cm} \textit{orthogonalize against prior vector} i-1;

\noindent (5) $\beta_i = \sqrt{ \langle w_i | w_i \rangle } $  \hspace{1cm}  \textit{find norm to get} $\beta_i$;

\noindent (6) $|v_{i+1}\rangle = \beta_i^{-1} | w_i \rangle $ \hspace{1cm} \textit{Normalize to get} i+1{th Lanczos vector}.

\smallskip 

If we had perfect arithmetic, this would be sufficient: the new Lanczos vector $ | v_{i+1} \rangle$ would be guaranteed to be orthogonal to 
all previous vectors. But we don't have perfect arithmetic, and due to round-off noise, small components of prior Lanczos vectors will creep in 
and eventuall grow exponentially.  

This requires us to enforce orthogonality against all prior Lanczos vectors:

\smallskip

\noindent (4)(alt.) \textit{For} j = 1 \textit{ to} i-1 : $ | w_i \rangle \leftarrow | w_i \rangle - | v_{j} \rangle \langle v_j | w_i  \rangle$

\smallskip

If one does not reorthogonalize, eventualy one gets `ghost eigenvalues', or 
repetitions of the same eigenvalues.  It is this need for reorthogonalization that keeps Lanczos from supplanting Householder as the go-to algorithm for 
full tridiagonalization of Hermitian matrices.

There has been much discussion and experimentation around partial reorthogonalization, but  no one clearly successful recipe.  {\tt BIGSTICK} 
fully reorthogonalizes against all prior vectors; in most cases (a few hundred iterations) reorthogonalization work does not overwhelm matvec work. 

You might notice that if one extended the \textit{for} loop in our alternate step (4), we would already get step (3).  Because of finite arithmetic, order matters. 
We find better results if we first compute $\alpha_i$ and then orthogonalize against all other vectors, rather than as a last step.  

It is possible for a user to experiment with these fine tweaks in {\tt BIGSTICK}.  The Lanczos iterations are found in subroutine  {\tt lanczos\_p} in 
file {\tt blanczoslib1.f90}. 

One can estimate the workload from reorthogonalization.  Again, let $N$ be the dimension of the vector space.  Each projection requires a dot product and a 
subtraction, or about $2N$ operations.  For $k$ Lanczos vectors, one has $k-1$ iterations. For the $j$th iteration one orthogonalizes against $j$ vectors or 
$2Nj$ operations; thus for $k-1$ iterations one has $2N$

\bigskip

Obviously with full reorthogonalization full Lanczos transformation to tridiagonal form becomes expensive; hence the dominance of the Householder algorithm for 
complete diagonalization.

\section{Thick-restart Lanczos}
\label{thickrestart}

Sometimes there is insufficient storage for the number of Lanczos vectors required for convergence.  An alternative is the \textit{thick-restart Lanczos}\citep{wu2000thick}.
In standard Lanczos there are essentially two dimensions, $N_\mathrm{keep}$, the number of converged states desired, and $N_\mathrm{iter}$, the 
number of iterations (typically $k$ above). But one must store $N_\mathrm{iter}+1$ Lanczos vector, which can be prohibitive.  For thick-restart Lanczos, 
there is an additional dimension $N_\mathrm{thick}$, with $N_\mathrm{keep} < N_\mathrm{thick} < N_\mathrm{iter}$, and is the dimension of the Krylov subspace
when restarting. In otherwords, one iterative creates a subspace of dimension $N_\mathrm{iter}+1$, but then truncates down to dimension $N_\mathrm{thick}$, and then
adds additional vectors back up to a subspace of dimension $N_\mathrm{iter}+1$, truncate back down, and repeat until convergence. 
The advantage is that  $N_\mathrm{iter}$, and the consequent number of vectors stored, is much 
smaller than would be needed for standard Lanczos. 

Thick-restart Lanczos follows this basic outline:

\begin{enumerate}

\item Start with some initial Lanczos pivot vector $| v_1 \rangle $ as usual.

\item Carry out $k$ Lanczos iterations so that you have $k+1$ Lanczos vectors $| v_i \rangle, i = 1,k+1$, and a 
truncated $k+1 \times k+1$ Hamiltonian matrix $\mathbf{T}^{k+1}$.

\item Diagonalize the $k \times k$ submatrix $\mathbf{T}^{k}$.

\item From the eigenpairs of step (3), choose the $N_\mathrm{thick}$ lowest states.  These will form the ``new'' Lanczos eigenvectors.  In addition, keep 
$|v_{k+1} \rangle$ and use this as the restarting vector for Lanczos. 

\item Now restart Lanczos, but instead of starting with $| v_1 \rangle$, start with $|v_{k+1} \rangle $ which is our new $|v_{N_\mathrm{thick} +1} \rangle$.  

\item Iterate until you have again $k+1$ Lanczos vectors and an truncated $k+1 \times k+1$ Hamiltonian matrix $\mathbf{T}^{k+1}$. This new matrix will no
longer be tridiagonal, but it will have a simple form, given below.

\end{enumerate}

Now let's describe this in more detail. 
Suppose we have carried out $k$ Lanczos iterations, so that we have a total of $k+1$ vectors $| v_i\rangle$, including the pivot. Then the transformed Hamiltonian, which 
is the Hamiltonian in the basis $\{ | v_i \rangle \}$, 
looks like 
\begin{equation}
\mathbf{T}^{k+1}=
\left ( 
\begin{array}{cccccc}
\alpha_1 & \beta_1 & 0           & 0              & \ldots   &  0 \\
\beta_1 & \alpha_2 & \beta_2 & 0              & \ldots   &  0 \\
0           & \beta_2  &  \alpha_3 & \beta_3   & \ldots   &  0 \\
              & \vdots   &                  &                &             &     \\
 0        &     0          &               \ldots  & \alpha_{k-1}  & \beta_{k-1} & 0 \\
 0        &     0          &               \ldots  &   \beta_{k-1} & \alpha_k  & \beta_k \\
 0        &     0          &               \ldots  &   0                 & \beta_{k} & [\alpha_{k+1} ] \\
\end{array}
\right )
\label{lanczos_matrix}
\end{equation}
(Actually, with $k$ iterations, although there are $k+1$ Lanczos vectors, the value of $\alpha_k$ hasn't yet been determined. It is not needed, however, at this point, and will 
be found later.)   Suppose, however, we only diagonalize $\mathbf{T}^k$, that is, stopping at the $k$th column and row, with $\mathbf{L}$ being the $k \times k$ unitary matrix 
of eigenvectors, that is,
\begin{equation}
\sum_{j=1}^k
\left ( \mathbf{T}^k \right)_{ij} L_{j\mu} = L_{i\mu}  \tilde{E}_\mu,
\end{equation}
for $\mu=1,k$. Here $\tilde{E}_\mu$ are the approximate eigenenergies. If we apply the unitary transform $\mathbf{L}$ to the first $N_\mathrm{thick}$ vectors, that is, 
introducing 
\begin{equation}
| v^\prime_\mu \rangle = \sum_{i=1}^k |v_i \rangle L_{i,\mu}, 
\label{transformthick}
\end{equation}
and $|v^\prime_{k+1} \rangle = | v_{k+1} \rangle $
the transformed matrix, which is the Hamiltonian the basis $\{ | v^\prime_i \rangle \}$, 
now becomes
\begin{equation}
\left ( 
\begin{array}{cccccc}
\tilde{E}_1 & 0& 0           & 0              & \ldots   &  \beta_k L_{k1} \\
0 & \tilde{E}_2  & 0 & 0              & \ldots   &  \beta_k L_{k2} \\
0           & 0 &  \tilde{E}_3 & 0 & \ldots   &  \beta_k L_{k3} \\
              & \vdots   &                  &                &             &     \\
 0        &     0          &               \ldots  &  \tilde{E}_{k-1} & 0  & \beta_k L_{k,k-1}\\
 0        &     0          &               \ldots  &  0 & \tilde{E}_k  & \beta_k L_{kk}\\
 \beta_k L_{k1}      &     \beta_k L_{k2}         &            \ldots  &   \beta_k L_{k,k-1}               & \beta_k L_{kk}& [ \alpha_{k+1} ]  \\
\end{array}
\right )
\end{equation}

The key to thick-restart Lanczos is to judiciously truncate this.  If you want to get the lowest $N_\mathrm{keep}$ states, 
truncate to  $N_\mathrm{thick}$ (with   $N_\mathrm{keep} < N_\mathrm{thick} < k$) vectors, that is, to take from  Eq.~(\ref{transformthick}) only the 
first $N_\mathrm{thick}$ new Lanczos vectors,
$$
|v^\prime_1 \rangle, \, |v^\prime_2 \rangle, \, |v^\prime_3 \rangle, \, \ldots |v^\prime_{N_\mathrm{thick}} \rangle
$$
\textit{plus} the last Lanczos vector, $| v_{k+1} \rangle$. Then the truncated Hamiltonian looks like
\begin{equation}
\left ( 
\begin{array}{cccccc}
\tilde{E}_1 & 0& 0           & 0              & \ldots   &  \beta_k L_{k1} \\
0 & \tilde{E}_2  & 0 & 0              & \ldots   &  \beta_k L_{k2} \\
0           & 0 &  \tilde{E}_3 & 0 & \ldots   &  \beta_k L_{k3} \\
              & \vdots   &                  &                &             &     \\
  0        &     0          &               \ldots  &  \tilde{E}_{N_\mathrm{thick}-1} & 0 & \beta_k L_{k {N_\mathrm{thick}}-1}\\
 0        &     0          &               \ldots  &  0 & \tilde{E}_{N_\mathrm{thick}} & \beta_k L_{k {N_\mathrm{thick}}}\\
 \beta_k L_{k1}      &     \beta_k L_{k2}         &            \ldots  &  \beta_k L_{k ,{N_\mathrm{thick}}-1}            & \beta_k L_{k {N_\mathrm{thick}}}& [ \alpha_{k+1} ] \\
\end{array}
\right )
\label{thicktransformed1}
\end{equation}
Now declare $| v_{k+1} \rangle$ to be the new $| v^\prime_{N_\mathrm{thick}+1} \rangle$ and start the Lanczos iterations on it:
\begin{eqnarray}
\mathrm{H} | v^\prime_{N_\mathrm{thick}+1} \rangle =  \\
\nonumber  \beta_k L_{k1}   | v^\prime_1\rangle + \beta_k L_{k2}   | v^\prime_2\rangle + \ldots 
+ \alpha_{N_\mathrm{thick} + 1} | v^\prime_{N_\mathrm{thick}+1} \rangle + \beta_{N_\mathrm{thick}+1}  | v^\prime_{N_\mathrm{thick}+2}\rangle
\end{eqnarray}
(This, incidentally, is when we find $\alpha_{k+1}$, only now rebranded as $\alpha_{N_\mathrm{thick} +1} $.) 
This first step is not a tridiagonal relation; furthermore , although our new  $| v^\prime_{N_\mathrm{thick}+1} \rangle $ is the same as our 
old $|v_{k+1} \rangle$, and our new  $\alpha_{N_\mathrm{thick} + 1}$ is the same as the old $\alpha_{k+1}$, the new vector 
$ | v^\prime_{N_\mathrm{thick}+2}\rangle$ is \textit{not} the same as $|v_{k+2} \rangle$ would have been had we continued the previous iteration,
although the former contains the latter as a component, because we orthogonalize $|v^\prime_{N_\mathrm{thick}+1} \rangle$ against a different set of vectors.

Now one continues iterations $N_\mathrm{thick}+2, N_\mathrm{thick}+3, \ldots, k+1$. Then one diagonalizes the approximate $\mathrm{T}^k$ again,although it is 
no longer a pure tridiagonal, and in fact looks like:
$$
\left ( 
\begin{array}{cccccccc}
\tilde{E}_1 & 0& 0           & 0              & \ldots   &  \beta_k L_{k1}  & 0 & \ldots \\
0 & \tilde{E}_2  & 0 & 0              & \ldots   &  \beta_k L_{k2}  & 0 & \ldots \\
0           & 0            &  \tilde{E}_3 & 0 & \ldots   &  \beta_k L_{k3} & 0 & \ldots \\
              & \vdots   &                  &    \ddots            &             &     & &  \\
  0        &     0          &               \ldots  &  \tilde{E}_{N_\mathrm{thick}-1} & 0 & \beta_k L_{k {N_\mathrm{thick}}-1} & 0 & \ldots \\
 0        &     0          &               \ldots  &  0 & \tilde{E}_{N_\mathrm{thick}} & \beta_k L_{k {N_\mathrm{thick}}}   & 0  \\
 \beta_k L_{k1}      &     \beta_k L_{k2}         &            \ldots  &  \beta_k L_{k ,{N_\mathrm{thick}}-1}            & \beta_k L_{k {N_\mathrm{thick}}}& \alpha_{N_\mathrm{thick}+1}  & \beta_{N_\mathrm{thick} + 1 } \\
 0                          &               0                     &   \ldots    &     0                                        & 0 & \beta_{N_\mathrm{thick} + 1} & \alpha_{N_\mathrm{thick}+2} \\
            \vdots        &                                     &               &                                                &     &                                              &  & \ddots
\end{array}
\right ),
$$
 and restarts as above, repeating under convergence. 

This thick-restart algorithm requires more matvec multiplications than standard Lanczos, because information is thrown away at each restart, but 
the storage and reorthogonalization of Lanczos vectors can be greatly reduced.   There is no recommended value of $N_\mathrm{thick}$ or $k$, but one should take $k$ 
as large as practical, and ``typical'' values of $N_\mathrm{thick} \approx 3 N_\mathrm{keep}$ or so.

Although the usual application to thick restart is to find low-lying states, it is conceivable to choose a slice of excited energy and to converge excited states. This will be investigated.

\subsection{Targeted thick-restart Lanczos: interior eigenvalues}


The Lanczos algorithm naturally leads to extremal eigenvalues, but 
sometimes one wants to obtain interior eigenvalues, that is, highly excited eigenstates. The targeted thick-restarted option, `{\tt (tx)},' is an approximate method to obtain such interior eigenvalues. The way it does so is by doing thick-started on 
$(\mathbf{H} - \bar{E} )^2$, where $\bar{E}$ is the energy target; then eigenstates near  to $\bar{E}$ are low in the synthetic spectrum.

\section{Block Lanczos}

\label{blocklanczos}

Block Lanczos is a variant where instead of carrying out $\mathbf{H} \vec{v}_i = \vec{v}_f$ on a single vector, one applies the Hamiltonian matrix to a \textit{block} of vectors. 
We have two motivations for introducing block Lanczos. 

The first motivation is the inherent inefficiencies of our factorization/on-the-fly algorithm. We do not store all the nonzero many-body matrix elements, but largely reconstruct them on the 
fly. This save us tremendously on storage of matrix elements, it takes time to reconstruct the matrix elements. (Numerical experiments suggest roughly a factor of two difference in 
matrix-multiplication time between storage and on-the-fly, which is surprisingly good.)  Furthermore, out of necessity the algorithms for reconstruction proton-proton, neutron-neutron, and 
proton-neutron matrix elements are different and can take different amounts of time; furthermore those times can have large fluctuations in them, which leads to difficulty in load-balancing. 

Furthermore, since the $M$-scheme matrix is very sparse, the indices of the vector elements accessed can be very far apart, leading to a loss of data locality. The complexity of the 
on-the-fly reconstruction algorithm only makes this worse. 

To be explict, consider the standard matrix-vector multiplication (matvec mult), $\mathbf{H} \vec{v} = \vec{w}$
\begin{equation}
w_i = \sum_j H_{ij} v_j.
\end{equation}
In our standard algorithm, the value of the matrix element $H_{ij}$ as well as the indices $i,j$ are reconstructed on the fly.  Only a small number of the $H_{ij}$ are nonzero, hence a very sparse matrix, and approximately, the indices are called randomly. 

(In fact, this work can be distributed over many 
MPI processes, ordered by sectors defined by quantum numbers,  and the final index $i$ is ordered to avoid race conditions in OpenMP parallelization, but these details are unimportant to the discussion here.) 

In block Lanczos, however, one applies the same equation to multiple vectors:
\begin{equation}
w_{i,a} = \sum_j H_{ij} v_{j,a},
\end{equation}
where $a$ labels the different vectors in the blocks. Because reconstruction $H_{ij}$ is relatively expensive, as well as variable in the cost in time, block Lanczos amortizes
the cost by applying it to multiple vectors.  Furthermore, for the matrix-matrix multiplication we store the vector blocks row-wise rather than column wise.  In pseudocode this 
looks like:
\begin{verbatim}
loop over matrix elements;
    fetch Hij,i,j
    loop over a
        w(a,i) = w(a,i) + Hij * v(a,j)
    end loop
end loop
\end{verbatim}
Because the elements of the block vector {\tt w(a,i)} and {\tt v(a,j)} are contiguous, i.e., \textit{local}  in memory with respect to {\tt a}, this dramatically reduces cache calls. 
The resulting factor of 2 speedup makes this part of the algorithm equal in speed to codes which store the matrix elements explicitly in memory.

The downside of block Lanczos is that it can require significantly more total iterations than standard Lanczos. 

Block Lanczos is well described by 
\citep{shimizu2019thick}, who focus on  thick-restart block Lanczos; for this option in {\tt BIGSTICK}, see Section 
\ref{TRBL}.

Running in block Lanczos mode is described above in section \ref{block}.  For important constraints when running in parallel using MPI, read section \ref{blockMPI}.

Here is how we carry out block Lanczos. Let $N_\mathrm{dim}$ be the $M$-scheme dimension, and let $N_\mathrm{block}$ be
the dimensions of the blocks (which is {\tt dimblock}  in the code), that is, the number of vectors in each block. Then  let $\mathbf{V}_n$ be the $n$th block of vectors, it is a 
$N_\mathrm{dim} \times N_\mathrm{block}$ rectangular matrix.   The column vectors of $\mathbf{V}_n$ should be orthonormal, not only to each other but also to all column 
vectors in all other block $\mathbf{V}_m$. 

The block $\mathbf{V}_n$  are generated iteratively by matrix multiplication, as in the standard (or vector) Lanczos. 
When carrying out the matrix multiplication, however, the transpose is stored, so as improve locality, that is, 
although we write
\begin{equation}
\mathbf{H \, V}_n = \mathbf{W}_n,
\label{blockiter}
\end{equation} 
where $\mathbf{W}_n$ is a temporary matrix, we actually do 
$$
\mathbf{ V_n}^T \mathbf{H }= \mathbf{W}^T_n,
$$
to reduce cache calls. In the rest of this discussion, however, we suppress this. 

The basic block Lanczos iteration is 
\begin{equation}
\mathbf{H \, V}_n = \mathbf{V}_{n-1} \mathbf{B}_{n-1} +  \mathbf{V}_{n} \mathbf{A}_{n}+ \mathbf{V}_{n+1} \mathbf{B}_{n},
\end{equation}
where $\mathbf{A}_n, \mathbf{B}_n$ are $N_\mathrm{block} \times N_\mathrm{block}$  square matrices; the $\mathbf{A}$ matrices are symmetric, but 
not the $\mathbf{B}$ matrices.
To remove the first term, we orthogonalize $\mathbf{W}_n$ against all previous blocks. Then we compute
\begin{equation}
\mathbf{V}_n^T \mathbf{W}_n = \mathbf{A}_n,
\end{equation}
and then subtract 
\begin{equation}
 \mathbf{W}_n - \mathbf{V}_n \mathbf{A}_n = \mathbf{W}^\prime_n = \mathbf{V}_{n+1} \mathbf{B}_n.
 \end{equation}
 To extract $ \mathbf{V}_{n+1}$ and $\mathbf{B}_n$, compute the $N_\mathrm{block} \times N_\mathrm{block}$  symmetric overlap matrix
 \begin{equation}
 \mathbf{O} = \mathbf{W}^{\prime T}_n \mathbf{W}_n^\prime,
 \end{equation}
 but this is equal to
 \begin{equation}
  = \mathbf{B}_n^T  \mathbf{V}_{n+1}^T \mathbf{V}_{n+1} \mathbf{B}_n =  \mathbf{B}_n^T   \mathbf{B}_n 
 \end{equation}
 because of the orthonormality of the column vectors of the $\mathbf{V}$ blocks. 
 
 Now we have to factor the overlap matrix. There are multiple options, but we choose to do a spectral decomposition (i.e., diagonalization); in all conceivable cases 
 $N_\mathrm{block}$ will be relatively small and diagaonalization quick, and by examining the eigenvalues one can easily find and remove singular or near-singular values. 
 (Note: as of this writing, version 7.9.8, that is not yet implemented in block Lanczos.) 
 
 Let $\mathbf{U}$ be the unitary matrix formed by the eigenvector of $\mathbf{O}$, which must have positive eigenvalues represented by the diagonal matrix $\mathbf{\Lambda}$. Then
 $\mathbf{O} = \mathbf{U \Lambda U}^T$.  From this we can conclude that 
 \begin{equation}
 \mathbf{B}_n = \sqrt{\mathbf{\Lambda}} \mathbf{U}^T
 \end{equation}
 (the square root is easy to carry out, because $\mathbf{\Lambda}$ is diagonal with positive-definite elements). Finally 
 \begin{equation}
 \mathbf{V}_{n+1} = \mathbf{W}^\prime_n \mathbf{B}_n^{-1} =  \mathbf{W}^\prime_n \mathbf{U}  {\mathbf{\Lambda}}^{-1/2}, 
 \end{equation}
 where, again, the inverse is easy to carry out. If one of the elements of $\mathbf{\Lambda}$ is near zero, then it is nearly singular, and one needs to `restart' with a new vector. 
 This will be implemented at a future date.
 
 We note that the $\mathbf{B}_n$ form the block-sub-diagonal, because $\mathbf{B}_n = \mathbf{V}^T_{n+1} \mathbf{H} \mathbf{V}_n$, that is, in the $n+1$th block-row and the $n$th column-block.
 
 Now the truncated Hamiltonian in the block-Lanczos representation is 
 \begin{equation}
 \mathbf{T} = \left ( 
 \begin{array}{cccccc}
 \mathbf{A}_1 & \mathbf{B}^T_1  & 0 & 0  & 0 & \ldots \\
\mathbf{B}_1 &  \mathbf{A}_2 & \mathbf{B}^T_2  & 0 & 0 &   \\
0 & \mathbf{B}_2 &  \mathbf{A}_3 & \mathbf{B}^T_3  & 0 & \\
0 & 0 & \mathbf{B}_3 &  \mathbf{A}_4 & \mathbf{B}^T_4  & \\
\vdots &               &                        &                       &     & \ddots
\end{array} \right )
\end{equation}
which is solved in the usual fashion. 

\subsection{Bootstrapped block Lanczos}

\label{bootstrap}

Although the matrix-matrix multiplication for block Lanczos is much more efficient than matrix-vector multiplication for vector Lanczos, experience shows that one often needs more Lanczos iterations. One can accelerate block Lanczos by starting from vectors that approximate the final vectors~\cite{zbikowski2023bootstrapped}. In order to do this, one must use option 
`{\tt (np)}' in the initial menu. 


\subsection{Block strength function}

\label{blockstrength}

An obvious useful application is a block strength function, that is, carrying out the strength function option but for a block of vectors.
The menu option for this is `{\tt (sb)}'.

We read in a block of pivot vectors, $| w_1 \rangle, | w_2 \rangle, | w_3 \rangle, | w_4 \rangle \ldots$ and we want the strength against 
(approximate) eigenvectors $| E_r \rangle$, that is, we want 
$$
\left | \langle w_i | E_r \rangle \right |^2.
$$
These initial vectors might  not be orthonormal, for example after applying a one-body operator (option `{\tt (o)}').  
After being read in, they must be orthonormalized. Let $\mathbf{W_0}$ be the initial block of vectors. As above, form the overlap matrix
$\mathbf{O} = \mathbf{W}_0^T \mathbf{W}_0 = \mathbf{U \Lambda U}^T$. Then 
\begin{equation} 
\mathbf{W}_0 \mathbf{U \Lambda}^{-1/2} = \mathbf{V}_1
\end{equation}
is now an orthonormal block, and the initial pivot block. Let's think what this means. The orthonormalized vectors, which form the pivot block for 
the Krylov space, are $| v_1 \rangle, | v_2 \rangle, | v_3 \rangle, | v_4 \rangle \ldots$ related by
\begin{equation}
| {v}_j \rangle = \sum_i | {w}_i \rangle U_{ij} \lambda_j^{-1/2},
\label{orthonormal_pivotblock} 
\end{equation}
where $\lambda_j$ is the $j$th eigenvalue of the overlap matrix. 


After a set number of iterations, we solve as before
\begin{equation}
\mathbf{T} = \mathbf{L \tilde{E} L}^T,
\end{equation}
where $\tilde\mathbf{E}$ is a diagonal containing the approximate eigenenergies of $\mathbf{T}$, and $\mathbf{L}$ are the eigenvectors represented in the Krylov space. 
Specifically, $L_{j, r}$ is the $j$th Krylov coefficient of the $r$th eigenvector, that is, 
$$
| E_r \rangle = \sum_j  | {v}_j \rangle L_{j,r}.
$$
By inverting (\ref{orthonormal_pivotblock}), that is, 
$$
| {w}_i \rangle = \sum_j  U_{ij} \lambda_j^{1/2}| {v}_j\rangle ,
$$
then
\begin{equation}
\langle w_i | E_r \rangle = \sum_j U_{ij} \lambda_j^{1/2} \langle v_i | E_r \rangle =  \sum_j U_{ij} \lambda_j^{1/2} L_{j,r}
\end{equation}
This we can do from the original, let  $\mathbf{B}_0 = \mathbf{\Lambda}^{1/2} \mathbf{U}^T$, and then 
$$
\langle w_i | E_r \rangle = \sum_j (B_0)_{ji} L_{j,r}
$$
that is, we get the overlaps from $\mathbf{B}_0^T \mathbf{L}$.

Note: One can also choose `{\tt (sbs)}', in which case the $J$, $T$ of the final states are not computed and the wave functions are not written to file. 
This can be a smart choice when carrying out a large large with many iterations. 

\subsection{Thick-restart block Lanczos}

\label{TRBL}

The thick-restart block Lanczos method (TRBL) combines the strengths and weaknesses of both the thick-restart and block Lanczos methods. TRBL has been shown to be an effective eigensolver for large-scale shell-model calculations where one desires large numbers of eigenstates \citep{shimizu2013nuclear}. The power of the thick-restart block Lanczos method is three-fold. First, time in matrix-matrix (or matrix-block vector) multiplication is 
reduced due to  improved data locality.  Matrix-block vector multiplication also amortizes the cost of the on-the-fly matrix element reconstruction algorithm.  Lastly, by restarting the block Lanczos process, reorthogonalization time is reduced by restricting the number of Lanczos vectors stored in memory.

In substance, TRBL follows the same iterative process of the conventional block Lanczos method, constructing a block tri-diagonal approximation of the  Hamiltonian and computing blocks of Lanczos vectors up to some chosen maximum number of block iterations $n_s$, resulting in $k$ total Lanczos vectors saved before block Lanczos is restarted. $N_{thick}$ total eigenvectors or $n_{thick}$ block Lanczos iterations worth of approximate eigenvectors are constructed and saved for restart. By construction, one takes a linear combination of Lanczos vectors weighted by the components of the eigenvectors of the truncated space. Lanczos vectors computed at the final block iteration before $n_s$ are used as the initial pivot after restarting,  analogous the single-vector thick-restart process. Eigenpairs of the reduced Hamiltonian are used to construct a restarted block Lanczos matrix, Eq.~(\ref{blockthicktransformed1}). 

No longer purely block tri-diagonal, the matrix $\mathbf{T}$ after restarting is 
\begin{equation}
 \mathbf{T} = \left ( 
 \begin{array}{cccccc}
 \mathbf{E}_{n_{thick}} & \mathbf{r}^T  & 0 & 0  & 0 & \ldots \\
\mathbf{r} &  \mathbf{A}_1 & \mathbf{B}^T_1  & 0 & 0 &   \\
0 & \mathbf{B}_1 &  \mathbf{A}_2 & \mathbf{B}^T_2  & 0 & \\
0 & 0 & \mathbf{B}_2 &  \mathbf{A}_3 & \mathbf{B}^T_3  & \\
\vdots &               &                        &                       &     & \ddots
\end{array} \right )
\label{blockthicktransformed1}
\end{equation}
where $\mathbf{E}_{n_{thick}}$ is a diagonal matrix containing the first ordered $N_{thick}$ eigenvalues of the reduced Hamiltonian. The sub-matrix $\mathbf{r}$ is  
\begin{equation}
\mathbf{r} := \mathbf{B}_n \mathbf{U}_{k-N_{block}+1:k,1:n_{thick}}.
\end{equation}
The matrix $\mathbf{U}$ contains the approximate $k$ eigenvectors of the Hamiltonian computed in the reduced space prior to restarting. TRBL works well in situations when one desires many eigenpairs, and one can achieve reasonable convergence restarting block Lanczos with approximately $N_\mathrm{block}$ eigenvectors or a couple multiples there of. When running the algorithm with large block dimensions, you are generally storing a larger proportion of eigenpairs you are interested in relative to the $N_\mathrm{thick}$ vectors saved in memory for a restart. One can further accelerate the convergence of the thick-restart block Lanczos method by bootstrapping the pivot, that is, loading in approximate eigenvectors projected in from a different model space as the starting pivot block. 

\section{Can I restart  standard Lanczos?} 

The standard Lanzos algorithm is an iterative algorithm. In principle, if you found the desired eigenpairs had not converged under the chosen number of
iterations, you could pick up and restart.  To do this you would need the Lanczos vectors created so far and the Lanczos coefficients. 

Although in prior version {\tt BIGSTICK} wrote the Lanczos vectors to disk,the current version stores all Lanczos vectors in RAM. In 
MPI parallelization the Lanczos vectors are stored across multiple processes. Therefore right now the restart option has been turned off. 
It is possible in future versions we may restore it, although it is not a high priority.

\chapter{Parallel computing and timing}

\label{parallel}

{\tt BIGSTICK} can run many non-trivial problems on modest desktop or even laptop computers. 
Because problems grow exponentially, however, single-processor calculations quickly reach limits. To overcome
these limits we invoke parallel processing.

Although many parts of the set up portion of the code have been parallelized, by far the 
most time-consuming part of the code is the matrix-vector multiplication, followed by reorthogonalization,
and it is these two portions it is most important to parallelize.  

For very large calculations, one needs to distribute both matvec operations (work load balance) and data 
(memory load balance).  Operations are parallelized using both MPI (distributed memory) and OpenMP (shared memory) while data can only
be distributed with MPI.

When {\tt BIGSTICK} starts, it tells you how many MPI processes and how many OpenMP threads per process it is using:
\begin{verbatim}
 Number of MPI processors =         512
 NUM_THREADS =            8
\end{verbatim}
This information is also written to the {\tt .log} file.  {\tt BIGSTICK} does not have any special requirements for setting up 
parallel runs, although to run in parallel one must use an executable compiled with parallel options, i.e. 
{\tt bigstick-mpi.x} compiled with {\tt make mpi}, {\tt bigstick-opemp.x} compiled with {\tt make openmp}, or the hybrid 
{\tt bigstick-omp-mpi.x} compiled with {\tt make openmp-mpi}.  Any user who wishes to use the parallel capability should already have some idea about 
submitting parallel jobs. For example, to set up the number of OpenMP threads on a desktop machine you typically
\begin{verbatim}
PROMPT>export OMP_NUM_THREADS=8
\end{verbatim}
and to submit an MPI job you may do
\begin{verbatim}
PROMPT>mpirun -n 512 bigstick-mpi.x
\end{verbatim}
Of course, the details will depend upon the local environment. Unfortunately, in our experience supercomputers do not have a uniform 
job submission protocol. 

\section{MPI}

\label{mpi}

 To carry out a Lanczos iteration, which includes a matvec followed by reorthogonalization, one needs the following data:

\noindent $\bullet$ an initial vector;

\noindent $\bullet$ an final vector;

\noindent $\bullet$ jump information used for on-the-fly construction of the many-body matrix elements; and

\noindent $\bullet$ the uncoupled two- (or, optionally, three-) body matrix used in construction of 
the many-body matrix elements;

\noindent $\bullet$ previously computed Lanczos vectors (used for calculation of the Lanczos coefficients, 
for reorthogonalization and, ultimately,
construction of the final eigenvectors which  represent  wavefunctions).

In large calculations some or all of these may need to be distributed via MPI.

To compute the distribution efficiently, 
{\tt BIGSTICK} goes throught the setup in two stages. First, it calculates the number of operations in each matvec, that is, 
in 
\begin{equation}
v^\mathrm{final}_i = \sum_j H_{ij} v^\mathrm{initial}_j,
\end{equation}
each update
\begin{equation}
v^\mathrm{final}_i \leftarrow v^\mathrm{final}_i + H_{ij} v^\mathrm{initial}_j.
\end{equation}
Because of 
factorization, {\tt BIGSTICK} does not have to actually generate every operation.
{\tt BIGSTICK} then generates the distribution, and each MPI process creates locally the data it needs.

{\tt BIGSTICK} attemps to distribute the operations across MPI processes as evenly as possible.  The operations are constructed from jumps, but the ratio of 
operations to jumps is not fixed.  We find it helpful to think of matvec operations as represented by the area of a rectangle, and the sides of the rectangle representing 
the jumps.  If the rectangle is nearly square, the reconstruction is efficient, but if one has a long, thin rectangle in either dimension, one requires considerably more 
storage of jumps relative to the number of operations.  Occasionally, an equitable distribution of operations will, on some small subset of MPI processes, so many 
jumps they cannot be stored. In that case, {\tt BIGSTICK} will distribute those jumps over multiple MPI processes; this of course leads to a load imbalance, but is 
necessary so as not to exhaust memory.

\subsection{Fragments}

\label{fragments}

 If the basis dimension is so large 
both initial and final vectors cannot be contained in core memory, they must be broken into \textit{fragments}.  When running in MPI, or in 
modeling mode, 
{\tt BIGSTICK}  will ask for this automatically:
\begin{verbatim}
  Enter desired limit on fragment size for breaking Lanczos vectors
  (Largest un-splittable block =              17 million basis states )
 ( Default =             500 million basis states )
 (  Enter fragment size in millions of states (NEW); enter 0 to use default )
\end{verbatim}
Note that the units for the fragment size is \textbf{millions} of basis states.  If the total basis dimension is less than one million, however, then
fragment size is given and read in as the number of basis state.

Somewhat counterintuitively, {\tt BIGSTICK} achieves best efficiency when the fragments are as large as possible.
The reason for this is that the factorization principle behind {\tt BIGSTICK} works most efficiently when 
combining large conjugate data.  

Fragments are generally combinations of contiguous sectors (a portion of the vector which is labeled by the 
proton quantum numbers), although, because the lengths of sectors can vary significantly, in some cases 
a fragment can be comprised of a single sector. In the most extreme cases {\tt BIGSTICK} will seek to divide a sector 
into two new `sectors,' although there are limitations to how finely this can be done.  Otherwise {\tt BIGSTICK} 
attempts to make the fragments as similar in size as practical. 
 
If you run {\tt BIGSTICK} in MPI mode, it will ask for the fragment size. The fragment size is approximately the length of the initial and final vectors 
stored on a given MPI process 
(because of the way the code chunks data, {\tt BIGSTICK} actually allows for a small overrun). Choosing a value of 0 will select the default value, currently 
500 million, which is actually on the small size. Because Lanczos vectors are stored in single precision, this requires roughly 1.6 Gb of RAM for the initial and final vector fragments.
On many machines you can choose this to be larger.

Matvec operations are now defined from an initial fragment (of a Lanczos vector) to a final fragment (of a Lanczos vector).   This work will generally be spread across 
mulitple MPI processes; hence one needs {\tt nproc} (the number of MPI processes) $\ge$ {\tt nfragments}$^2$ (the number of fragments). In fact, 
{\tt BIGSTICK} will complain if {\tt nproc}  $< $ $2 \times $ {\tt nfragments}$^2$.

\bigskip

NB: In versions prior to 7.9.6, the fragment size was given in number of basis states; starting with 7.9.6, it is given in millions of basis states, as described above.

\subsection{Block Lanczos}

\label{blockMPI}

Running block Lanczos on large cases on parallel machines using MPI brings additional constraints.  For the sake of efficiency, the blocks are actually stored as long vectors. If the 
vectors are not broken up, then the size of each vector is {\tt dimblock} $\times$ {\tt dimbasis}.  If the basis is broken into fragments, the size is 
{\tt dimblock} $\times$ {\tt dimfragment}.  Where a problem can occur is if the dimensions are very large. These blocks-as-vectors are passed via MPI, but standard MPI has a limit 
of the size of vectors that can be passed, approximately 2 billion.  This limit can be breached if the basis is very large.  You should choose a fragment size so that 
{\tt dimblock} $\times$ {\tt dimfragment} is less than 2 billion.  Smaller fragments can sometimes require more MPI processes or ranks, as one should at a minimum have 
$2 \times (${\tt nfragments}$)^2$ processes. 

The code will attempt to warn you about problems, but you will have to read the output as well as the log file to find the warning.

(For people who like to fiddle with code: the relevant subroutines are those such as {\tt br\_pull\_block1\_from\_reg} and  {\tt br\_push\_block2\_to\_reg} 
in file {\tt bblock\_algebra.f90}; there could be others.)

\subsection{Opbundles and optypes}

The operations are organized by a derived {\tt type} (Fortran's designation for a bundle of data, very much like a 
{\tt struct} in C) called \textit{opbundles}, or bundles of operations. Opbundles are the `natural' way to divide up 
work in {\tt BIGSTICK}.  Opbundles orchestrate the application of matvec operations, and {\tt BIGSTICK} provide information 
about opbundles. Most users will not need this information. 

Each opbundle has an associated `optype,' which classifies the physical origin of the matrix elements being reconstructed. For example, the 
`PP' optype is for interactions betwen two protons, with neutrons as spectators. There are also NN and PN optypes, and for three-body forces, 
PPP, NNN, PNN, and PPN.  Finally there has been an optype SPE for single-particle energies and related single-particle potentials.  However this 
has been absorbed into PP, PN, and NN optypes. We do this by multiplying any one-body term by $(\hat{N}-1)/(A-1)$, with $\hat{N}$ the number 
operator and $A$ the (valence) mass number.  In the same way, if one runs with three-body forces, any two-body forces are subsumed into 
three-body by multiplying the two-body operators by $(\hat{N}-2)(A-2)$. 

Optypes signal different operations and invoke different methods of reconstructing the matrix elements. PP optypes use proton `two-body jumps' and loop
over spectator neutron Slater determinants, NN optypes use neutron two-body jumps and loop over spectator proton Slater determinants, and 
PN optypes use both proton one-body jumps and neutron one-body jumps. Not only do these invoke different subroutines, the time per operation 
is different for different optype, because the loops are different, and may be different on different machines. This  information in turn is used to 
calculation the distribution of work.  Information on the timing of these operations 
is found in the file {\tt timinginfo.bigstick}.  In many cases, by carrying out a short run to establish the time per operation, written to {\tt timinginfo.bigstick}, 
and then running the desired run using this information, can lead to significantly greater efficiency. 

Unfortunately, the time per operation is not as fixed for an optype as we originally hope, and detailed investigations show a great deal of 
fluctuations. We are still investigating this issue and attempt to arrive at better weighting and distribution algorithms.

\subsection{Jump storage and `greedy' opbundles}

{\tt BIGSTICK} works by factorizing both the basis and the interaction into separate proton and neutron components. For the interaction, the 
action of operators are stored as `jumps.'   {\tt BIGSTICK} will boast about its efficiency:
\begin{verbatim}
  total # operations           40912499332 , ~ ops/jump =    163.303177    
  Effective storage per operation =   0.14767923276870706       bytes 
\end{verbatim}
for $^{12}$C at $N_\mathrm{max}=6$, or
\begin{verbatim}
  total # operations         5140214153296 , ~ ops/jump =    52443.1055    
  Effective storage per operation =    4.5794092421044884E-004  bytes 
\end{verbatim}
for $^{60}$Zn in the $pf$ shell. 
Remember that an operation is approximately a matrix element. (Each operation actually represents the action of an operator, and diagonal matrix elements 
can be the sum of several operations.) You can see that this is very efficient. Model spaces without many-body truncations on $W$, as in in the second case, 
are significantly more efficient. 

Despite this efficiency, jump storage can become a problem. This is especially true for large no-core shell-model like calculations, and in particular when there is 
a large difference between $N$ and $Z$.  This can be understood through an apt analogy: imagine a rectangle, with one side representing proton information 
and the other side neutron information, and the interior representing the combined information. {\tt BIGSTICK} works by storing the perimeter and not the area. 
However, this is most efficient when the rectangle is square, that is, has sides of equal length.  Long, thin rectangles, conversely, are much less efficient. 

When running in MPI mode, {\tt BIGSTICK} divides up the work by time.  The jumps themselves are controlled by opbundles, and the opbundles are split to 
divide up the work. 
In very large calculations with large differences between $N$ and $Z$, however, this can run into a problem, because in order to divide up the word more memory 
may be required on a particular MPI rank than is available. (This is still less memory than would be required by simple storing the non-zero matrix elements.) 
This leads to `greedy' opbundles, which must be handled separately.  When modeling or running in MPI, {\tt BIGSTICK} will provide information about these greedy opbundles in the logfile.

Most users will not need to worry about this. if you do, the main options are: 
\begin{itemize}

\item Increase the number of MPI ranks;

\item Increase the memory available to each MPI rank, for example, by assigning more OpenMP threads. This is advanced parallel work and if you do not know how to do this, 
you should discuss it with a consultant for your HPC machine;

\item You can also change the variable {\tt maxjumpmemory\_default} in module {\tt flagger} in the file {\tt bmodules\_flags.f90}.  It is typically set at either 32 or 64 (Gb). 
This will depend upon the amount of memory available per MPI rank. Don't forget you need to set aside memory for the Lanczos vectors as well as storing the uncoupled 
matrix elements, all of which are also provided when {\tt BIGSTICK} models a run.

\item If you make multiple runs that crash or halt, the file {\tt timinginfo.bigstick} may become corrupted. {\tt BIGSTICK} may try to alert you to this fact. Try 
deleting  {\tt timinginfo.bigstick} and re-running. 

Most recently (version 7.9.12) a modified distribution algorithm mostly addresses this. ({\tt BIGSTICK} distributes work by assigning weights to different operations;
when faced with a greedy opbundle, the code now simply inflates the weighting by the excess memory requirement. This is not guaranteed to work universally, 
but has solved several previously intractable cases.) 

And remember--some problems may simply be too large, or too large for the machine available, no matter what.

\end{itemize}

\subsection{Modeling}

\label{modeling}
One menu option {\tt BIGSTICK} offers is modeling, or choice `{\tt m}' on the main menu. This will run mostly like a normal run, with the following differences:

\begin{itemize}

\item No interaction file information will be requested (although if three-body forces are enables, it will ask if you want to model the use of three-body forces);

\item  Prompt for mandatory information on fragments;

\item  Prompt for mandatory information on the number of MPI processes; information on the number of OpenMP threads is not needed;

\item  Prompt for the number of Lanczos vectors.

\end{itemize}

You can model a run using a different number of MPI processes than the modelled number.

The modeling option will calculate the distribution of work and data. This is useful because you can find out if the number of MPI processes requested 
is insufficient, or if {\tt BIGSTICK} can find a distribution solution at all. (In some rare cases the algorithm currently fails.)

\section{OpenMP}

{\tt BIGSTICK} uses OpenMP where it can, in particular in matvec. Unfortunately due to the nature of the problem, there are limitations to the speedup form 
OMP.  Because the matrix elements are very sparse, one tends to lose locality. 
Modern computers have at least three levels of storage: disk storage, RAM storage, and cache storage.  These three kinds of memory are increasingly 
close to the CPU and thus are increasingly faster; they are also increasing smaller in size. 
When data is fetched from disk or even from RAM, 
the CPU also fetches nearby data and leaves it in the cache. If the program needs that cached data next, it is handily nearby and thus faster to 
be accessed. Because of the highly nonlocal nature of the data, however, {\tt BIGSTICK} has trouble reaching maximum efficiency. While we continue to work on this issue, by the very nature of the sparse matrix this is difficult.  Some of the work we have carried out is described 
in \citet{shan2015parallel,shan2017locality}.

\section{Timing}

In order to improve efficiency, {\tt BIGSTICK} contains a number of built-in variables and routines for tracking and reporting timing.  
When running in serial, {\tt BIGSTICK} uses the FORTRAN routines {\tt date\_and\_time} or {\tt cpu\_time}. Unfortunately these do not provide very accurate
timing, on the order of 0.01 second, so some information is not accurate. When running in MPI, {\tt BIGSTICK} uses {\tt BMPI\_Wtime}, which is 
much more accurate. 

{\tt BIGSTICK}  will give an estimate of the time to run,
\begin{verbatim}
  Approximate time per iterations estimated :  2112  sec, or  35.2 min
\end{verbatim}
but keep in mind this is a rough estimate. This uses information in {\tt timinginfo.bigstick} which contains timing from previous runs. 
If you previously ran a similar problem, this estimate is likely reliable, but if the problem changes, or if you are using the default assumption, 
when {\tt timinginfo.bigstick} does not exist, then the results may vary.

\subsection{Mode times}

The main timing in {\tt BIGSTICK} is to measure the amount of time the code spends in various modes of operation, i.e., in generating the basis, 
computing jumps, matvec (matrix-vector multiplication), reorthognalization, and so on.  
At the end of a run, {\tt BIGSTICK} prints out the culmulative time. These times are written to the terminal as well as the {\tt .res}  results file.
The output looks something like this:
\begin{verbatim}
  Total time to run :    58.7889999998733     
  Time to compute basis :   3.999999724328518E-003
  Time to count up jumps :   1.099999994039536E-002
  Time to decouple matrix elements :   1.999999862164259E-003
  Time to compute jumps :   1.899999985471368E-002
  Time to compute lanczos :    42.0250000003725     
  Time total in H mat-vec multiply  :    30.9959999998100     
  Time to apply sp energies :   4.599999962374568E-002
  Time in pn :    13.1739999908023     
  Time in pn(back) :    8.17700000526384     
  Time in 2-body (pp) :    2.46199999982491     
  Time in 2-body (pp)(back) :    2.45600000442937     
  Time in 2-body (nn) :    4.66199999954551     
  Time in reorthogonalization :    10.8760000029579     
  Time to compute J^2, T^2 :   9.499999973922968E-002
  Time in applyobs :   0.950999999884516     
  Time spent diagonalizing. :   7.299999939277768E-002
\end{verbatim}

\subsection{Timing for parallel runs}

In addition to timing various modes during a run, {\tt BIGSTICK} provides timing data useful for load balancing MPI parallel runs. 
As discussed elsewhere, {\tt BIGSTICK} attempts to distribute work across MPI processes by counting up the number of 
operations and distributing the work. Operations are  managed by opbundles, and each opbundle is associated with a particular 
Hamiltonian mode:  proton-proton (PP), neutron-neutron (NN), proton-neutron (PN), and so on. 
Therefore {\tt BIGSTICK} tracks the time spent on each MPI process, on each Hamiltonian mode on each MPI process, and 
finally on each opbundle. 

\chapter{Recent additions}

In this brief chapter, we list new modifications to be integrated into the rest of the manual.

\section{Density matrix output}

As of 7.9.10, density matrices are now written out exclusive to the {\tt .dres} file. An explicit listing of the single particle orbitals is 
included at the beginning.


\appendix

\chapter{Matrix elements and operators}

\section{Reduced matrix elements}

\label{reduced}

The Wigner-Eckart theorem states that a matrix element which depends upon 
$J_z$ is proportional to a Clebsch-Gordan coefficient, that is, 
\begin{eqnarray}
\langle J_f M_f | \hat{O}_{K M} | J_i M_i \rangle 
= [J_f ]^{-1} ( J_i M_i,  K M | J_f M_f ) ( J_f || \hat{O}_K || J_i )
\label{WignerEckart} \\
= (-1)^{J_f -M_f} 
\left ( \begin{array}{ccc}
J_f & K & J_i \\
-M_f & M_K & M_i 
\end{array}
\right )( J_f || \hat{O}_K || J_i )
\nonumber 
\end{eqnarray}
where $( J_f || \hat{O}_K || J_i )$ is the \textit{reduced matrix element}, which 
encapuslates the fundamental matrix element independent of orientation, and 
which in \ref{transitionstrengthsfromdensities} is related to a sum over \textit{all} 
orientations. 

Eq.~(\ref{WignerEckart}) can also be thought of as the definition 
of the reduced matrix element (and the Wigner-Eckart theorem a statement 
that this definition is consistent using any set of $M$s). Note that it is possible to 
have a variant definition with different pre-factors, that is, the phase and factors 
like $\sqrt{2J_f+1}$ are conventions. Only the Clebsch-Gordan coefficients are dictated 
by the theorem. The choices of (\ref{WignerEckart}), taken from  \citet{edmonds1996angular}
 are the most widely used ones. 

The Wigner-Eckart theorem applies not just to angular momentum but any SU(2) algebra; 
hence one can reduce in isospin as well, and a \textit{doubly}-reduced matrix element
follows naturally:
\begin{eqnarray}
\label{doublyreduced}
\langle J_f M_f; T_f M_{Tf} | \hat{O}_{K M; T M_T} | J_i M_i; T_i M_{Ti} \rangle =
\\
 \frac{  ( J_i M_i,  K M | J_f M_f )} { \left [ J_f \right ]}
\frac{ (T_i M_{Ti},  T M_T | T_f M_{Tf} )}{  \left [T_f \right  ]}
 ( J_f, T_F || \hat{O}_{K,T} || J_i, T_i ). \nonumber
\end{eqnarray}

\section{The Hamiltonian and other operators in second quantization}

\label{ops2ndquant}

Here we carefully define our operators in second quantization, that is, using fermion creation and annihilation operators and coupled up to 
good angular momentum. 
To denote generic operators $\hat{\alpha}, \hat{\beta}$ coupled up to good total angular momentum $J$ and total $z$-component $M$, we use 
the notation 
\begin{equation}
( \hat{\alpha} \times \hat{\beta} )_{JM}
= \sum_{m_\alpha, m_\beta} (j_\alpha m_\alpha, j_\beta m_\beta | J M) \hat{\alpha}_{j_\alpha m_\alpha} \hat{\beta}_{j_\beta m_\beta} ,
\label{angmomcoupling}
\end{equation}
where $ (j_\alpha m_\alpha, j_\beta m_\beta | J M) $ is a Clebsch-Gordan coefficient (here and throughout we use the conventions of 
\citet{edmonds1996angular}). 

Hence we can define the general fermion pair creation operator
\begin{eqnarray}
\hat{A}^\dagger_{JM} (ab) = (\hat{a}^\dagger \times\hat{b}^\dagger)_{JM}
\label{Adagger}
\end{eqnarray}
with  two particles in orbits $a$ and $b$. 
We also introduce the time-reverse of $A^\dagger_{JM} (ab)$, the pair annihilation operator,
\begin{eqnarray}
\tilde{A}_{JM} (cd) = - (\tilde{c} \times \tilde{d})_{JM}
\label{tildeA}
\end{eqnarray}
Here we use the standard convention $\tilde{c}_{m_c}= (-1)^{j_c +m_c}\hat{c}_{-m_c}$, where $m_c$ is the $z$-component of angular momentum. 
An alternate notation is 
\begin{equation}
\hat{A}_{JM}(cd) = \left ( \hat{A}^\dagger_{JM}(cd) \right)^\dagger = (-1)^{J+M} \tilde{A}_{J,-M}(cd)
\end{equation}
A normalized pair operator is 
\begin{equation}
\frac{1}{\sqrt{1+\delta_{ab} }}\hat{A}^\dagger_{JM} (ab) 
\end{equation}

With this we can write down a standard form for any one- plus two-body Hamiltonian or Hamiltonian-like operator, which are angular momentum 
scalars.  To simplify we use
\begin{eqnarray}
\hat{H} = \sum_{ab} e_{ab} \hat{n}_{ab}  \nonumber \\
+ \frac{1}{4} \sum_{abcd}
\zeta_{ab} \zeta_{cd}
\sum_J V_J(ab,cd) \sum_{M} \hat{A}^\dagger_{JM}(ab) \hat{A}_{JM} (cd) , \label{hamdef}
\end{eqnarray} 
where $\hat{n}_{ab} = \sum_m \hat{a}^\dagger_m \hat{b}_m$ and $\zeta_{ab} = \sqrt{1+\delta_{ab}}$. 
Here $V_J(ab,cd) = \langle a b;J | \hat{V} | cd; J \rangle $ is the matrix element of the purely two-body part of $\hat{H}$ between normalized two-body states with good angular momentum $J$; because it is a 
scalar it is independent of the $z$-component $M$. 
To make our results as broadly interpretable as possible, we also write this as
\begin{eqnarray}
\sum_{ab} e_{ab} [j_a] \left ( \hat{a}^\dagger \times \tilde{b} \right )_{0,0} \nonumber \\
+ \frac{1}{4} \sum_{abcd} \zeta_{ab} \zeta_{cd}
 \sum_J V_J(ab,cd)\, [J]  \, \left (  \hat{A}^\dagger_{J}(ab) \times \tilde{A}_{J} (cd) \right)_{0,0}  
\end{eqnarray}
where we use the notatation $[x] = \sqrt{2x+1}$, which some authors write as $\hat{x}$; we use the former to avoid 
getting confused with operators which always are denoted by either $\hat{a}$ or $\tilde{a}$.

Finally we also can introduce one-body transition operators with good angular momentum rank $K$ and $z$-component of angular momentum $M$,
\begin{equation}
\hat{F}_{K,M} = \sum_{ab} F_{ab} \frac{1}{[K]}\left ( \hat{a}^\dagger \times \tilde{b} \right)_{K,M} \label{transopdef}
\end{equation}
Here $F_{ab} = \langle a || \hat{F}_K || b \rangle$ is the reduced one-body matrix element.  

\section{Symmetries of matrix elements}

Two-body matrix elements satisfy the following symmetries:
\begin{eqnarray}
V_J(ab,cd) = -(-1)^{j_a +j_b +J}V_J(ba,cd) \\ = -(-1)^{j_c +j_d +J}V_J(ab,dc)  \
\nonumber =
(-1)^{j_a +j_b +j_c + j_d}V_J(ba,dc).
\end{eqnarray}
Including isospin, 
\begin{eqnarray}
V_{JT}(ab,cd) = -(-1)^{j_a +j_b +J+1+T}V_{JT}(ba,cd) \\
\nonumber 
= -(-1)^{j_c +j_d +J+1+T}V_{JT}(ab,dc) =
(-1)^{j_a +j_b +j_c + j_d}V_{JT}(ba,dc).
\end{eqnarray}
Because we assume real-valued matrix elements, $V_{JT}(ab,cd) = V_{JT}(cd,ab)$.
Although internally {\tt BIGSTICK} has a specified order for storing matrix elements, 
the code can read in matrix elements in any order and with the indices $a,b,c,d$ in any 
order.

Non-scalar spherical tensors  should satisfy \citep{edmonds1996angular}:
\begin{equation}
\left( \hat{F}_{KM} \right)^\dagger = (-1)^{M} \hat{F}_{K,-M}.
\label{sphtensoradjoint}
\end{equation}

For non-charge-changing transitions, Eq.~(\ref{sphtensoradjoint}) implies $F_{ab} = (-1)^{j_a -j_b} F_{ba}^*$.

\chapter{A summary of options}

\label{menu}

{\tt BIGSTICK} is a menu-driven code, to make it easier for novices.
Here we write out the options (as of \textbf{version 7.11.4}), and point to where to find more information.

\section{Main menu}

When you initiate {\tt BIGSTICK}, the initial menu provides you with the most used options. 

\begin{verbatim}
 * * * * * * * * * * * * * * * * * * * * * * * * * * * * * * * * * * * * * * 
  *                                                                         * 
  *               OPTIONS (choose one)                                      * 
  * (i) Input automatically read from "autoinput.bigstick" file             * 
  *  (note: autoinput.bigstick file created with each nonauto run)          * 
  * (n) Compute spectrum (default); (ns) to suppress eigenvector write up   * 
  * (d) Densities: Compute spectrum + all one-body densities (isospin fmt)  * 
  * (2) Two-body density from previous wfn (default p-n format)             * 
  * (x) eXpectation value of a scalar Hamiltonian (from previous wfn)       * 
  * (o) Apply a one-body (transition) operator to previous wfn and write out* 
  * (s) Strength function (using starting pivot )                           * 
  * (g) Apply the resolvent 1/(E-H) to a previous wfn and write out         * 
  * (m) print information for Modeling parallel distribution                * 
  * (l) print license and copyright information                             * 
  * (?) Print out all options                                               * 
  *                                                                         * 
  * * * * * * * * * * * * * * * * * * * * * * * * * * * * * * * * * * * * * * 
    
\end{verbatim}

For more information on:

\noindent `{\tt(i)}' see Section \ref{autoinput};

\noindent `{\tt(n)}' see Section~\ref{normal} and throughout this manual;

\noindent `{\tt(d)}' see Sections~\ref{densitymatrix},\ref{densities} ;

\noindent `{\tt(2)}' see Section~\ref{2bdens} ;

\noindent `{\tt(x)}' see Section~\ref{expectation} ;

\noindent `{\tt(o)}' see Section~\ref{apply1body}, \ref{apply1bodybasic} ;

\noindent `{\tt(s)}' see Section~\ref{strength} ;

\noindent `{\tt(g)}' see Section~\ref{green};

\noindent `{\tt(m)}' see Section~\ref{modeling};

\noindent Option `{\tt(l)}' simply prints out the license information from Section \ref{CClicense}.

\bigskip

\bigskip

To give the full and exhaustive menu, enter `{\tt(?)}' to get:

\begin{verbatim}
 * * * * * * * * * * * * * * * * * * * * * * * * * * * * * * * * * * * * * * 
  *                                                                         * 
  *               OPTIONS (choose 1)                                        * 
  * (i) Input automatically read from "autoinput.bigstick" file             * 
  *  (note: autoinput.bigstick file created with each nonauto run)          * 
  * (n) Compute spectrum (default); (ns) to suppress eigenvector write up   * 
  * (ne) Compute energies but NO observable (i.e. J or T)                   * 
  * (np) Compute spectrum starting from prior pivot                         * 
  * (d) Densities: Compute spectrum + all one-body densities (isospin fmt)  * 
  * (dx[m]) Densities: Compute one-body densities from previous run (.wfn)  * 
  *     optional m enables mathematica output                               * 
  * (dp) Densities in proton-neutron format                                 * 
  * (dxp) Compute one-body densities from prior run (.wfn) in p-n format.   * 
  * (db) Write one-body densities to a binary file                          * 
  * (dxb) Compute one-body densities from prior run, write to a binary file * 
  * (2) Two-body density from previous wfn (default p-n format)             * 
  * (2d) Two-body density from previous wfn, only initial=final, Jt=0       * 
  * (2i) Two-body density from previous wfn ( isopin format)       * 
  * (3) Normal spectrum but using three-body forces (beta version)          * 
  * (p) Compute spectrum + single-particle occupations,(ps) to supress wfn  * 
  * (occ) single-particle occupations (from previous wfn)                   * 
  * (x) eXpectation value of a scalar Hamiltonian (from previous wfn)       * 
  * (o) Apply a one-body (transition) operator to previous wfn and write out* 
  * (s),(sn) Strength function (using starting pivot ) wfn out normalized   * 
  * (ss) Strength function (using starting pivot ), but no output wfn or J,T* 
  * (su) Strength function (using starting pivot ) wfn out unnormalized     * 
  * (sb) Strength function (using block of starting pivots )                * 
  * (sbs) Strength function (using block of starting pivots ) no wfn out    * 
  * (a) Apply a scalar Hamiltonian to a previous wfn and write out          * 
  * (h) Compute matrix elements of a scalar Hamiltonian (inputs as basis)  * 
  * (g) Apply resolvent 1/(E-H) to a previous wfn and write out             * 
  * (gv) Apply resolvent 1/(E-H) to a previous wfn, then take dot prod      * 
  * (gc) Apply resolvent 1/(E-H) to a previous wfn, E complex, and write out* 
  * (v) Overlap of initial states with final states                         * 
  * (pv) Read in previous vector, write out Lanczos coef, take dot prod     * 
  * (m) print information for Modeling parallel distribution                * 
  * (md) Modeling parallel distribution for 1-body densities                * 
  * (m0) Compute dimensions only                                            * 
  * (t) create TRDENS-readable file for post processing                     * 
  * (tx) create TRDENS-readable file for post processing from previous wfn  * 
  * (tw) from TRDENS-readable file create standard BIGSTICK .wfn file       * 
  * (b) Create binary file with full basis information (for postprocessing) * 
  * (ba) Create ASCII file with full basis information (for postprocessing) * 
  * (wh) write out Hamiltonian matrix to a file and stop                    * 
  * (wo) write out one-body transition matrix to a file and stop            * 
  * (co) Compute configurations (partitions)                                * 
  * (cx) Compute configurations (partitions) from prior wfn                 * 
  * (jp) Project states of good J from prior wfns and normalize             * 
  * (ro) Read in multiple files of wfns and orthonormalize                  * 
  * (ru) Read in multiple files of wfns but DO NOT orthonormalize           * 
  * (c) Compute traces                                                      * 
  * (l) print license and copyright information                             * 
  * (?) Print out all options                                               * 
  *                                                                         * 
  * * * * * * * * * * * * * * * * * * * * * * * * * * * * * * * * * * * * * * 
    
\end{verbatim}

\noindent `{\tt(ne)}', `{\tt(ns)}', see section~\ref{normal};

\noindent `{\tt(np)}',  see Sections~\ref{block}, \ref{postprocess}, \ref{bootstrap};

\noindent `{\tt(dx[m])}', `{\tt(dxp)}' see Section~\ref{densitymatrix};

\noindent `{\tt(db)}', `{\tt(dxb)}' see Section~\ref{density_binary};


\noindent `{\tt(p)}',  `{\tt(ps)}' see Section ~\ref{occ1b};

\noindent `{\tt(occ)}',  see Section~\ref{occ1b};;


\noindent `{\tt(sn)}', `{\tt(ss)}',  see Section~\ref{introstrength};

\noindent  `{\tt(sb)}', `{\tt(sbs)}'  see Section~\ref{introstrength}, \ref{blockstrength};

\noindent `{\tt(a)}',  see Section~\ref{apply2body};

\noindent `{\tt(h)}',  see Section~\ref{xme}

\noindent `{\tt(gv)}',`{\tt(gc)}', see Section~\ref{green};

\noindent `{\tt(v)}', see Section~\ref{overlap}; 

\noindent `{\tt(pv)}', see Section~\ref{green};

\noindent `{\tt(t)}', `{\tt(tw)}', `{\tt(tx)}', see Section~\ref{trdens}; 

\noindent `{\tt(b)}', `{\tt(ba)}', see Section~\ref{altbasis}; 

\noindent `{\tt(wh)}', `{\tt(wo)}', see Section~\ref{printham};

\noindent `{\tt(cv)}',`{\tt(cx)}' see Section~\ref{config};

\noindent `{\tt(jp)}', see Section~\ref{jproject};

\noindent `{\tt(ro)}', `{\tt(ru)}', see Section~\ref{combine} ; 

\noindent `{\tt(c)}', see Section~\ref{trace}.

\section{Diagonalization menu}

If one is finding eigenpairs, then 
after the initial setup, which include constructing the basis and reading in the Hamiltonian,  one chooses a method of finding the eigenpairs. See also 
Sec.~\label{diagonalmenu}.

\begin{verbatim}
  / ------------------------------------------------------------------------\ 
  |                                                                         | 
  |    DIAGONALIZATION OPTIONS (choose one)                                 | 
  |                                                                         | 
  | (ex) Exact and full diagonalization (use for small dimensions only)     | 
  |                                                                         | 
  | (ld) Lanczos with default convergence (STANDARD)                        | 
  | (lf) Lanczos with fixed (user-chosen) iterations                        | 
  | (lc) Lanczos with user-defined convergence                              | 
  |                                                                         | 
  | (bd) Block Lanczos with default convergence (STANDARD)                  | 
  | (bf) Block Lanczos with fixed (user-chosen) iterations                  | 
  | (bc) Block Lanczos with user-defined convergence                        | 
  |                                                                         | 
  | (td) Thick-restart Lanczos with default convergence                     | 
  | (tf) Thick-restart Lanczos with fixed iterations                        | 
  | (tc) Thick-restart Lanczos with user-defined convergence                | 
  | (tx) Thick-restart Lanczos targeting states near specified energy       | 
  | (tb) Thick-restart block Lanczos with default convergence               | 
  |                                                                         | 
  | (sk) Skip Lanczos (only used for timing set up)                         | 
  | (li) Lanczos iterations only, no further eigensolutions                 | 
  |                                                                         | 
  \ ------------------------------------------------------------------------/     
\end{verbatim}

\noindent `{\tt (ex)}' This option will use Householder to find all eigenpairs, although you can choose to write out only the lowest $N$.  Recommended for basis dimensions of small numbers ($< 100$), can work easily for up to dimensions of around 1000, and can be applied, with increasing time, up to dimensions $< 10,000$. 
Note that Householder scales like (dimension)$^3$.

\bigskip

The most common options to use are standard Lanczos (or vector Lanczos, to distinguish from block Lanczos).  

\medskip

\noindent `{\tt (ld)}' is the most common choice. For a discussion of the 
default convergence criteria, see Sec.~\ref{converge}.
\begin{verbatim}
  Enter nkeep, max # iterations for lanczos 
  (nkeep = # of states printed out )    
\end{verbatim}
One can also set a fixed number of iterations; this can be useful for timing purposes and for cases of tricky convergence.

\noindent `{\tt (lf)}'
\begin{verbatim}
  Enter nkeep, # iterations for lanczos  
  (nkeep = # of states printed out )    
\end{verbatim}
It is possible to choose your own convergence criteria. See also Sec.~\ref{converge}.

\noindent `{\tt (lc)}'
\begin{verbatim}
      Enter nkeep, max # iterations for lanczos      
\end{verbatim}
\begin{verbatim}
  Enter how many ADDITIONAL states for convergence test 
  ( Defaul t=            5 ; you may choose 0 )     
\end{verbatim}
\begin{verbatim}
     Enter one of the following choices for convergence control :
 (0) Average difference in energies between one iteration and the last; 
 (1) Max difference in energies between one iteration and the last; 
 (2) Average difference in wavefunctions between one iteration and the last; 
 (3) Min difference in wavefunctions between one iteration and the last;     
\end{verbatim}
\begin{verbatim}
  Enter desired tolerance 
 (default tol =  0.100E-02 )    
\end{verbatim}

\bigskip

The next set of options are block Lanczos. Instead of the Hamiltonian matrix acting on a single Lanczos vectors, it acts on a block of Lanczos vectors. This leads to efficiencies, as the cost of constructing a Hamiltonian matrix element is amortized across the application to more than one vector. On the other hand, if one starts with a random block, one needs more iterations than in standard Lanczos. The solution is to use bootstrapped block Lanczos (\cite{zbikowski2023bootstrapped}), reading in an approximate solution. To read in a block of vectors, use option `{\tt (p)}' in the initial menu.  For more details see Sec.~\ref{block}.

\bigskip

Yet another but useful alternative is thick-restart Lanczos(\cite{wu2000thick}), described in Sec.~\ref{thickrestart}.  Thick-restart requires more iterations, but the amount of storage for Lanczos vectors, as well as reorthogonalization time, is greatly reduced. This is particularly useful for very large dimension cases on limited systems.

\bigskip

The final options are rather specialized:

\noindent `{\tt (sk)}' This option is only used if you want to know how much time is being used in set-up.

\noindent `{\tt (li)}' Again, for timing purposes, will carry out Lanczos iterations but not find the eigenpairs. The Lanczos coefficients will be written out to a file with extension {\tt .lcoef}.
\begin{verbatim}
  Lanczos iterations ONLY to get Lanczos coefficients 
  Enter # iterations for lanczos     
\end{verbatim}

\chapter{Troubleshooting}

 {\tt BIGSTICK} is a large and complex code, designed to run flexibly on platforms from laptops up to leading-edge supercomputers. 
 While we have tried to make it as robust and user-friendly as practical, given that it is a code primarily for cutting-edge research, and only secondarily 
 for pedagogy, it is easy to make mistakes or get confused. 

\section{Overall}

\begin{itemize}

\item  Read this manual!  It contains much valuable information and many valuable hints.

\item Try the sample runs provided.

\item We strongly encourage you to try some small, simple cases and gradually build your way up.  No one goes directly from integrating $f(x)= x$ to performing 
contour integrals in the complex plane. By taking the time to run cases of increasing complexity you will build up your familiarity with the capabilities of {\tt BIGSTICK}.

\item Read the output \textit{carefully}. If it fails to recognize an input, or an input file, it will try to tell you.  Also read the log file (either {\tt XXX.log} or 
{\tt bigstick.logfile} depending whether or not you gave an output name {\tt XXX}).  The log file  also contains much valuable information and not infrequently warnings of problems or 
potential problems. 

\end{itemize}

\section{Compilation}
 
 \begin{itemize}
 
 \item Note that the default compiler is current the Intel {\tt ifort} compiler. This is the compiler for {\tt make serial}, {\tt make openmp}, etc.  There are compiler options 
 for {\tt gfortran} which is widely available, including {\tt make gfortran-openmp}.  To the best of our working knowledge, however, {\tt gfortran} does not have a straightforward
 implementation with MPI. 
 
 \item If you switch compiler options, i.e., go from {\tt make serial} to {\tt make openmp}, or from {\tt make gfortran} to {\tt make gfortran-openmp}, you 
 must recompile from scratch. Do a {\tt make clean} to remove all intermediate object and module files.  
 
 \end{itemize}
 
 \section{Inputs}
 
 \begin{itemize}
 
 \item \textbf{Interaction files}.  It is fairly easy to make a mistake entering the interaction file information. Remember that, broadly speaking, we have two main types of 
 interaction files, which we chose to follow the format of other, widely used codes: {\tt OXBASH/NuShell}-type files, and {\tt MFDn}-style files. In general, the 
  {\tt OXBASH/NuShell}-type files \textbf{must have a name with the extension} {\tt .int}, \textbf{but you only enter the name, not the extension}. Conversely, for 
   {\tt MFDn}-style files, \textbf{you must enter the full name, even if it has a .int extension}.  
   
   For example, if you have an interaction file {\tt usdb.int}, which is in {\tt OXBASH/NuShell}-type format, you must enter 
\begin{verbatim}
usdb
\end{verbatim}
when asked for the file name. If you enter {\tt usdb.int}, the code will attempt to interpret it as an {\tt MFDn}-format.  It is acceptable to have a {\tt MFDn}-file format 
with an extension, for example, {\tt MyLittleInteraction.int}, but you \textbf{must enter the full name for the file to be read correctly}.

\item Note that {\tt OXBASH-NuShell}-format files include isospin-format, and both normalized and unnormalized proton-neutron formats. See section \ref{detailinput} for more details.
{\tt OXBASH-NuShell}-format includes single-particle energies, while {\tt MFDn}-format do not. 

\item  For best results, make sure all input files refer to exactly the same single-particle space. In particular, the interaction file single particle spaces should match 
those from the {\tt .sps} file or the {\tt auto} option when defining the single-particle space. If you are not sure, read the output from {\tt BIGSTICK} as it attempts to 
read the file; it will tell you what it thinks the model space is.

 \end{itemize}
 
 \section{Large cases and parallel computing}
 
 \begin{itemize}
 
 \item  We encourage you to use the modeling option (option `{\tt m}') from the main menu before beginning a large parallel run, to determine the memory requirements. 
 You can run this in MPI mode itself, on a small number of processors, which for large no-core shell model calculations can speed up the modeling. 
 
 \item if you fail to get a distribution, either in modeling or in an actual run, try increasing the number of processors. You may also need to increase the memory 
 available, for example by using more OpenMP threads; this makes more memory available to each MPI process.  Also, you may try deleting the file {\tt timinginfo.bigstick}; this file keeps track of the timing of the code on a particular machine and problem, but can in some cases become corrupted. 
 
 \item The real computational burden for any configuration interaction code is not the basis dimension, but the number of nonzero matrix elements. This is roughly the 
 number of operations as computed by {\tt BIGSTICK}, though not exactly, as diagonal matrix elements can require multiple operations.  The factorization algorithm used
  by {\tt BIGSTICK} and similar codes reduces the memory burden relative to codes that store the nonzero matrix elements, such as {\tt MFDn}, 
  but the time for a matrix-vector multiplication is still the same. 
 
 \end{itemize}

\subsection{Proton-neutron imbalance}

{\tt BIGSTICK}'s efficiency stems from factorizing the proton-neutron partitioning. 
The downside is that if one has a large problem with a significant proton-neutron imbalance, the efficiency of {\tt BIGSTICK} is impaired and, for the largest cases, may not even run. So, for example, a no-core calculation of $^{29}$F, with $Z=9$ and $N=20$, at high $N_\mathrm{max}$ becomes bogged down.   We are continuing to research approaches to mitigate such problems, but the solution is not simple.  Unfortunately, at this time, simply adding MPI ranks does not provide an easy solution. 

If, for example, you are carrying out a single-species calculation, such as computing tin isotopes with a $^{100}$Sn core, {\tt BIGSTICK} cannot factorize the problem. Instead, effectively all matrix elements are stored in memory. (Technically, the problem is even worse, because diagonal matrix elements have many contributing operations; see Section \ref{particlehole}.) If you are doing a full-configuration calculation, however, one can mitigate the problem by artificially breaking up the model space using the weighting truncation (cf.~section \ref{truncation}).  Assign one of the single-particle orbits a different value of $W$ from the rest; allow for truncations, but then choose the maximum allowed truncation.  This is not a perfect solution, but it does seem to prevent certain error messages, notably about `non-contiguous introns' (a strategy to minimize the storage of jumps).  One should not overdo an artificial break-up of the space.  Using $W$ to break up the space slows down the code, and too fine-grained a break-up will make it run unacceptably slow, or not at all. Experience suggests a reasonable strategy is to select the orbital with the largest $j$ to have a different $W$.

{\tt BIGSTICK} is both flexible and complex, so the best solution is not always obvious.  In particular, valence-space calculations, where one has many particles in a limited single-particle space, are very different from no-core calculation, where one typically has a few particles in a much larger single-particle space.  You are encouraged to experiment with different configurations to find an appropriate one.

\subsection{Important clues in a failure}

Sometimes a run will fail. 
While experience is the best guide to what has gone wrong in a calculation, 
here are some things  you can look at.  An experienced computational physicist will know that the final error message is seldom the entire story, and that one typically needs to look upstream to discover the problem.

The two main failures are, first, the program crashes or otherwise does not run, and second, 
the results are strange in some way.

\begin{itemize}
    \item \textbf{Basis dimension}. Though not the most important 
quantity, certainly the basis dimension can give signals as to potential problems.  The basis dimension is always printed out:
\begin{verbatim}
  Total basis =             177070720    
\end{verbatim}
If the dimension is small, say fewer than a few hundred, it is 
better to avoid Lanczos and use ``exact'' (i.e., full) diagonalization. 
If the basis dimension is less than a few million, you do not need to 
use MPI.  (In any case, using many MPI ranks for small cases 
is counterproductive.) If the basis dimension is greater than a few tens of millions, MPI is often more efficient, and if greater than 
100 million, likely necessary. Note that you can quickly compute 
the basis dimension \textit{only} through the option `{\tt (m0)}.'
Furthermore, dimension of the order of ten billion starts to reach the limits of the code on modern supercomputers.

There is no rigid rule as to how many MPI ranks to assign. It depends upon many factor: how close is $N$ to $Z$ ($N=Z$ yields the most efficient factorization in terms of matrix element storage), the choice of many-body truncation (large `{\tt max excite}' tend to be 
less efficient), and so on.  Our recommendation is always to try some smaller cases first and work your way up to your target.

    \item \textbf{Jump storage}. More important than basis dimension is the storage of the data for non-zero Hamiltonian matrix elements. These are stored as `jumps' and again this information is output:
\begin{verbatim}
 RAM for jumps in storage    (total)  :   675.022 Mb     
\end{verbatim}
In MPI mode, the code will attempt to distribute these jumps. 
\begin{verbatim}
RAM for jumps in storage    (total)     :   32586.691 Mb 
 Max RAM for local storage of jumps     :    4693.353 Mb     
\end{verbatim}
{\tt BIGSTICK} sets a ceiling for how much memory per MPI rank will be assigned to jumps. This is the constant {\tt maxjumpmemory\_default} 
which is set in the file {\tt bmodules\_flags.f90}
\begin{verbatim}
   real    :: maxjumpmemory_default  = 64.0       
\end{verbatim}
You can change this (and recompile), but be sure you system has this much memory per MPI rank!  This is where hybrid MPI+OpenMP is particularly useful: you can assign multiple OpenMP cores (threads) to each MPI rank in order to build up sufficient memory.  Other solutions include 
changing the number of MPI ranks as well as changing the size of the fragments. In some cases, 
a slight change will allow the code to find a better distribution of the work.

\item \textbf{Bad values of $J$}. {\tt BIGSTICK} is an $M$-scheme code, which means that total $M$ or $J_z$ of the basis is fixed. Because it is assumed that the Hamiltonian is rotationally invariant--and, indeed, significant changes to the code would be necessary to violate this assumption--one can have simultaneous eigenstates of the Hamiltonian and of $\hat{J}^2$. In general, therefore, converged states should have ``good'' values of $J$, that is, integer for even numbers of particles and half-integer for odd-numbers. The only exception is when one has degeneracies, states of different $J$ with the same energy. ({\tt BIGSTICK} does not automatically 
separate there.) Unless one is working with, for example, an algebraic or schematic interaction, 
such as pure pairing/seniority or $Q \cdot Q$, this should not happen. 

    \item \textbf{Missing/wrong format Hamiltonian file}.   If the 
input Hamiltonian file is formatted incorrectly, clashes with the single-particle space, or simply not read in, you can get strange answers, such as non-integer/non-half-integer values of $J$.
This can particularly happen if there is a mismatch between the defined single-particle space 
and the Hamiltonian file, if one is using either the default `{\tt iso}' format (which is inherited from previous codes) or the similar proton-neutron formats `{\tt xpn}' or `{\tt upn}.'
(The `{\tt mfd}' format does  not have this problem, but it is still possible to have such 
files with matrix elements missing.) The code will try to check, and you should pay attention. You will see such information as
\begin{verbatim}
  * * NOTICE: I expect single-particles space with  3  orbits 
\end{verbatim}
and
\begin{verbatim}
  As a check, first two-body matrix element in list is   -1.3796   
\end{verbatim}

    \item \textbf{Too few/too many Lanczos iterations}.  If you have too few Lanczos iterations, the solutions may not converge. One should avoid requesting too many Lanczos iterations. For one, this can overwhelm the storage of Lanczos vectors. When you initiate Lanczos, you should see
\begin{verbatim}
  Internal storage of all lanczos vectors =    2.79135983E-02  Gb     
\end{verbatim}
this is the storage per MPI rank. You will have to gauge how much storage per MPI rank you have available. 

Another, rare situation can occur in cases with highly degenerate spectra. In that case, 
one can have the Lanczos parameter $\beta_i \approx 0$. This can lead to a divide-by-zero problem. (See section ~\ref{vectorlanczos} 
for more on the Lanczos algorithm.) {\tt BIGSTICK} tries to restart, but if $\beta_i$ is merely very small, and the ratio $\beta_i/ \beta_{i-1}$ is not in a designated range, it may fail to do. In that case the new Lanczos vector is incorrect and one can get very strange values. There is no simple solution for this; either carefully monitor the Lanczos procedure, or add a small random Hamiltonian to slightly split degenerate solutions.

\end{itemize}
 
 \section{If you want to contact us with a problem}
 
 We welcome feedback on {\tt BIGSTICK}, including bugs. If you are having difficulty, it is best to send us as much information as possible: send us a complete copy of the output (not 
 just an error message out of context--{\tt BIGSTICK} will often print out information which may be helpful), the log file (either {\tt XXX.log} or 
{\tt bigstick.logfile} depending whether or not you gave an output name {\tt XXX}), as well as your input files and the input commands or scripts you used. 
Most of the time the mistakes made are simple ones, arising from simply not understanding the inputs; as noted, this is a complex code, so it is easy to make a mistake.

\chapter{Glossary}

\label{glossary}

{\tt BIGSTICK} is a big code with complex algorithms, and in the code itself and while running, one can find some unusual terms of art.  While you do not need to know all these words, we explain here some of the specialized terms

\begin{itemize}
    \item \textbf{Jumps}. These are the basic data to enable reduced storage of the Hamiltonian. Initially devised by Caurier and collaborators, jump arrays store the action of particle-number conserving $n$-body operators, for $n=1$ ($\hat{a}^\dagger \hat{a}$),
    or 
    $n=2$ ($\hat{a}^\dagger \hat{a}^\dagger \hat{a} \hat{a}$). ($n=3$ 
    exists, but is currently under refurbishing.)  Jumps are used to reconstruct the Hamiltonian. For proton-proton or neutron-neutron interactions, jumps are just the matrix elements; but for proton-neutron interactions, one combines a proton one-body jump with a neutron one-body jump to get a two-body proton-neutron two-body interaction. This is the source of {\tt BIGSTICK}'s efficiency.

    \item \textbf{Fragments}. (Only used in MPI, that is, distributed memory.)  For large dimension spaces, one breaks up the active vectors for matrix-vector (or matrix-matrix) multiplication into fragments. Contrary to common assumptions, one should try to keep the fragments as large as possible; this makes the reconstruction of the Hamiltonian matrix elements from jumps more efficient.

    \item \textbf{Sterile orbitals}. A weight of 99 in the {\tt .sps} file signals a `sterile' orbital which is not used. This is a away to help define different proton and neutron valence spaces.

    \item \textbf{Species}. {\tt BIGSTICK} assumes fermions come in two distinguishable species, usually protons and neutrons, although it can also be spin-up and spin-down fermions.
    
\end{itemize}

Here are some additional terms which, while you are unlikely to ever need to know, help explain some of how {\tt BIGSTICK} works.

\begin{itemize}

\item \textbf{Opbundles}. Opbundles are collections of jumps between the same initial and 
final sectors.

    \item \textbf{Hops}.  These are similar to jumps, but are more primitive particle addition $\hat{a}^\dagger$ or removal $\hat{a}$. 
    Used to construct jumps. Hops are between \textit{haikus}.

    \item \textbf{Haikus}. The basis is constructed from occupation-number representations of Slater determinants. However the only actual storage is in `half-Slater determinants,' or \textit{haikus}. 
    One has `left' haikus, constructed from single-particle states 
    with $m < 0$, and `right' haikus, constructed from single-particle states with $m \geq 0$. (In the nuclear case, values of $m$ are half-integers, divided into protons and neutrons. In atomic cases, however, one has spin-up and spin-down particles, and then $m$ refers to the orbital angular momentum and is an integer. This latter is not used very much.)

    \item \textbf{Pieces}.  (Only used in MPI, that is, distributed memory.) When not active, that is, part of the Hamiltonian multiplication process, the Lanczos vectors are stored across multiple MPI ranks.  These vectors are broken up as `pieces.' The same piece of each Lanczos vector is stored on the same MPI rank. Pieces are much smaller than fragments and function differently.

    \item \textbf{Sectors}. A sector is a grouping of Slater determinants of a given species that have the same quantum numbers, namely $M$, parity, and $W$.
    
    \item \textbf{Blocks}. A block is a grouping of haikus (of a given species) that that the same quantum numbers, namely $M$, parity, $W$,
    and particle number.

\item \textbf{Conjugate}. {\tt BIGSTICK} achieves efficiency through quantum numbers which control how to combine data. For example, 
in generating the basis, the quantum numbers of the proton Slater determinants and the quantum numbers of the neutron Slater determinants must combine to a fixed result, such as total $M$, parity, or up to some maximum total $W$.  By grouping Slater determinants into sectors, and at a lower level, haikus into blocks, one can set up relatively simple and efficient loops. For a sector or block with some set of quantum numbers, the conjugate sectors or blocks are those that are allowed to combine with it. 
    
\end{itemize}

 \chapter{Highlighted references}
 
 \label{shellmodelrefs}
 
 There are a number of books and review articles on the configuration-interaction shell model.  We focus on those in nuclear physics. 
 One of the best, but nowadays difficult to 
 get, is  \citet{BG77}.  Some other useful references, in historical order, are \citet{de2013nuclear},
 \citet{towner1977shell}, \citet{lawson1980theory} (thorough, but be aware his phase conventions differ from most others), 
  \citet{talmi1993simple},  \citet{heyde1994nuclear},  \citet{suhonen2007nucleons}, and others.
A particular useful review article touching on many of the ideas here \citet{ca05}; the review article \citet{br88} is older but has useful information on applications of the shell model. 
The no-core shell model and other \textit{ab initio} methods are a rapidly evolving field, but good overviews of the topic are \citet{navratil2000large}
and \citet{barrett2013ab}.

For angular momentum coupling a widely used reference is the slim volume by \citet{edmonds1996angular}. If you can't find what you need in 
Edmonds, you can almost certainly find it in \citet{varshalovich1988quantum}. Sadly, neither are good pedagogical 
introductions to the topic of angular momentum algebra.

Several papers and conference proceedings describe our work on {\tt BIGSTICK}: 
\citep{BIGSTICK,shan2015parallel,shan2017locality,shan2018improving}, as well as this manual, whose original citation is \citep{johnson2018bigstick}.

Some groups besides ours  use  {\tt BIGSTICK} in their research; for some recent examples see ~\cite{kruppa2021entanglement,PhysRevC.104.044323,PhysRevC.104.054317,PhysRevC.105.055504,romero2022solving,haxton2024effective,PhysRevC.111.034317}.

\bibliographystyle{abbrvnat}
\bibliography{johnsonmaster}

\begin{thebibliography}{56}
\providecommand{\natexlab}[1]{#1}
\providecommand{\url}[1]{\texttt{#1}}
\expandafter\ifx\csname urlstyle\endcsname\relax
  \providecommand{\doi}[1]{doi: #1}\else
  \providecommand{\doi}{doi: \begingroup \urlstyle{rm}\Url}\fi

\bibitem[Andreozzi and Porrino(2001)]{AP01}
F.~Andreozzi and A.~Porrino.
\newblock \emph{J. Phys. G: Nucl. Part. Phys}, 27:\penalty0 845, 2001.

\bibitem[Barrett et~al.(2013)Barrett, Navr{\'a}til, and Vary]{barrett2013ab}
B.~R. Barrett, P.~Navr{\'a}til, and J.~P. Vary.
\newblock Ab initio no core shell model.
\newblock \emph{Progress in Particle and Nuclear Physics}, 69:\penalty0
  131--181, 2013.

\bibitem[Br\"okemeier et~al.(2025)Br\"okemeier, Hengstenberg, Keeble, Robin,
  Rocco, and Savage]{PhysRevC.111.034317}
F.~Br\"okemeier, S.~M. Hengstenberg, J.~W.~T. Keeble, C.~E.~P. Robin, F.~Rocco,
  and M.~J. Savage.
\newblock Quantum magic and multipartite entanglement in the structure of
  nuclei.
\newblock \emph{Phys. Rev. C}, 111:\penalty0 034317, Mar 2025.

\bibitem[Brown et~al.(1985)Brown, Etchegoyen, and Rae]{OXBASH}
B.~Brown, A.~Etchegoyen, and W.~Rae.
\newblock {Computer code OXBASH: the Oxford University-Buenos Aires-MSU shell
  model code}.
\newblock \emph{Michigan State University Cyclotron Laboratory Report No. 524},
  1985.

\bibitem[Brown and Rae(2014)]{NuShellX}
B.~A. Brown and W.~D.~M. Rae.
\newblock {The Shell-Model Code NuShellX@MSU}.
\newblock \emph{{Nuclear Data Sheets}}, 120:\penalty0 115--118, 2014.

\bibitem[Brown and Richter(2006)]{PhysRevC.74.034315}
B.~A. Brown and W.~A. Richter.
\newblock New ``usd'' hamiltonians for the $\mathit{sd}$ shell.
\newblock \emph{Phys. Rev. C}, 74:\penalty0 034315, Sep 2006.

\bibitem[Brown and Wildenthal(1988)]{br88}
B.~A. Brown and B.~H. Wildenthal.
\newblock Status of the nuclear shell model.
\newblock \emph{{Annual Review of Nuclear and Particle Science}}, 38:\penalty0
  29--66, 1988.

\bibitem[Brussard and Glaudemans(1977)]{BG77}
P.~Brussard and P.~Glaudemans.
\newblock \emph{Shell-model applications in nuclear spectroscopy}.
\newblock North-Holland Publishing Company, Amsterdam, 1977.

\bibitem[Caurier and Nowacki(1999)]{ANTOINE}
E.~Caurier and F.~Nowacki.
\newblock Present status of shell model techniques.
\newblock \emph{Acta Physica Polonica B}, 30:\penalty0 705--714, 1999.

\bibitem[Caurier et~al.(1999)Caurier, Mart\'{\i}nez-Pinedo, Nowacki, Poves,
  Retamosa, and Zuker]{PhysRevC.59.2033}
E.~Caurier, G.~Mart\'{\i}nez-Pinedo, F.~Nowacki, A.~Poves, J.~Retamosa, and
  A.~P. Zuker.
\newblock Full 0$\hbar\omega$ shell model calculation of the binding energies
  of the ${1f}_{7/2}$ nuclei.
\newblock \emph{Phys. Rev. C}, 59:\penalty0 2033--2039, Apr 1999.

\bibitem[Caurier et~al.(2005)Caurier, Martinez-Pinedo, Nowacki, Poves, and
  Zuker]{ca05}
E.~Caurier, G.~Martinez-Pinedo, F.~Nowacki, A.~Poves, and A.~P. Zuker.
\newblock The shell model as a unified view of nuclear structure.
\newblock \emph{{Reviews of Modern Physics}}, 77:\penalty0 427--488, 2005.

\bibitem[Cirigliano et~al.(2022)Cirigliano, Fuyuto, Ramsey-Musolf, and
  Rule]{PhysRevC.105.055504}
V.~Cirigliano, K.~Fuyuto, M.~J. Ramsey-Musolf, and E.~Rule.
\newblock Next-to-leading order scalar contributions to
  $\ensuremath{\mu}\ensuremath{\rightarrow}e$ conversion.
\newblock \emph{Phys. Rev. C}, 105:\penalty0 055504, May 2022.

\bibitem[Cook(1998)]{cook1998handbook}
D.~B. Cook.
\newblock \emph{Handbook of computational quantum chemistry}.
\newblock Oxford University Press, 1998.

\bibitem[De-Shalit and Talmi(2013)]{de2013nuclear}
A.~De-Shalit and I.~Talmi.
\newblock \emph{Nuclear shell theory}, volume~14.
\newblock Academic Press, 2013.

\bibitem[Draayer et~al.(2012)Draayer, Dytrych, Launey, and
  Langr]{draayer2012symmetry}
J.~Draayer, T.~Dytrych, K.~Launey, and D.~Langr.
\newblock Symmetry-adapted no-core shell model applications for light nuclei
  with qcd-inspired interactions.
\newblock \emph{Progress in Particle and Nuclear Physics}, 67\penalty0
  (2):\penalty0 516--520, 2012.

\bibitem[Edmonds(1996)]{edmonds1996angular}
A.~R. Edmonds.
\newblock \emph{Angular momentum in quantum mechanics}.
\newblock Princeton University Press, 1996.

\bibitem[Gloeckner and Lawson(1974)]{gloeckner1974spurious}
D.~Gloeckner and R.~Lawson.
\newblock Spurious center-of-mass motion.
\newblock \emph{Physics Letters B}, 53\penalty0 (4):\penalty0 313--318, 1974.

\bibitem[Gorton(2024)]{gorton2024shell}
O.~C. Gorton.
\newblock \emph{Shell Model Methods, Statistical Nuclear Reactions, and
  Beta-delayed Neutron Emission}.
\newblock PhD thesis, University of California, Irvine, 2024.

\bibitem[Haxton et~al.(2024)Haxton, McElvain, Menzo, Rule, and
  Zupan]{haxton2024effective}
W.~Haxton, K.~McElvain, T.~Menzo, E.~Rule, and J.~Zupan.
\newblock Effective theory tower for $\mu$→ e conversion.
\newblock \emph{Journal of High Energy Physics}, 2024\penalty0 (11):\penalty0
  1--59, 2024.

\bibitem[Heyde(1994)]{heyde1994nuclear}
K.~L. Heyde.
\newblock \emph{The nuclear shell model}.
\newblock Springer, 1994.

\bibitem[Jensen(2017)]{jensen2017introduction}
F.~Jensen.
\newblock \emph{Introduction to computational chemistry}.
\newblock John Wiley \& Sons, 2017.

\bibitem[Johnson(2015)]{PhysRevC.91.034313}
C.~W. Johnson.
\newblock Spin-orbit decomposition of \textit{ab initio} nuclear wave
  functions.
\newblock \emph{Phys. Rev. C}, 91:\penalty0 034313, Mar 2015.

\bibitem[Johnson et~al.(2013)Johnson, Ormand, and Krastev]{BIGSTICK}
C.~W. Johnson, W.~E. Ormand, and P.~G. Krastev.
\newblock Factorization in large-scale many-body calculations.
\newblock \emph{Computer Physics Communications}, 184:\penalty0 2761--2774,
  2013.

\bibitem[Johnson et~al.(2018)Johnson, Ormand, McElvain, and
  Shan]{johnson2018bigstick}
C.~W. Johnson, W.~E. Ormand, K.~S. McElvain, and H.~Shan.
\newblock Bigstick: A flexible configuration-interaction shell-model code.
\newblock \emph{arXiv preprint arXiv:1801.08432}, 2018.

\bibitem[Knowles and Handy(1984)]{knowles1984new}
P.~Knowles and N.~Handy.
\newblock A new determinant-based full configuration interaction method.
\newblock \emph{Chemical physics letters}, 111\penalty0 (4-5):\penalty0
  315--321, 1984.

\bibitem[Kruppa et~al.(2021)Kruppa, Kov{\'a}cs, Salamon, and
  Legeza]{kruppa2021entanglement}
A.~Kruppa, J.~Kov{\'a}cs, P.~Salamon, and {\"O}.~Legeza.
\newblock Entanglement and correlation in two-nucleon systems.
\newblock \emph{Journal of Physics G: Nuclear and Particle Physics},
  48\penalty0 (2):\penalty0 025107, 2021.

\bibitem[Lawson(1980)]{lawson1980theory}
R.~Lawson.
\newblock \emph{Theory of the nuclear shell model}.
\newblock Clarendon Press Oxford, 1980.

\bibitem[L\"owdin(1955)]{PhysRev.97.1474}
P.-O. L\"owdin.
\newblock Quantum theory of many-particle systems. {I}. {P}hysical
  interpretations by means of density matrices, natural spin-orbitals, and
  convergence problems in the method of configurational interaction.
\newblock \emph{Phys. Rev.}, 97:\penalty0 1474--1489, Mar 1955.

\bibitem[Navr{\'a}til et~al.(2000)Navr{\'a}til, Vary, and
  Barrett]{navratil2000large}
P.~Navr{\'a}til, J.~Vary, and B.~Barrett.
\newblock Large-basis ab initio no-core shell model and its application to 12
  c.
\newblock \emph{Physical Review C}, 62\penalty0 (5):\penalty0 054311, 2000.

\bibitem[Palumbo(1967)]{palumbo1967intrinsic}
F.~Palumbo.
\newblock Intrinsic motion and translational invariance in shell-model
  calculations.
\newblock \emph{Nuclear Physics A}, 99\penalty0 (1):\penalty0 100--112, 1967.

\bibitem[Palumbo and Prosperi(1968)]{palumbo1968effects}
F.~Palumbo and D.~Prosperi.
\newblock Effects of translational invariance violation in particle-hole
  calculations. application to 208pb.
\newblock \emph{Nuclear Physics A}, 115\penalty0 (2):\penalty0 296--308, 1968.

\bibitem[Papenbrock and Dean(2003)]{PD03}
T.~Papenbrock and D.~J. Dean.
\newblock Factorization of shell-model ground states.
\newblock \emph{Phys. Rev. C}, 67:\penalty0 051303, May 2003.

\bibitem[Papenbrock and Dean(2005)]{PD05}
T.~Papenbrock and D.~J. Dean.
\newblock Density matrix renormalization group and wavefunction factorization
  for nuclei.
\newblock \emph{Journal of Physics G: Nuclear and Particle Physics},
  31\penalty0 (8):\penalty0 S1377, 2005.

\bibitem[Papenbrock et~al.(2004)Papenbrock, Juodagalvis, and Dean]{PJD04}
T.~Papenbrock, A.~Juodagalvis, and D.~J. Dean.
\newblock Solution of large scale nuclear structure problems by wave function
  factorization.
\newblock \emph{Phys. Rev. C}, 69:\penalty0 024312, Feb 2004.

\bibitem[Parlett(1980)]{parlett1980symmetric}
B.~N. Parlett.
\newblock \emph{The symmetric eigenvalue problem}, volume~7.
\newblock SIAM, 1980.

\bibitem[Press et~al.(1992)Press, Flannery, Teukolsky, and
  Vetterling]{numericalrecipesfortran}
W.~H. Press, B.~Flannery, S.~Teukolsky, and W.~Vetterling.
\newblock \emph{Numerical recipes in fortran}.
\newblock Cambridge university press, 1992.

\bibitem[Rodkin and Tchuvil'sky(2021)]{PhysRevC.104.044323}
D.~M. Rodkin and Y.~M. Tchuvil'sky.
\newblock Detailed theoretical study of the decay properties of states in the
  $^{7}\mathrm{He}$ nucleus within an ab initio approach.
\newblock \emph{Phys. Rev. C}, 104:\penalty0 044323, Oct 2021.

\bibitem[Romero et~al.(2022)Romero, Engel, Tang, and
  Economou]{romero2022solving}
A.~Romero, J.~Engel, H.~L. Tang, and S.~E. Economou.
\newblock Solving nuclear structure problems with the adaptive variational
  quantum algorithm.
\newblock \emph{arXiv preprint arXiv:2203.01619}, 2022.

\bibitem[Romero et~al.(2021)Romero, Yao, Bally, Rodr\'{\i}guez, and
  Engel]{PhysRevC.104.054317}
A.~M. Romero, J.~M. Yao, B.~Bally, T.~R. Rodr\'{\i}guez, and J.~Engel.
\newblock Application of an efficient generator-coordinate subspace-selection
  algorithm to neutrinoless double-$\ensuremath{\beta}$ decay.
\newblock \emph{Phys. Rev. C}, 104:\penalty0 054317, Nov 2021.

\bibitem[Shan et~al.(2015)Shan, Williams, Johnson, McElvain, and
  Ormand]{shan2015parallel}
H.~Shan, S.~Williams, C.~Johnson, K.~McElvain, and W.~E. Ormand.
\newblock Parallel implementation and performance optimization of the
  configuration-interaction method.
\newblock In \emph{Proceedings of the International Conference for High
  Performance Computing, Networking, Storage and Analysis}, page~9. ACM, 2015.

\bibitem[Shan et~al.(2017)Shan, Williams, Johnson, and
  McElvain]{shan2017locality}
H.~Shan, S.~Williams, C.~Johnson, and K.~McElvain.
\newblock A locality-based threading algorithm for the
  configuration-interaction method.
\newblock 2017.
\newblock URL \url{http://escholarship.org/uc/item/9sf515zf}.

\bibitem[Shan et~al.(2018)Shan, Williams, and Johnson]{shan2018improving}
H.~Shan, S.~Williams, and C.~W. Johnson.
\newblock Improving mpi reduction performance for manycore architectures with
  openmp and data compression.
\newblock In \emph{2018 IEEE/ACM Performance Modeling, Benchmarking and
  Simulation of High Performance Computer Systems (PMBS)}, pages 1--11. IEEE,
  2018.

\bibitem[Shavitt(1998)]{Sh98}
I.~Shavitt.
\newblock The history and evolution of configuration interaction.
\newblock \emph{{Molecular Physics}}, 94:\penalty0 3--17, 1998.

\bibitem[Sherrill and Schaefer(1999)]{sherrill1999configuration}
C.~D. Sherrill and H.~F. Schaefer.
\newblock The configuration interaction method: Advances in highly correlated
  approaches.
\newblock \emph{Advances in quantum chemistry}, 34:\penalty0 143--269, 1999.

\bibitem[Shimizu(2013)]{shimizu2013nuclear}
N.~Shimizu.
\newblock Nuclear shell-model code for massive parallel computation," kshell".
\newblock \emph{arXiv preprint arXiv:1310.5431}, 2013.

\bibitem[Shimizu et~al.(2019)Shimizu, Mizusaki, Utsuno, and
  Tsunoda]{shimizu2019thick}
N.~Shimizu, T.~Mizusaki, Y.~Utsuno, and Y.~Tsunoda.
\newblock Thick-restart block {L}anczos method for large-scale shell-model
  calculations.
\newblock \emph{Computer Physics Communications}, 244:\penalty0 372--384, 2019.

\bibitem[Sternberg et~al.(2008)Sternberg, Ng, Yang, Maris, Vary, Sosonkina, and
  Le]{MFDN}
P.~Sternberg, E.~Ng, C.~Yang, P.~Maris, J.~Vary, M.~Sosonkina, and H.~V. Le.
\newblock {Accelerating configuration interaction calculations for nuclear
  structure}.
\newblock \emph{The Proceedings of the 2008 ACM/IEEE Conference on
  Supercomputing}, 2008.

\bibitem[Suhonen(2007)]{suhonen2007nucleons}
J.~Suhonen.
\newblock \emph{From Nucleons to Nucleus: Concepts of Microscopic Nuclear
  Theory}.
\newblock Springer Science \& Business Media, 2007.

\bibitem[Talmi(1993)]{talmi1993simple}
I.~Talmi.
\newblock \emph{Simple models of complex nuclei}.
\newblock CRC Press, 1993.

\bibitem[Toivanen(2006)]{Toi06}
J.~Toivanen.
\newblock Efficient matrix-vector products for large-scale nuclear shell-model
  calculations.
\newblock \emph{arXiv preprint arXiv:nucl-th/0610028}, 2006.

\bibitem[Towner(1977)]{towner1977shell}
I.~S. Towner.
\newblock \emph{A shell model description of light nuclei}.
\newblock Clarendon Press Oxford, 1977.

\bibitem[Varshalovich et~al.(1988)Varshalovich, Moskalev, and
  Khersonskii]{varshalovich1988quantum}
D.~A. Varshalovich, A.~N. Moskalev, and V.~K. Khersonskii.
\newblock \emph{Quantum theory of angular momentum}.
\newblock World scientific, 1988.

\bibitem[Weiss(1961)]{PhysRev.122.1826}
A.~W. Weiss.
\newblock Configuration interaction in simple atomic systems.
\newblock \emph{Phys. Rev.}, 122:\penalty0 1826--1836, Jun 1961.

\bibitem[Whitehead et~al.(1977)Whitehead, Watt, Cole, and Morrison]{lanczos}
R.~R. Whitehead, A.~Watt, B.~J. Cole, and I.~Morrison.
\newblock Computational methods for shell model calculations.
\newblock \emph{{Advances in Nuclear Physics}}, 9:\penalty0 123--176, 1977.

\bibitem[Wu and Simon(2000)]{wu2000thick}
K.~Wu and H.~Simon.
\newblock Thick-restart {L}anczos method for large symmetric eigenvalue
  problems.
\newblock \emph{SIAM Journal on Matrix Analysis and Applications}, 22\penalty0
  (2):\penalty0 602--616, 2000.

\bibitem[Zbikowski and Johnson(2023)]{zbikowski2023bootstrapped}
R.~M. Zbikowski and C.~W. Johnson.
\newblock Bootstrapped block {L}anczos for large-dimension eigenvalue problems.
\newblock \emph{Computer Physics Communications}, 291:\penalty0 108835, 2023.

\end{thebibliography}
\end{document}